\pdfoutput=1
\documentclass[11pt,twoside,a4paper,cmspaper,final,collab]{cms-tdr}
\def\svnVersion{40dd93f}\def\svnDate{2023/07/20}\def\cmsCernNoTag{CERN-EP-2023-019}\def\cmsCernDate{\today}\def\cmsMessage{Published in the Journal of High Energy Physics as \href{http://dx.doi.org/10.1007/JHEP07(2023)110}{\doi{10.1007/JHEP07(2023)110}.}}
\begin{document}\cmsNoteHeader{SUS-21-004}

\newcommand{\taul}{\ensuremath{\PGt_{\Pell}}\xspace}
\newcommand{\emu}{\ensuremath{\Pe\PGm}\xspace}
\newcommand{\mumu}{\ensuremath{\PGm\PGm}\xspace}
\newcommand{\tauhtauh}{\ensuremath{\tauh\tauh}\xspace}
\newcommand{\taultauh}{\ensuremath{\taul\tauh}\xspace}
\newcommand{\ltauh}{\ensuremath{\Pell\tauh}\xspace}
\newcommand{\etauh}{\ensuremath{\Pe\tauh}\xspace}
\newcommand{\mutauh}{\ensuremath{\PGm\tauh}\xspace}
\newcommand{\cmsTable}[1]{\resizebox{\textwidth}{!}{#1}}
\newcommand{\ST}{\ensuremath{S_{\mathrm{T}}}\xspace}
\newcommand{\tW}{\ensuremath{{\PQt}{\PW}}\xspace}
\newlength\cmsTabSkip\setlength{\cmsTabSkip}{1ex}
\newlength\cmsTabSkipSmall\setlength{\cmsTabSkipSmall}{0.1ex}
\newlength\cmsTabSkipLarge\setlength{\cmsTabSkipLarge}{4ex}
\DeclareRobustCommand{\PSGtpDo}{{\HepSusyParticle{\PGt}{1}{+}}\Xspace}
\DeclareRobustCommand{\PSGtpmDo}{{\HepSusyParticle{\PGt}{1}{\pm}}\Xspace}
\hyphenation{cal-or-i-me-ters}

\cmsNoteHeader{SUS-21-004}

\title{Search for top squark pair production in a final state with at least one hadronically decaying tau lepton in proton-proton collisions at \texorpdfstring{$\sqrt{s} = 13\TeV$}{sqrt(s) = 13 TeV}}

\date{\today}

\abstract{A search for pair production of the supersymmetric partner of the top quark, the top squark, in proton-proton collisions at $\sqrt{s} = 13\TeV$ is presented in final states containing at least one hadronically decaying tau lepton and large missing transverse momentum. This final state is highly sensitive to scenarios of supersymmetry in which the decay of the top squark to tau leptons is enhanced. The search uses a data sample corresponding to an integrated luminosity of 138\fbinv, which was recorded with the CMS detector during 2016--2018. No significant excess is observed with respect to the standard model predictions. Exclusion limits at 95\% confidence level on the masses of the top squark and the lightest neutralino are presented under the assumptions of simplified models. The results probe top squark masses up to 1150\GeV for a nearly massless neutralino. This search covers a relatively less explored parameter space in the context of supersymmetry, and the exclusion limit is the most stringent to date for the model considered here.
}

\hypersetup{
pdfauthor={CMS Collaboration},
pdftitle={Search for top squark pair production in a final state with at least one hadronically decaying tau lepton in proton-proton collisions at sqrt(s) = 13 TeV},
pdfsubject={CMS},
pdfkeywords={CMS, supersymmetry, top squark}}

\maketitle 

\section{Introduction}\label{sec:intro}
The standard model (SM) of particle physics is the most successful theoretical description of the fundamental particles of nature and their interactions.
However, it has various shortcomings.
Several theories have been proposed to address these deficiencies, among which Supersymmetry (SUSY)~\cite{Ramond:1971gb, Golfand:1971iw, Neveu:1971rx, Wess:1973kz, Fayet:1974pd, tHooft:1979rat, Kaul:1981hi, Nilles:1983ge, Martin:1997ns} is one of the most widely studied.
It assumes a new symmetry between bosons and fermions, thereby introducing a bosonic (fermionic) superpartner for every SM fermion (boson).
The fermionic superpartners of the SU(2)$\times$U(1) gauge fields of the SM, known as gauginos and higgsinos, are combined resulting in mass eigenstates that are referred to as charginos and neutralinos, or collectively as electroweakinos.
In $R$-parity conserving SUSY models~\cite{Farrar:1978xj}, the weakly interacting lightest neutralino \PSGczDo can be interpreted as a dark matter candidate.
The superpartners of the left- and right-handed top quarks are the top squarks, \PSQtL and \PSQtR, respectively.
The combination of these bosonic fields results in mass eigenstates \PSQtDo and \PSQtDt, where \PSQtDo is defined to be the lighter of the two.
The top squarks play an important role in stabilizing the Higgs boson (\PH) mass calculation by canceling the dominant top quark loop corrections~\cite{Witten:1981nf, Dimopoulos:1981zb, Sakai:1981gr}.
Depending on the mixing scenario~\cite{Hall:2011aa, Arbey:2011ab}, the mass of \PSQtDo can be within the reach of the CERN LHC for the top squark to effectively cancel the divergent contributions of the top quark to the Higgs boson mass.
Hence it is important to search for top squark production at the LHC.

The minimal supersymmetric standard model (MSSM) is the simplest SUSY extension of the SM, and it incorporates a wide variety of SUSY phenomenologies.
Both the gauge and Yukawa components~\cite{Martin:1997ns} of the chargino \PSGcpmDo and neutralino are involved in their interaction with fermion-sfermion pairs.
As a result, higgsino-like chargino and neutralino preferentially couple to third-generation fermion-sfermion pairs through the large Yukawa coupling.
Moreover, the Yukawa coupling to tau lepton-slepton pairs can be large for a high value of \tanb even if the higgsino component is relatively small.
Here the quantity \tanb is defined as the ratio of the vacuum expectation values of the two Higgs doublets in the MSSM.
For a large value of \tanb, the lighter state of the superpartner \PSGtDo of the tau lepton can be significantly less massive than the superpartners of the first and second generation leptons.
Consequently, the chargino decays predominantly as $ \PSGcpmDo \to \PSGtpmDo \PGnGt $ or $ \PGtpm \PSGnGt $, where \PSGnGt is the superpartner of the tau neutrino.
The decay probabilities of the electron and muon channels are thus greatly reduced~\cite{Baer:1997yi, Guchait:2002xh}.
Throughout this paper, charge conjugation symmetry is assumed and equal branching fractions are considered for \PSGcpDo decays to $\PSGtpDo\PGnGt$ and $ \PGtp \PSGnGt $.
Since the tau lepton decays to hadrons more often than to electrons and muons~\cite{Workman:2022ynf}, and the hadronic decay mode has lower background contribution relative to the signal, searches for SUSY signals in electron and muon channels are less sensitive to higgsino-like and high \tanb scenarios.
In this search, we focus on the signal of top squark pair production in a final state with two tau leptons, at least one of them decaying hadronically, probing part of the MSSM parameter space where the lightest charginos and neutralino preferentially couple to third-generation fermions.

\begin{figure}[!htbp]
	\centering
	
	\includegraphics[width=0.45\textwidth]{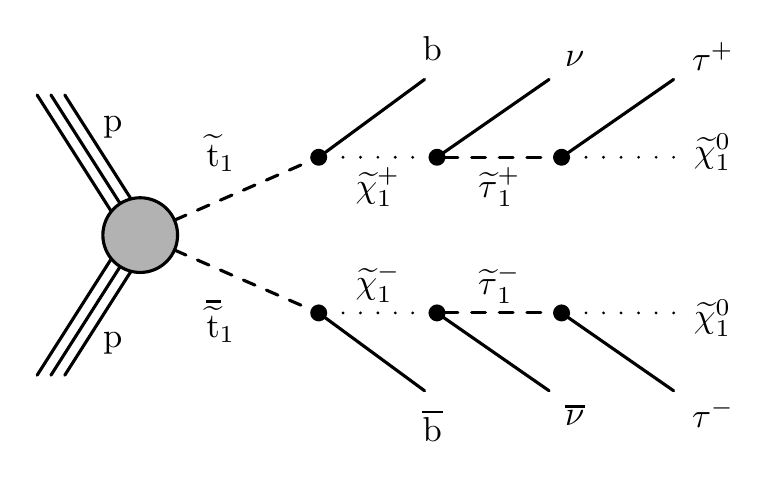}
	\includegraphics[width=0.45\textwidth]{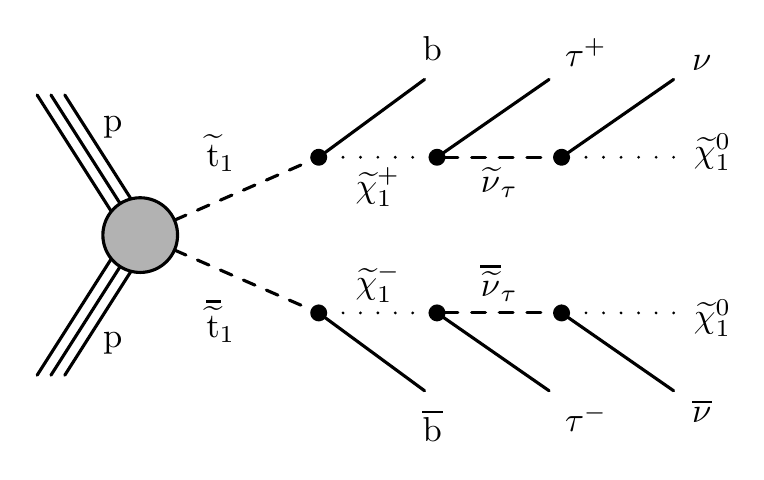} \\
	\includegraphics[width=0.45\textwidth]{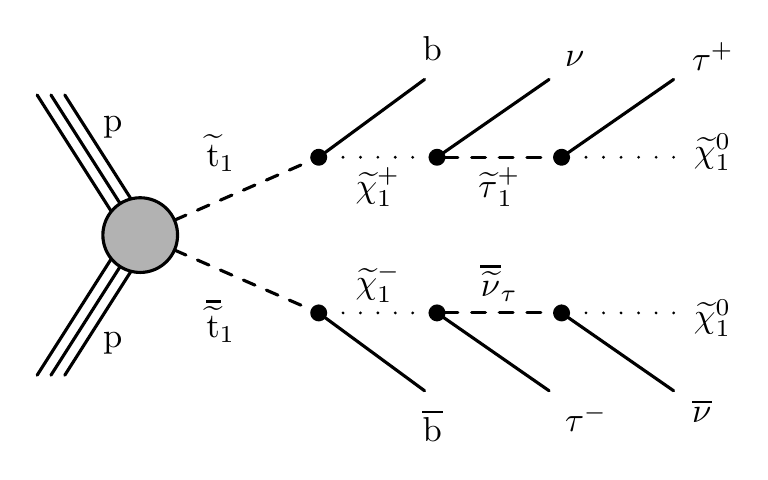}
	\includegraphics[width=0.45\textwidth]{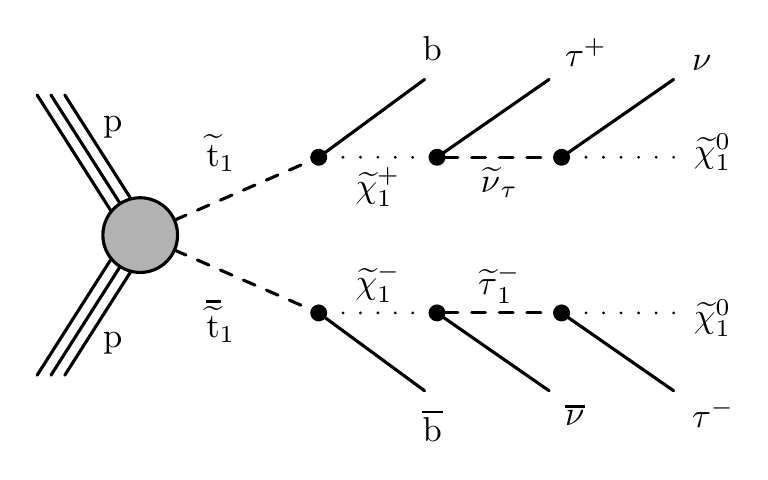}
	
	\caption{Diagrams of top squark pair production in $\Pp\Pp$ collisions at the LHC, and the decays that lead to final states with pairs of \PQb quarks and tau leptons accompanied by neutrinos and LSPs, within the framework of SMS.
	}
	\label{fig:sigFeyn}
\end{figure}

To address both the gauge hierarchy problem and the possibility of preferential couplings between electroweakinos and third-generation particles, we focus on the top squark decay chains
$\PSQtDo \to \PQb \PSGcpDo \to \PQb \PSGtpDo \PGnGt \to  \PQb \PGtp \PSGczDo \PGnGt$
~and~
$\PSQtDo \to \PQb \PSGcpDo \to \PQb \PGtp \PSGnGt \to \PQb \PGtp \PSGczDo \PGnGt$ and their charge conjugate reactions.
We assume $R$-parity conservation and consider \PSGczDo to be the lightest SUSY particle (LSP).
Being neutral and weakly interacting, \PSGczDo leaves no recorded signal in the detector.
The decay chains are depicted by the four diagrams in Fig.~\ref{fig:sigFeyn} within the framework of simplified model spectra (SMS)~\cite{Alwall:2008ag, Alves:2011wf}, where a branching ratio of 50\% is assumed for both $ \PSGcpDo \to \PSGtpDo\PGnGt$ and $ \PSGcpDo \to \PGtp\PSGnGt$.

This search is performed using proton-proton ($\Pp\Pp$) collision events at a center-of-mass energy of 13\TeV, recorded by the CMS experiment at the LHC.
The data sample corresponds to integrated luminosities of 36.3, 41.5, and 59.8\fbinv collected during the 2016, 2017, and 2018 operating periods of the LHC, respectively.
Signal-like events are characterized by the presence of at least one hadronically decaying tau lepton \tauh, jets identified as likely to have originated from the fragmentation of \PQb quarks, \ie, \PQb-tagged jets, and a large missing transverse momentum.
The other tau lepton decays either hadronically or leptonically, \taul, to an electron or a muon.
Events with both tau leptons decaying leptonically are not considered as they constitute about only 13\% of the final states.
The semileptonic final states are referred to as \etauh and \mutauh (or collectively as \ltauh) categories, and the fully hadronic final state as the \tauhtauh category.
Contributions from SM processes are estimated using a combination of Monte Carlo (MC) simulated samples and control samples in data.

Both the CMS~\cite{Sirunyan:2017xse, Sirunyan:2017leh, Chatrchyan:2013xna, Khachatryan:2016pup, Khachatryan:2016pxa, Sirunyan:2016jpr, Sirunyan:2017wif, Sirunyan:2017pjw} and ATLAS \cite{Aaboud:2017nfd, Aad:2015pfx, Aad:2014kra, Aad:2014qaa, Aaboud:2016lwz} Collaborations have performed searches for top squark pair production in leptonic and hadronic final states, establishing limits on top squark masses in the framework of SMS models.
The final state used in this search has also been studied by the ATLAS Collaboration~\cite{PhysRevD.98.032008, ATLAS:2021jyv}.
However the results are not directly comparable as the ATLAS searches were optimized for a gauge-mediated SUSY-breaking scenario, where the top squark decays as $\PSQtDo \to \PQb \PSGtpDo \PGnGt$, resulting in different kinematics of the final state particles.
A search performed by the CMS Collaboration~\cite{CMS:2019lrh} using 2016 and 2017 data has studied the same signal model as this study, albeit in the \tauhtauh final state alone.
This study expands upon the search in Ref.~\cite{CMS:2019lrh} by including the 2018 data and the \ltauh final states, and employs improved \tauh identification and \PQb tagging algorithms.
Since we interpret the results in the framework of simplified models, previous constraints from direct chargino and tau slepton production searches by LEP, and LHC experiments have not been imposed in this analysis.
The \textsc{HEPData} record for the analysis can be found in Ref.~\cite{hepdata}.

\section{The CMS detector}\label{sec:CMS_det}
The central feature of the CMS apparatus is a superconducting solenoid of 6\unit{m} internal diameter that provides a magnetic field of 3.8\unit{T}. Within the solenoid volume are a silicon pixel and strip tracker, a lead tungstate crystal electromagnetic calorimeter (ECAL), and a brass and scintillator hadron calorimeter (HCAL), each composed of a barrel and two endcap sections. Forward calorimeters extend the pseudorapidity ($\eta$) coverage provided by the barrel and endcap detectors. Muons are detected in gas-ionization chambers embedded in the steel flux-return yoke outside the solenoid.
A more detailed description of the CMS detector, together with a definition of the coordinate system used and the relevant kinematic variables, is reported in Ref.~\cite{Chatrchyan:2008zzk}.

Events of interest are selected using a two-tiered trigger system~\cite{Khachatryan:2016bia}. The first level, composed of custom hardware processors, uses information from the calorimeters and muon detectors to select events at a rate of around 100\unit{kHz} within a fixed latency of less than 4\mus~\cite{CMS:2020cmk}. The second level, known as the high-level trigger, consists of a farm of processors running a version of the full event reconstruction software optimized for fast processing, and it reduces the event rate to around 1\unit{kHz} before data storage.

\section{Event simulation}\label{sec:mc_sim}
Simulated samples are used to estimate several SM backgrounds as well as to predict signal rates.
The background and signal samples are generated with representative distributions of additional $\Pp\Pp$ interactions per bunch crossing, referred to as pileup.
These samples are produced for each year of data taking separately to account for different pileup and detector conditions in the three years.
Additionally, the simulated samples for each year are reweighted such that their pileup profiles match that measured in the data of the corresponding year.

For background processes, the \POWHEG~v2 \cite{Oleari:2010nx,Nason:2004rx,Frixione:2007vw,Alioli:2010xd,Frixione:2007nw} MC event generator is used for the pair production of top quarks ($\ttbar$) and the single top quark $t$-channel process, whereas \POWHEG~v1 \cite{Alioli_2009} is used for the {\PQt}{\PW} process. The \MGvATNLO (v2.2.2 for 2016, v2.4.2 for 2017 and 2018)~\cite{Alwall:2014hca} event generator is used at leading order (LO) for modeling the Drell--Yan+jets (DY+jets) and {\PW}+jets backgrounds;
these two LO MC samples are normalized to cross sections calculated with the \textsc{fewz} v3.1 program~\cite{Li:2012wna} at NNLO order in pQCD.
The \MGvATNLO event generator is also used to simulate the diboson (VV and VH) and $\ttbar$V (V = \PW or \PZ) processes at NLO in pQCD.
For the 2016 analysis, the parton showering and hadronization are simulated with \PYTHIA~v8.212 \cite{Sjostrand:2014zea}.
All samples use the CUETP8M1~\cite{CMS:2015wcf} underlying event tune, except for \ttbar simulation, which uses the CUETP8M2T4~\cite{CMS-PAS-TOP-16-021} tune. For the 2017 and 2018 analyses, \PYTHIA~v8.230 with the CP5 \cite{CMS:2019csb} tune is used. The CMS detector response is modeled using \GEANTfour \cite{GEANT4:2002zbu}, and simulated events are then reconstructed in the same way as collision data.

The signal is simulated based on simplified SUSY models.
The signal process of top squark production, shown in Fig.~\ref{fig:sigFeyn}, is simulated at LO using \MGvATNLO  followed by \PYTHIA v8.212 with the tune CUETP8M1 for 2016 and tune CP2 \cite{CMS:2019csb} for the 2017 and 2018 analyses.
The signal cross sections are evaluated using NNLO plus next-to-leading logarithmic (NLL) calculations in QCD~\cite{bib-nlo-nll-01, bib-nlo-nll-02, bib-nlo-nll-03, bib-nlo-nll-04, bib-nlo-nll-05}.
The detector response for the signal sample is simulated using the fast CMS detector simulation (\textsc{FastSim})~\cite{Giammanco:2014bza}.
For all simulated signal and background events, small discrepancies observed between simulation and data are corrected by adding several scale factors, as discussed in Section~\ref{sec:syst_unc}.
Additional corrections are applied to the signal to account for differences between \textsc{FastSim} and \GEANTfour simulations.

\begin{figure}[!htbp]
	\centering
	
	\includegraphics[width=0.5\textwidth]{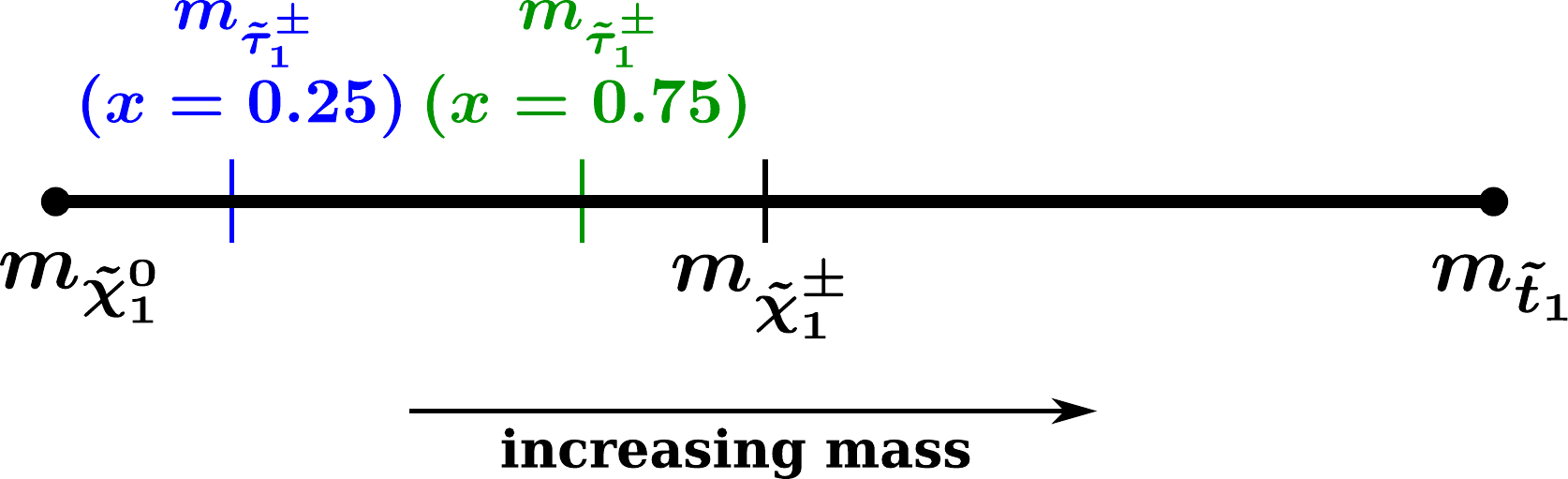}
	
	\caption{A graphical representation of the mass parameterization described in Eq.~\ref{eq:mass}.}
	\label{fig:sigMassParam}
\end{figure}

We assume a branching fraction of 50\% for each of the two decay modes of the chargino, $ \PSGcpDo \to \PSGtpDo\PGnGt$ and $ \PSGcpDo \to \PGtp\PSGnGt$. Each of the four diagrams in Fig.~\ref{fig:sigFeyn} therefore contributes to 25\% of the generated signal events.
The masses of SUSY particles appearing in the decay chain are parameterized as
\begin{equation}
\begin{aligned}
m_{\PSGcmDo} - m_{\PSGczDo} &= 0.5\ (m_{\PSQtDo} - m_{\PSGczDo} ) , \\
m_{\PSGtDo} - m_{\PSGczDo} &= x\ (m_{\PSGcmDo} - m_{\PSGczDo} ) , \\
\text{where } x &\in [0.25,\ 0.5,\ 0.75] , \\
\text{and } m_{\PSGnGt} & = m_{\PSGtDo} .
\end{aligned}
\label{eq:mass}
\end{equation}
In this parameterization, the chargino mass is fixed to have the mean of the top squark and $\PSGczDo$ masses. The masses of the leptonic superpartners are set by the value of $ x $ for a given pair of top squark and $\PSGczDo$ masses. A graphical representation of the mass parameterization is presented in Fig.~\ref{fig:sigMassParam}.

Therefore, the kinematic properties of the final state particles in each of the decay chains depicted in Fig.~\ref{fig:sigFeyn} depend on the choice of $x$:
\begin{itemize}
	
	\item $ x = 0.25 $: The mass of the lepton superpartner is
	closer to that of the \PSGczDo than to that of the
	\PSGcpmDo. Hence, the upper left diagram in
	Fig.~\ref{fig:sigFeyn} produces tau leptons with lower energy with respect to the ones produced in the  upper right. The lower two diagrams both typically
	produce two tau leptons with a large difference in energy.
	
	\item $ x = 0.75 $: The masses of the \PSGtpmDo
	and the \PSGcpmDo are relatively close, so the upper left
	diagram in Fig.~\ref{fig:sigFeyn} produces more energetic
	tau leptons than the upper right. The lower two diagrams
	produce the same energy asymmetry as in the case of $ x=0.25 $.
	
	\item $ x = 0.5 $: The tau leptons in all four diagrams have similar energies.
\end{itemize}

In fact, when all four diagrams are taken into account, the distributions of the kinematic properties are found to be very similar for the three different values of $ x $, for a given set of chargino and LSP masses.
It is important to note, however, that parameterizing the chargino mass to a point other than halfway between the $\PSQtDo$ and $\PSGczDo$ masses does affect the overall sensitivity of this search.
These cases are not explored in this paper.

\section{Event reconstruction}
\label{sec:evt_reco}

The particle-flow (PF) algorithm~\cite{CMS-PRF-14-001} reconstructs each individual particle in an event, with an optimized combination of information from various components of the CMS detector.
The energies of electrons and photons are measured in the ECAL.
The momentum of electrons is determined by a combined measurement of the track momentum in the tracker, the energy of the matching ECAL supercluster, and the energy of all bremsstrahlung photons consistent with originating from the track.
The momentum of muons is obtained from the bending of the corresponding tracks in the tracker and the muon spectrometer, which comprises three technologies: drift tubes, cathode strip chambers, and resistive-plate chambers.
The energy of charged hadrons, which is corrected for zero-suppression effects and for the response function of the calorimeters to hadronic showers, is determined from a combination of the momentum measured in the tracker and the energy of the matching ECAL and HCAL clusters.
Finally, the energy of neutral hadrons is obtained from the corrected energies in the corresponding ECAL and HCAL clusters.

Vertices reconstructed in an event are required to be within 24\unit{cm} of the center of the detector in the $ z $ direction (along the beam), and to have a transverse displacement from the beam line of less than 2\unit{cm}.
The primary vertex (PV) is taken to be the vertex corresponding to the hardest scattering in the event, evaluated using tracking information alone, as described in Section 9.4.1 of Ref.~\cite{CMS-TDR-15-02}.

Reconstruction of jets is performed by clustering PF objects using the anti-\kt algorithm~\cite{Cacciari:2008gp,Cacciari:2011ma} with a distance parameter of $R=0.4$.
Jet momentum is determined as the vector sum of all particle momenta in the jet, and is typically within 5--10\% of the momentum of the particle level jet over the entire \pt spectrum within the detector acceptance.
Pileup interactions contribute to spurious tracks and calorimetric energy deposits, increasing the apparent jet momentum.
To mitigate this effect, tracks identified as originating from pileup vertices are discarded, and an offset is applied to correct for the
remaining contributions~\cite{Khachatryan:2016kdb}.
Jets are calibrated using information from both simulation and data~\cite{Khachatryan:2016kdb}.
Additional selection criteria are applied to remove jets that are potentially dominated by instrumental effects or reconstruction failures~\cite{CMS-PAS-JME-16-003}.
Only jets with $\pt>20\GeV$ and $\abs{\eta}<2.4$ are considered in this analysis.

Jets originating from the fragmentation of \PQb quarks are identified as \PQb-tagged jets \cite{CMS:2017wtu} using the \textsc{DeepJet} algorithm \cite{CMS-DP-2018-058, Bols:2020bkb}. The algorithm employs properties of reconstructed secondary vertices and charged and neutral particle constituents of the jet as inputs to a convolutional deep neural network.
The ``medium" (``loose") selection or working point (WP) of this algorithm corresponds to a signal efficiency of about 80 (90)\%, with a mistagging probability of about 1 (10)\% for light jets (from gluons and up, down and strange quarks) and about 11 (50)\% for jets originating from charm quarks.	
The medium \textsc{DeepJet} WP is used to identify \PQb-tagged jets in the search regions, whereas the loose WP is used to veto events in control regions (CRs), as described in Section~\ref{sec:fakeBkg_LepTauh}.

Electrons are identified using the ``tight" WP of a boosted decision tree  algorithm~\cite{CMS:2020uim} that uses inputs based on the spatial distribution of the shower, track--cluster matching criteria, and consistency between the cluster energy and the track momentum.
This WP corresponds to a signal efficiency of 80\%, with a mistagging probability of about 1.0 (1.8)\% for hadrons in the barrel (endcaps).
The relative energy resolution ranges 0.8--5.2\% for electrons with \pt between 10 and 300\GeV; it is generally better in the barrel region than in the endcaps, and also depends on the bremsstrahlung energy emitted by the electron as it traverses the material in front of the ECAL~\cite{CMS:2020uim, Khachatryan:2015hwa}.
Only electrons with $ \pt > 30 (36)\GeV$ and $ \abs{\eta} < 2.4 $ are considered in this analysis of the 2016 (2017 and 2018) samples.
The stricter requirement on the electron \pt in 2017 and 2018 samples is because of increased pileup in those years, which necessitated a higher single-electron trigger threshold.

Muon reconstruction uses a global fit combining information from the tracker and muon spectrometers.
Muon candidates are required to pass the ``medium" WP of the algorithm that uses criteria on the geometrical matching between the tracks in the tracker and the muon spectrometers, and on the quality of the global fit.
This WP corresponds to a signal efficiency of more than 98\%, with a misidentification probability of about 0.15 (0.40)\% for pions (kaons)~\cite{Sirunyan:2018_1804.04528}.
Muons with \pt between 2 and 100\GeV, matched to the tracks measured in the silicon tracker, results in a \pt resolution of 1\% in the barrel and 3\% in the endcaps.
The muons are measured with a \pt resolution better than 7\% in the barrel with a \pt of up to 1\TeV~\cite{Sirunyan:2018_1804.04528}.

The present search considers only muons with $ \pt > 28 \GeV$ and $ \abs{\eta} < 2.4 $.
The requirement on the muon \pt is determined by the single-muon trigger threshold.

Isolation criteria are imposed on the lepton (electron and muon) candidates to reject leptons originating from hadronic decays.
The isolation variable used for this purpose is defined as the scalar \pt sum of reconstructed charged and neutral particles, excluding the lepton candidate, within a
cone of radius $ \Delta R = \sqrt{\smash[b]{(\Delta\eta)^2+(\Delta\phi)^2}}$ = 0.3\ (0.4) around the electron (muon) candidate track, divided by the \pt of the lepton candidate, where $\phi$ is the azimuthal angle in radians.
Charged particles not originating from the primary vertex are excluded from this sum, and a correction is applied to account for the neutral components originating from pileup, as described in Ref.~\cite{Khachatryan:2015hwa}.
This relative isolation is required to be less than 15 (20)\% for electrons (muons).

Hadronic tau lepton candidates are reconstructed from one charged hadron and up to two neutral pions, or three charged hadrons and up to one neutral pion, consistent with originating from the decay of a tau lepton, using the hadrons-plus-strips algorithm~\cite{Sirunyan:CMS-TAU-16-003}.
To distinguish between jets originating from quarks or gluons and genuine hadronic tau lepton decays, the discriminant of a deep neural network  algorithm called \textsc{DeepTau}~\cite{CMS:2022prd} is used.
The \tauh candidates are selected with the ``tight" WP of the above discriminant, which has an efficiency of ${\approx}$60\% and a misidentification probability of ${\approx}$0.5\%.
The ``loose" (``very loose") WP, which has an efficiency of ${\approx}$80 (${\approx}$90)\% and a misidentification probability of ${\approx}$1.5 (${\approx}$3.5)\% is used for estimating the background from misidentified \tauh candidates in the \tauhtauh ($\ltauh$) category.
In this analysis, only \tauh with $\pt>30$ (40)\GeV and $\abs{\eta}<2.3$ (2.1) are used for the $\ltauh$ (\tauhtauh) category. The stricter requirement on the \pt and $\abs\eta$ of the \tauh candidate in the \tauhtauh category is because of a higher double-\tauh trigger threshold.

The missing transverse momentum vector, \ptvecmiss, is computed as the negative vector \ptvec sum of all the PF objects in an event, and its magnitude is denoted as \ptmiss~\cite{Sirunyan:2019kia}.
The \ptvecmiss is modified to account for the energy calibration of all the PF candidates in an event, clustered into jets or not.

\section{Event selection}
\label{sec:evt_sel}
The sources of \ptmiss in the signal events are the neutrinos and the weakly interacting neutralinos, whose kinematic properties are correlated with those of the visible objects (in particular the \tauh and $\taul$ candidates).
In contrast, \ptmiss in the SM background processes is primarily due to neutrinos.
This difference can be exploited by first constructing the transverse mass \mT, defined as follows:
\begin{linenomath}
	\begin{equation}
	\begin{aligned}
	\mT^{2}(\vec{\pt}^\text{vis},\vec{\pt}^\text{inv}) &=
	m^{2}_\text{vis} + m^{2}_\text{inv} + 2(\et^\text{vis} \et^\text{inv}
	- \vec{\pt}^\text{vis} \cdot \vec{\pt}^\text{inv}) , \\
	\end{aligned}
	\label{eq:mT}
	\end{equation}
\end{linenomath}
where $\et^{2} = m^{2} + \pt^{2}$ for either visible or invisible particles.
Here the masses of the visible (vis) and invisible (inv) particles are denoted by $ m_{\text{vis}} $ and $ m_{\text{inv}} $, respectively.
The value of \mT has a maximum at the mass of the parent of the visible and the invisible particles when there is only one source of missing momentum in the system.
To account for pair-produced particles where both have visible and invisible decay products, the ``stransverse mass (\mTii)" \cite{Lester:1999tx, Barr:2003rg} is defined as:
\begin{linenomath}
	\begin{equation}
	\begin{aligned}
	\mTii^{2} (\text{vis1}, \text{vis2}, \ptmiss) &= \min_{\vec{\pt}^\text{\text{inv1}} + \vec{\pt}^\text{\text{inv2}} = \ptvecmiss} [\max \{ \mT^{2}(\vec{\pt}^\text{\text{vis1}},\vec{\pt}^\text{\text{inv1}}), \mT^{2}(\vec{\pt}^\text{\text{vis2}},\vec{\pt}^\text{\text{inv2}}) \}] .
	\end{aligned}
	\label{eq:mT2}
	\end{equation}
\end{linenomath}
Since the momenta of the individual invisible particles in Eq.~(\ref{eq:mT2}) are unknown, \ptvecmiss is divided into two components ($ \vec{\pt}^\text{\text{inv1}} $ and $\vec{\pt}^\text{\text{inv2}} $) in such a way that the value of \mTii is minimized.
If \mTii is computed using the two $\tauh$ candidates (or the $\taul$ and $\tauh$ candidates for the $\ltauh$ category) as the visible objects, ``vis1'' and ``vis2'', then its upper limit in the signal will be at the chargino mass.
This is different from the SM background processes.
For example in \ttbar events, the upper limit is at the \PW boson mass.
In searches where the masses of the invisible particles are unknown, the calculation of \mTii requires an assumption on their masses.
For this analysis, it was chosen to consider them as massless~\cite{Barr:2009wu}.

The signal and background processes can be further separated by utilizing the total visible momentum of the system.
This is characterized using the quantity \HT for the \tauhtauh category, defined as the scalar sum of the \pt of all jets and the \tauh candidates in the event.
Jets lying within a cone of $ \Delta R = 0.3 $ around either of the two selected $\tauh$ candidates are excluded from this sum to avoid double counting.
Since \HT is a measure of the visible transverse momentum of the system, it is sensitive to the mass of the top squark.
For the $\ltauh$ category, we construct an analogous quantity \ST, which includes the additional contribution from the lepton \pt.

Events in the \tauhtauh category are selected using $\tauhtauh$ triggers where both \tauh candidates are required to have $\abs{\eta} < 2.1$, and $\pt > 35$ or 40\GeV depending on the trigger logic.
The \tauh\tauh trigger has an efficiency of ${\approx}$95\% for \tauh candidates that pass the offline selection.
For the \etauh (\mutauh) category, single-electron (single-muon) trigger is used, with the trigger efficiency of ${\sim}90$ (${\sim}95$)\% for electron (muon) candidates that pass the offline selection.
The single-electron and single-muon triggers have \pt thresholds of 27 (34)\GeV and 24 (27)\GeV respectively, in 2016 (2017 and 2018).
The triggers described above are emulated using simulation, and the efficiencies measured therein are corrected to match those measured in the data.

For the offline selection, events are required to have $\ptmiss> 50$\GeV, $\HT > 100$\GeV (for the \tauhtauh category only), $\ST > 100$\GeV  (for the $\ltauh$ category only), and at least one \PQb-tagged jet with $\abs{\eta} < 2.4$ and $\pt > 25$ (20)\GeV for the $\ltauh$ (\tauhtauh) category.
The $\etauh$ ($\mutauh$) categories require exactly one electron (muon) and exactly one \tauh of opposite-sign charge, while the \tauhtauh category requires two \tauh of opposite-sign charges.
Additionally, events in the \tauhtauh category having $40 < m(\tauhtauh) < 90\GeV$ are vetoed in order to suppress the contribution from DY+jets events.
Here $m(\tauhtauh)$ is the invariant mass of the two \tauh candidate system.
Events in the $\ltauh$ category are vetoed if they have any extra \Pe, \PGm, or \tauh to avoid any overlap between the $\etauh$, $\mutauh$, and $\tauhtauh$ categories.
This veto also helps to reduce the contribution from rare SM background processes like VV and $\ttbar$V.
The requirements on \ptmiss and the number of \PQb-tagged jets ($ n_{\PQb} $) help to reduce the contributions from DY+jets and SM events comprised uniquely of jets produced through the strong interaction, referred to as multijet events.

\begin{figure}[!htbp]
	\centering
	
	\includegraphics[width=0.495\textwidth]{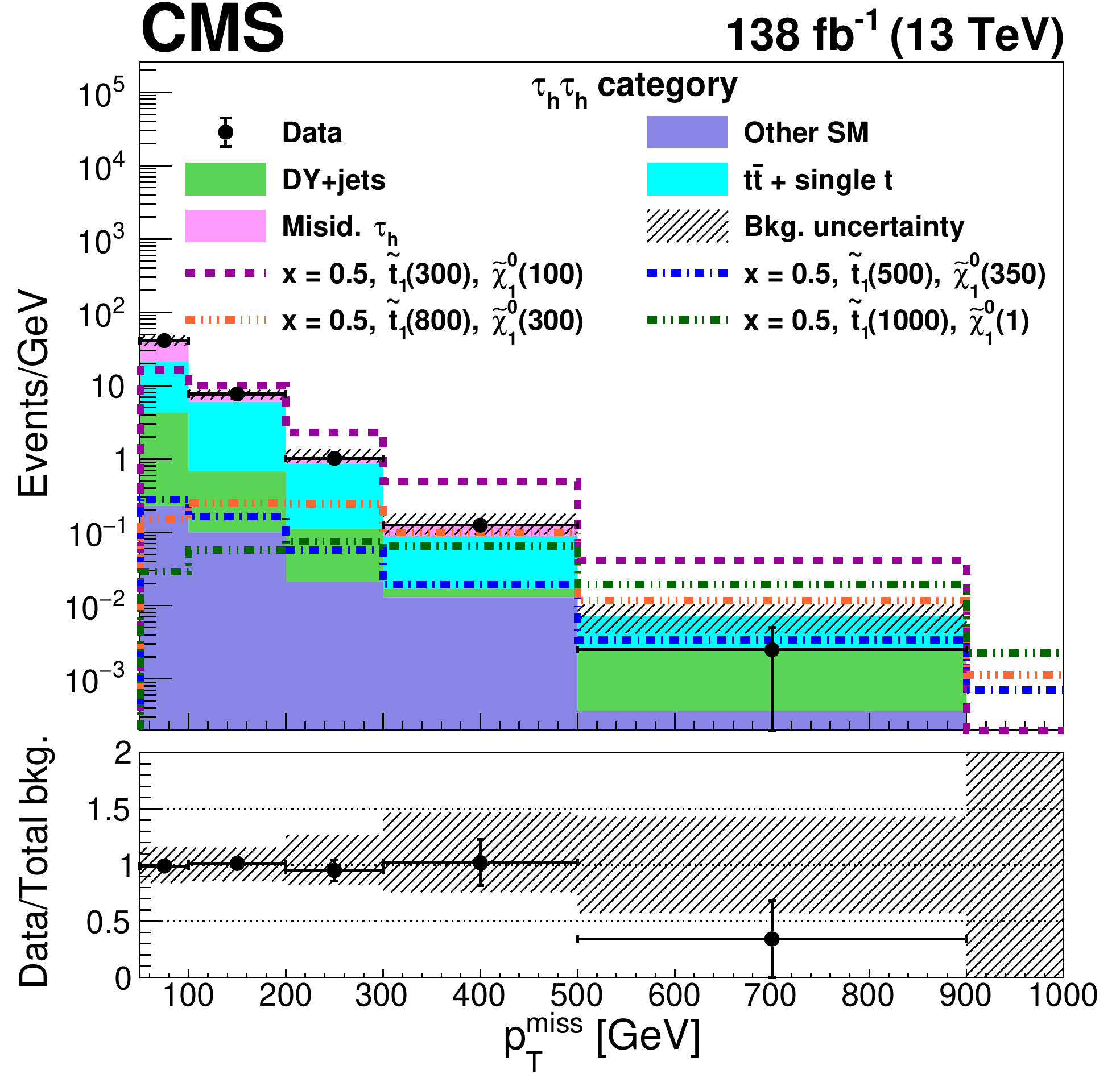}
	\includegraphics[width=0.495\textwidth]{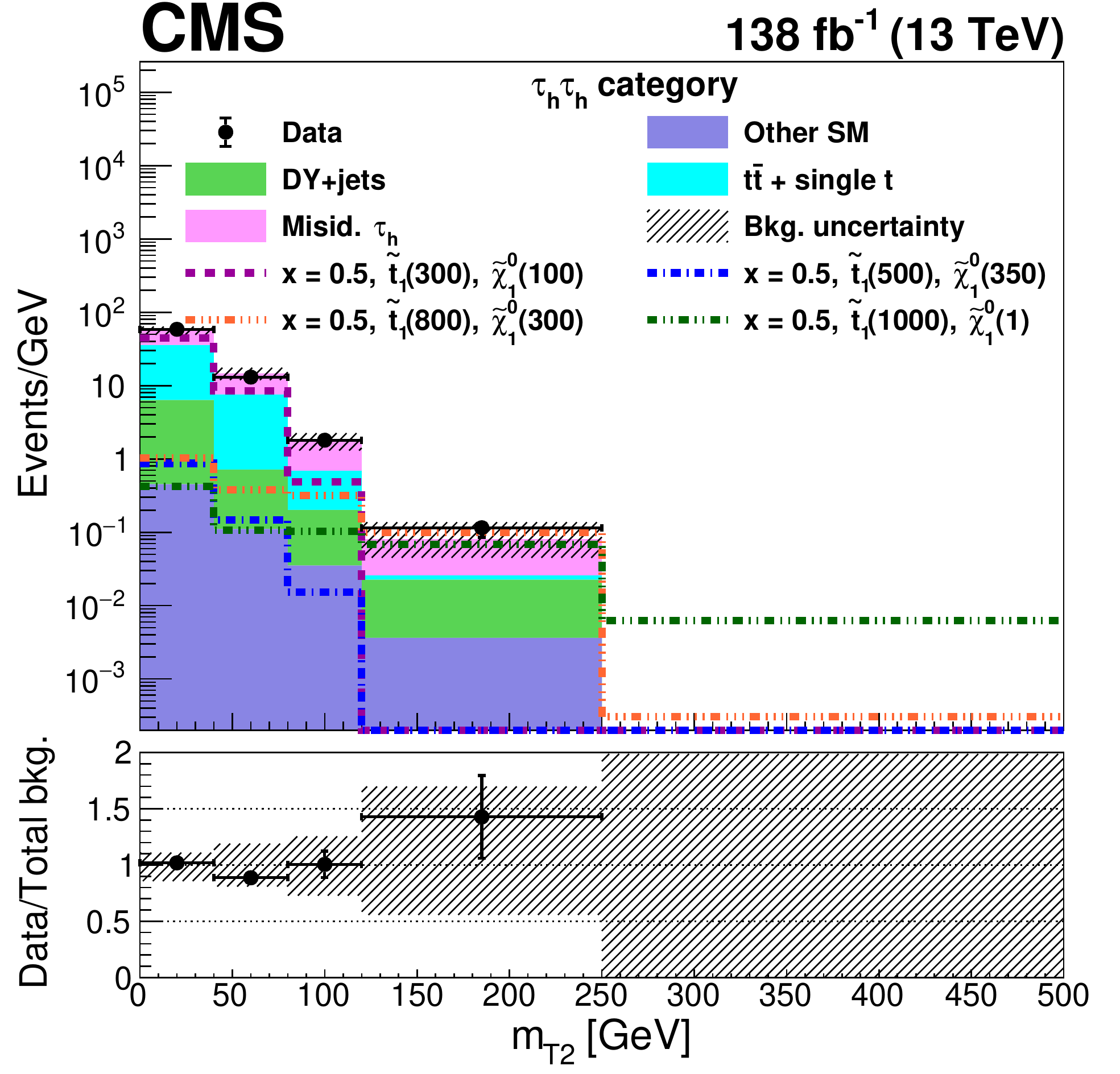} \\
	\includegraphics[width=0.495\textwidth]{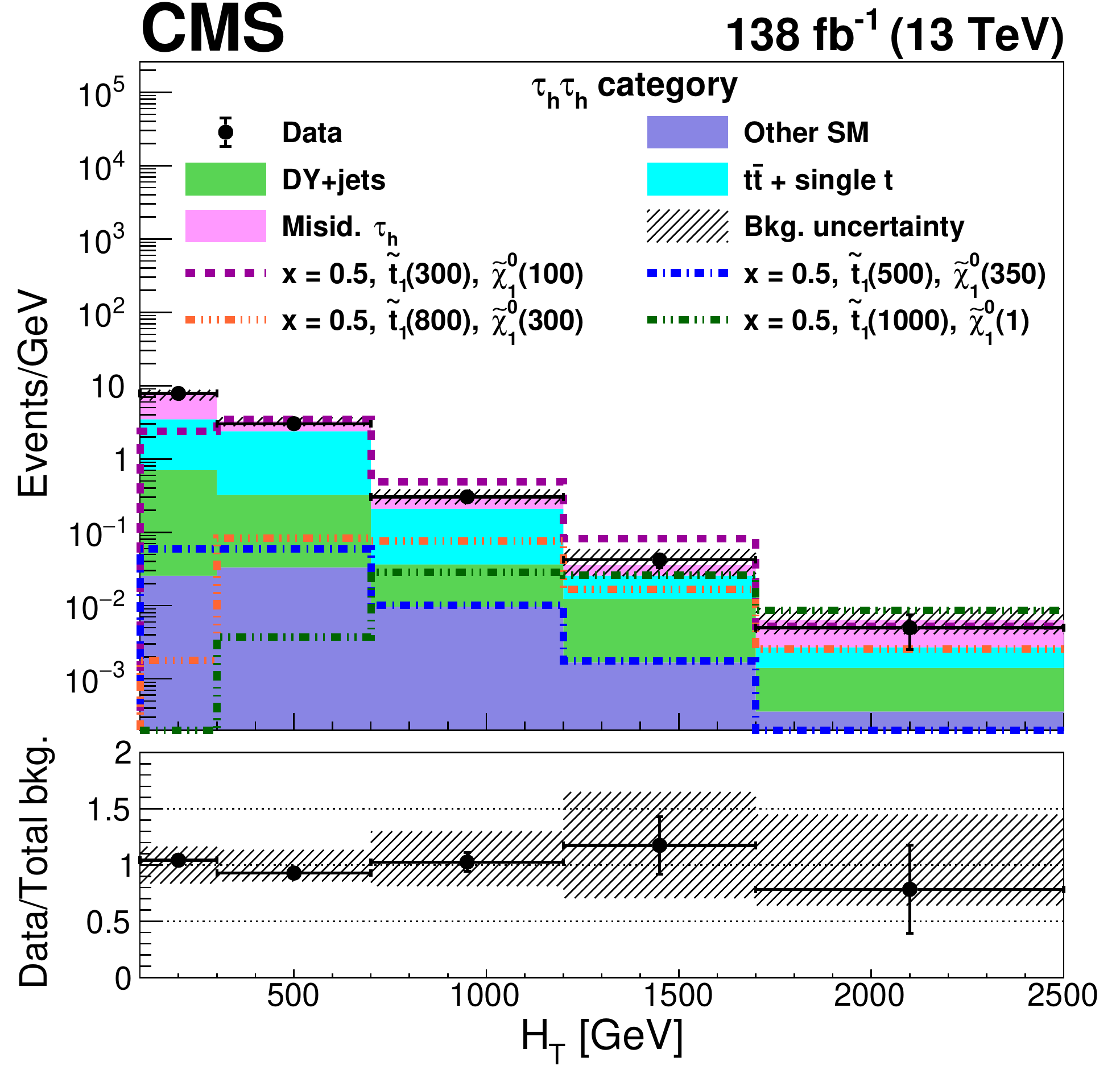}
	
	\caption{
		Distributions of the search variables \ptmiss, \mTii,
		and \HT after event selection described in Sec.~\ref{sec:evt_sel} for data and predicted backgrounds, corresponding to the \tauhtauh category.
		The histograms for the background processes are stacked, and the signal distributions expected for a few representative sets of model parameter values are overlaid: $ x = 0.5 $ and [$ m_{\PSQtDo} $, $ m_{\PSGczDo} $] = [300, 100], [500, 350], [800, 300], and [1000, 1]\GeV.
		The lower panel indicates the ratio of the observed number of events to the total predicted number of background events.
		The shaded bands indicate the statistical and systematic uncertainties in the predicted backgrounds, added in quadrature.
		The last bin includes the overflow.
	}
	\label{fig:SRvariableDataMC_TauTau}
	
\end{figure}

\begin{figure}[!htbp]
	\centering
	
	\includegraphics[width=0.495\textwidth]{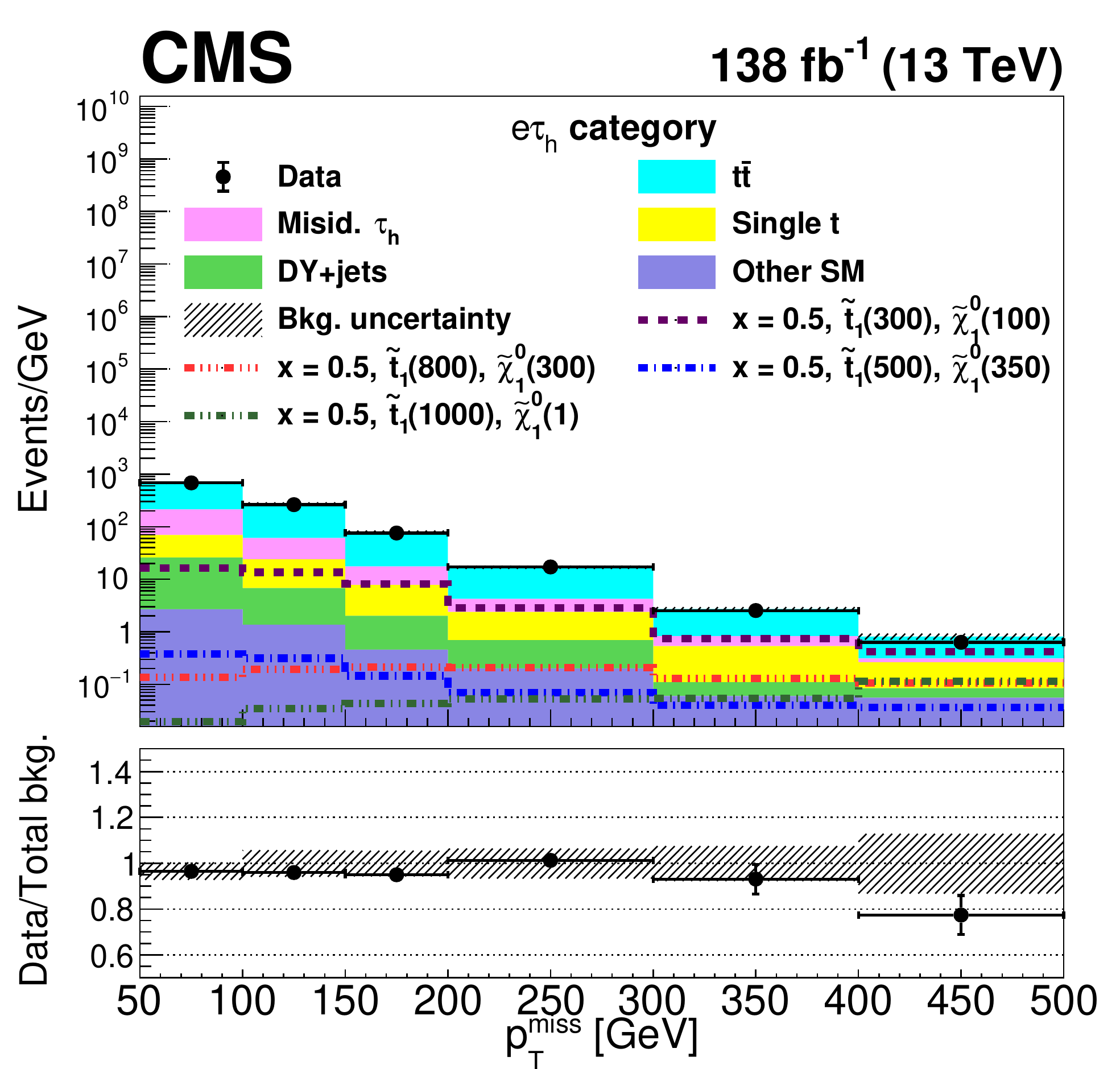}
	\includegraphics[width=0.495\textwidth]{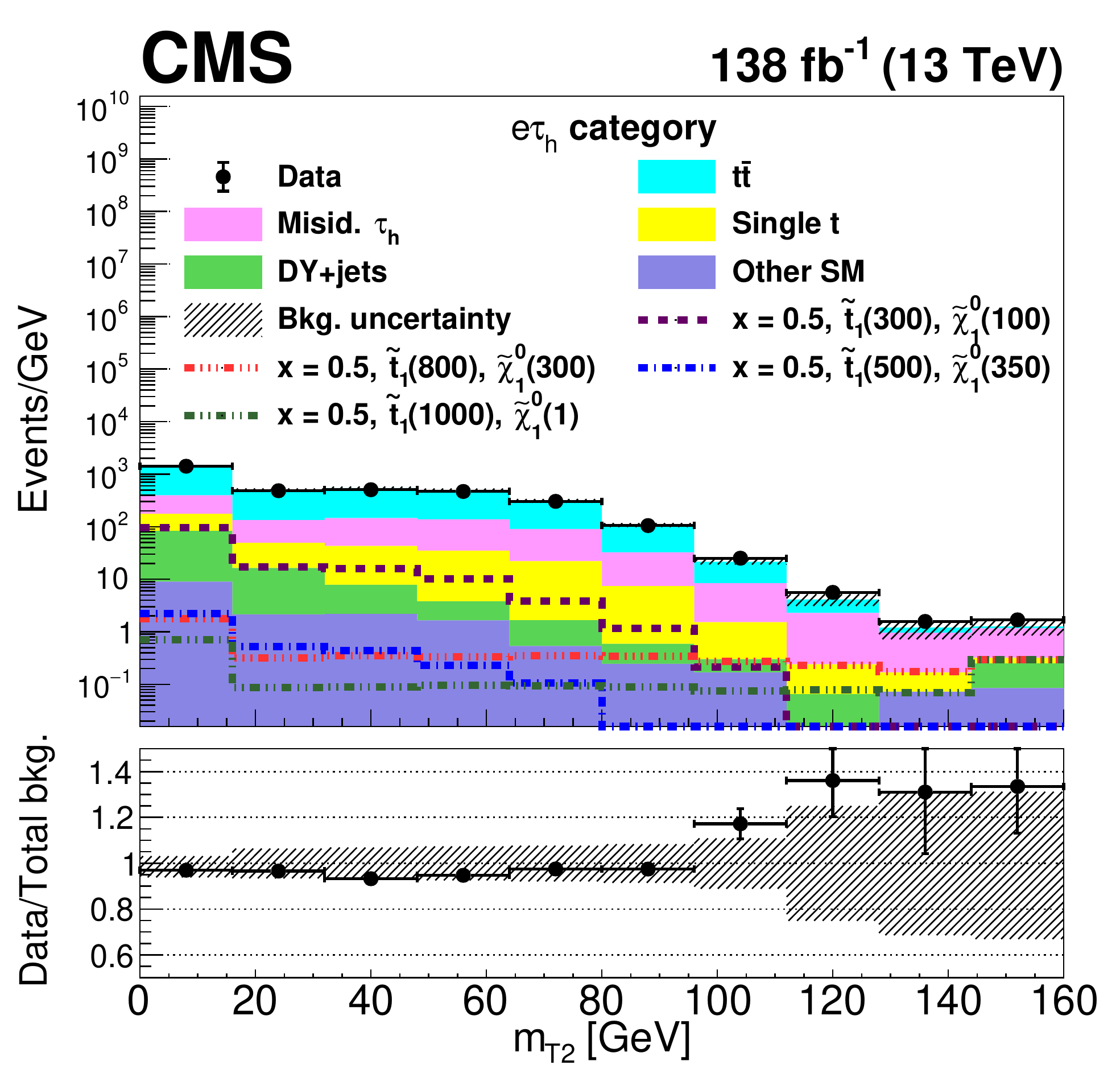} \\
	\includegraphics[width=0.495\textwidth]{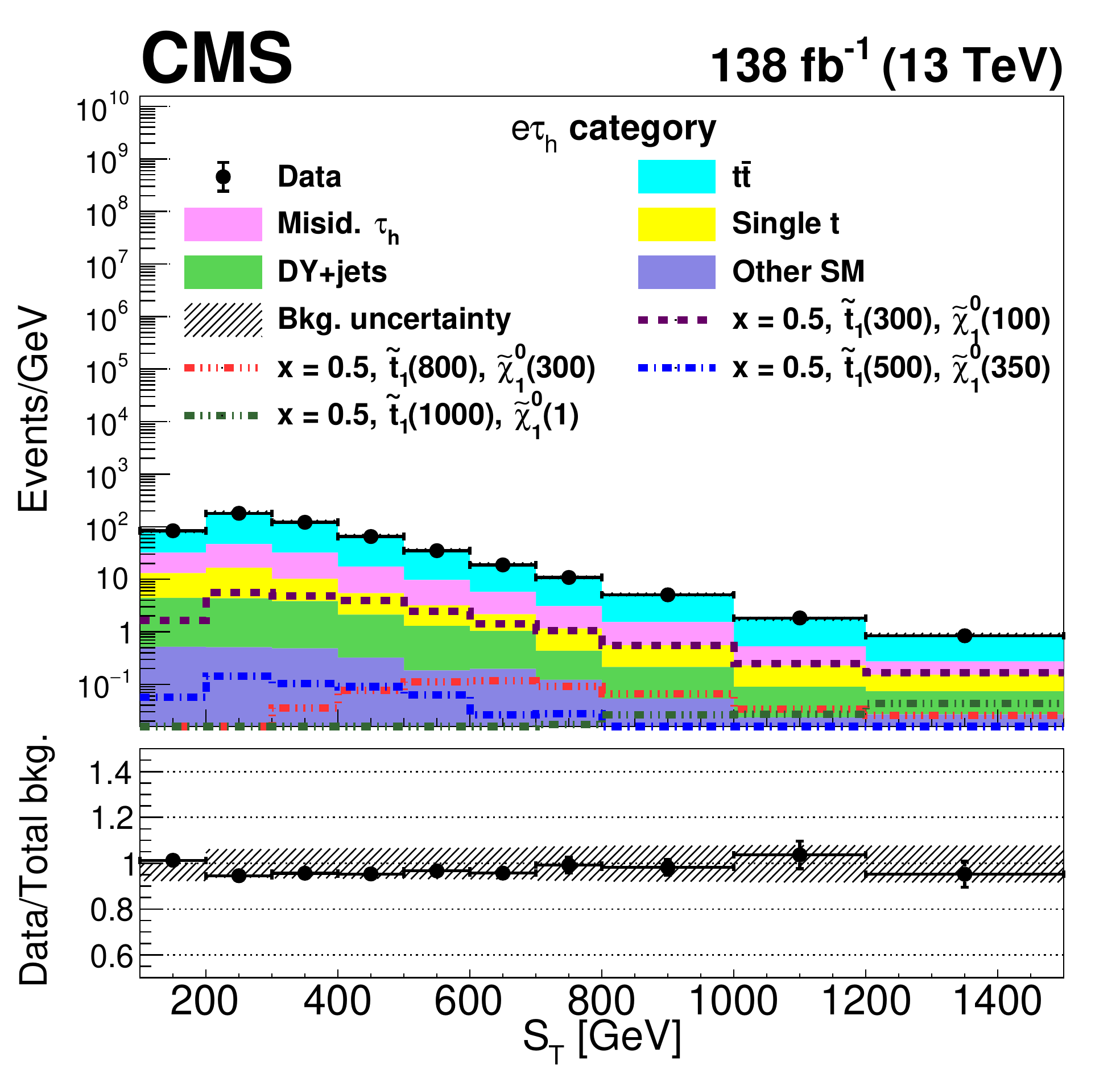}
	\caption{
		Distributions of the search variables \ptmiss, \mTii,
		and \ST after event selection described in Sec.~\ref{sec:evt_sel} for data and predicted backgrounds, corresponding to the \etauh category.
		The histograms for the background processes are stacked, and the signal distributions expected for a few representative sets of model parameter values are overlaid: $ x = 0.5 $ and [$ m_{\PSQtDo} $, $ m_{\PSGczDo} $] = [300, 100], [500, 350], [800, 300], and [1000, 1]\GeV.
		The lower panel indicates the ratio of the observed number of events to the total predicted number of background events.
		The shaded bands indicate the statistical and systematic uncertainties in the predicted backgrounds, added in quadrature.
		The last bin includes the overflow.
	}
	\label{fig:SRvariableDataMC_ETau}
	
\end{figure}

\begin{figure}[!htbp]
	\centering
	
	\includegraphics[width=0.495\textwidth]{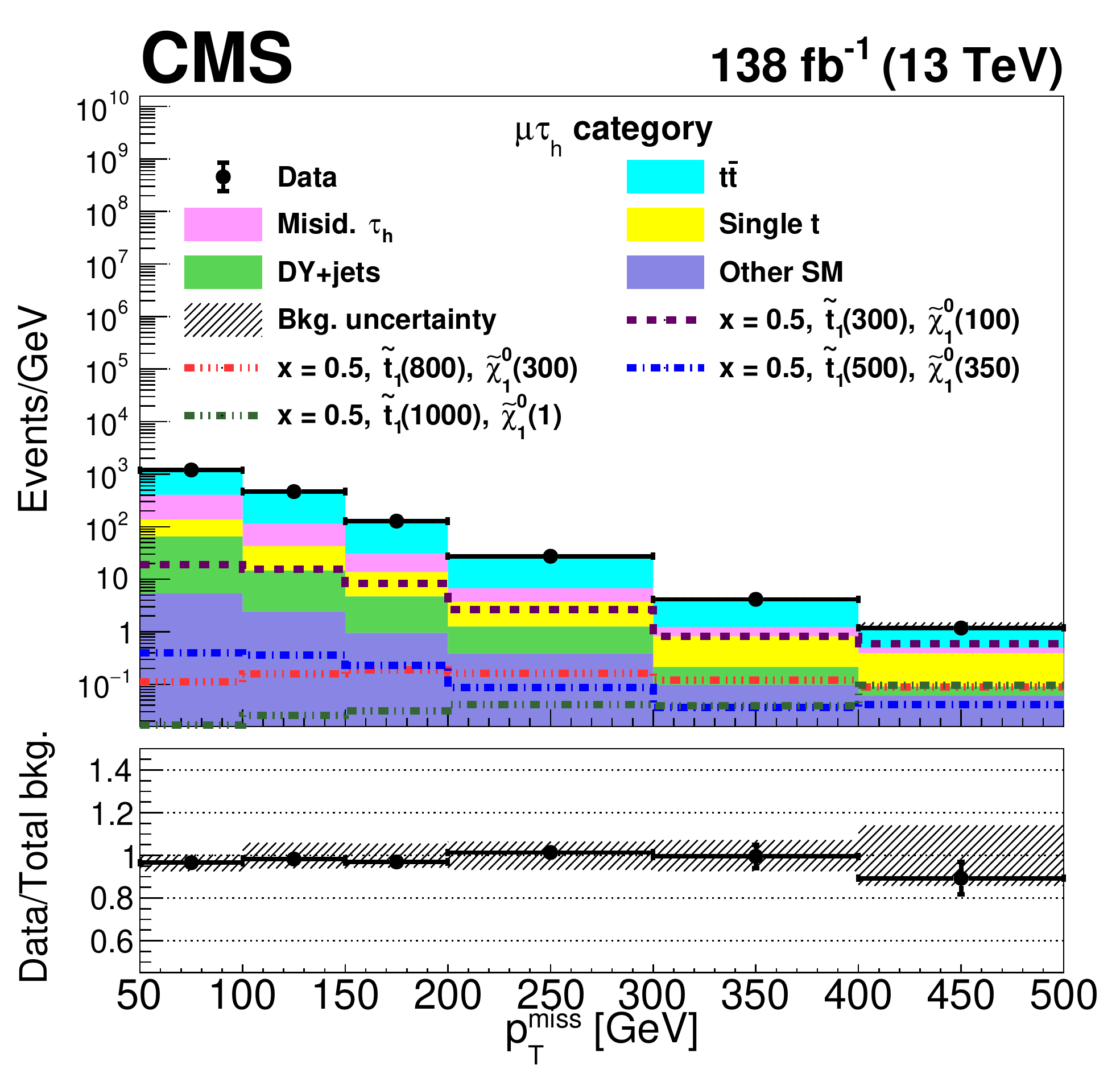}
	\includegraphics[width=0.495\textwidth]{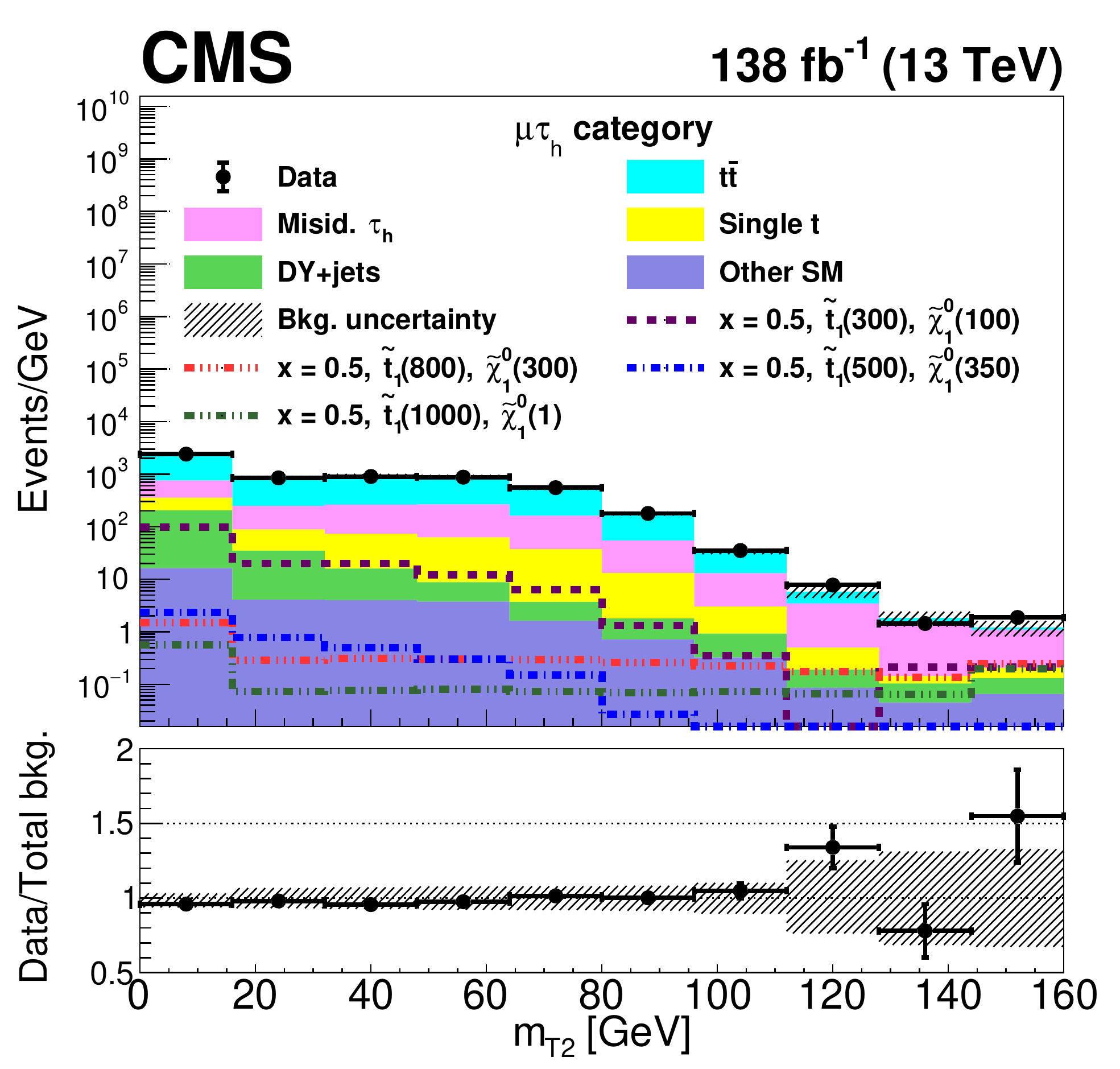} \\
	\includegraphics[width=0.495\textwidth]{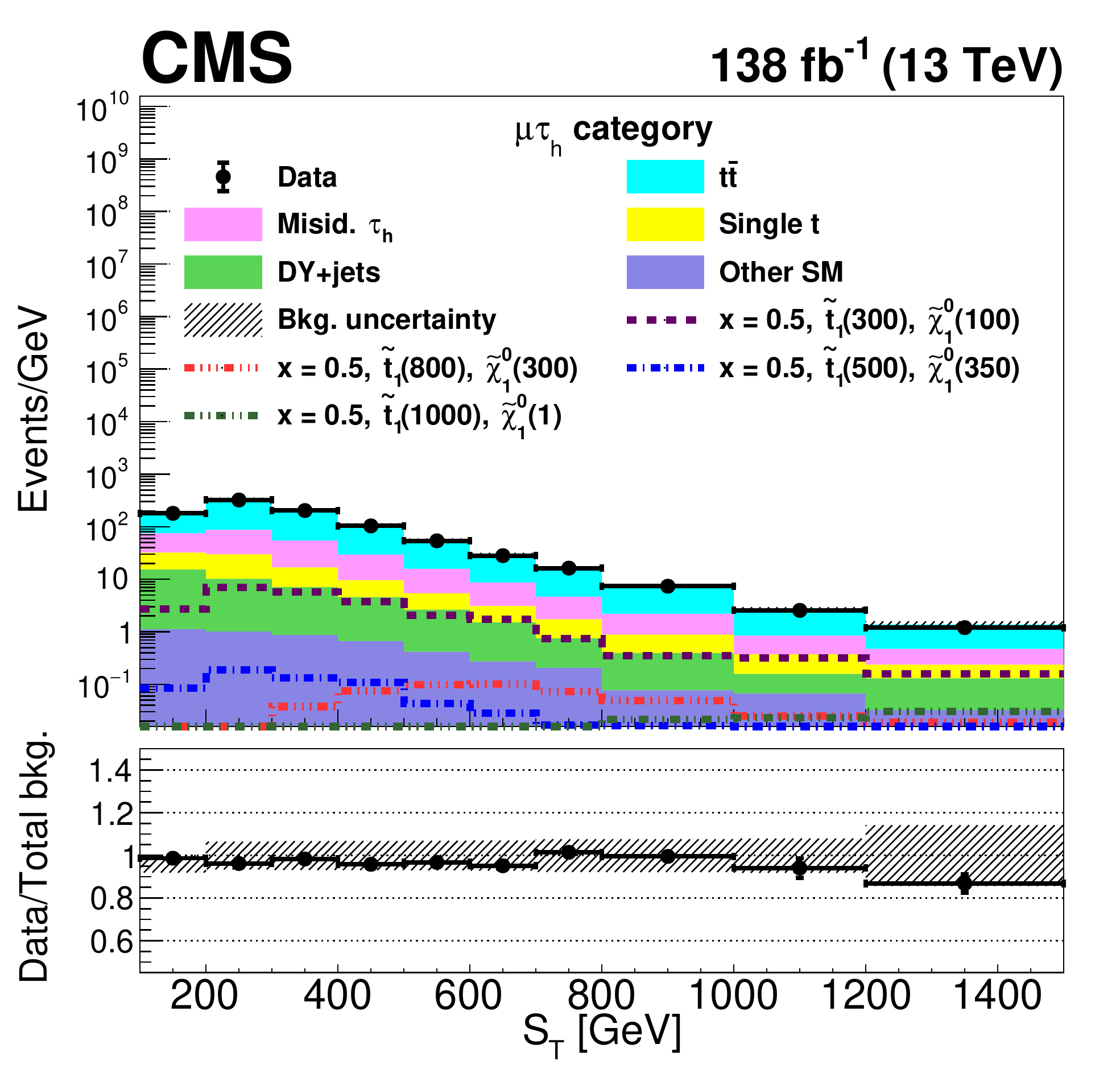}
	\caption{
		Distributions of the search variables \ptmiss, \mTii,
		and \ST after event selection described in Sec.~\ref{sec:evt_sel} for data and predicted backgrounds, corresponding to the \mutauh category.
		The histograms for the background processes are stacked, and the signal distributions expected for a few representative sets of model parameter values are overlaid: $ x = 0.5 $ and [$ m_{\PSQtDo} $, $ m_{\PSGczDo} $] = [300, 100], [500, 350], [800, 300], and [1000, 1]\GeV.
		The lower panel indicates the ratio of the observed number of events to the total predicted number of background events.
		The shaded bands indicate the statistical and systematic uncertainties in the predicted backgrounds, added in quadrature.
		The last bin includes the overflow.
	}
	\label{fig:SRvariableDataMC_MuTau}
	
\end{figure}

Distributions of the variables \ptmiss, \mTii, and \HT (or \ST) after this selection are shown in Figs.~\ref{fig:SRvariableDataMC_TauTau}--\ref{fig:SRvariableDataMC_MuTau} for data and the predicted background, along with representative signal distributions.
The SM backgrounds are estimated using the methods described in Section~\ref{sec:BkgEstimation}.

\begin{figure}[!htbp]
	\centering
	
	\includegraphics[width=0.65\textwidth]{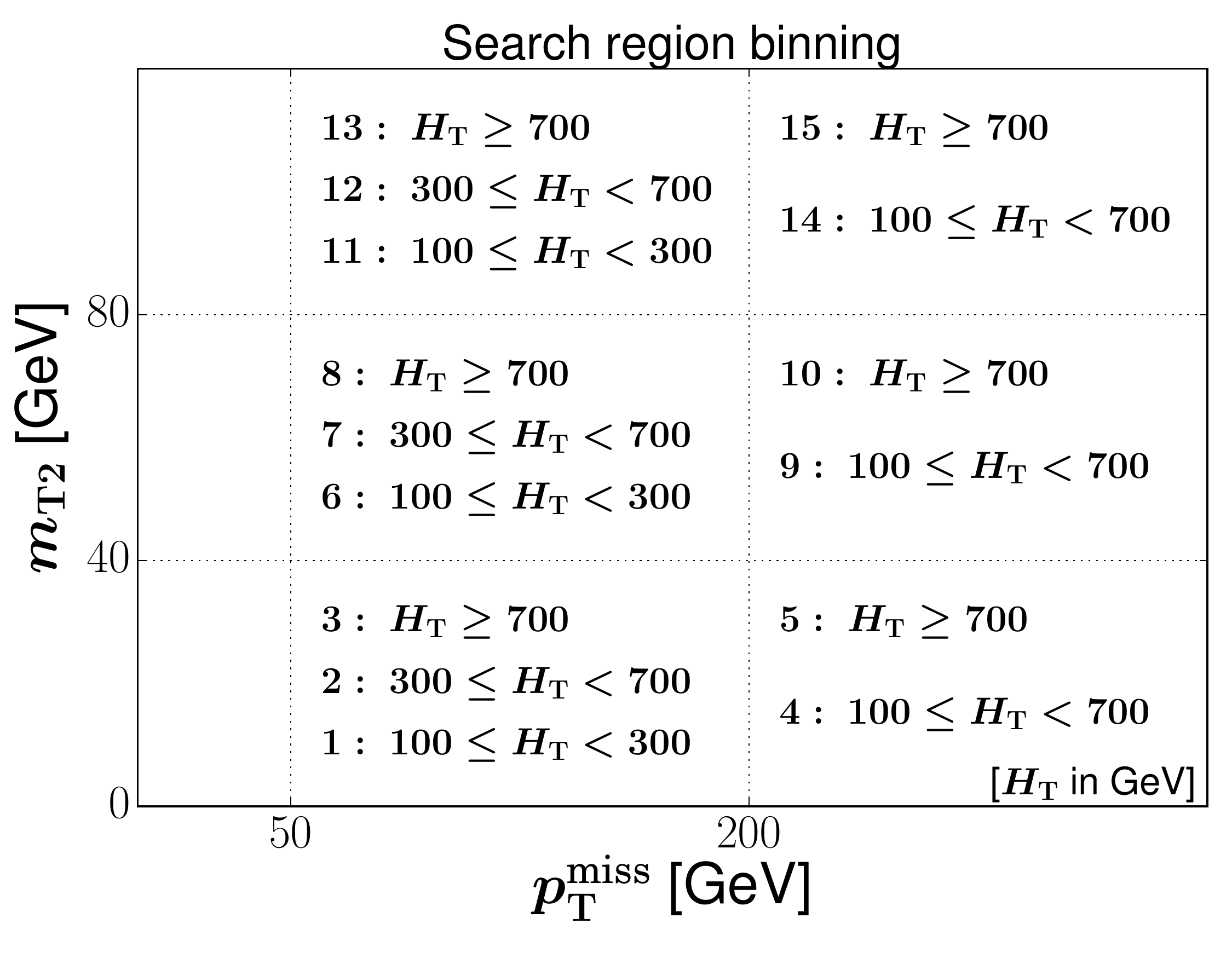}
	
	\caption{The 15 signal regions defined in bins of \ptmiss, \mTii, and \HT. The bin boundaries for \ST are the same as those for \HT.}
	\label{fig:sigBin}
\end{figure}

Signal events with different top squark and LSP masses have decay products with different kinematics and populate different regions of the phase space.
For example, regions with low \ptmiss, \mTii, and \HT (or \ST) are sensitive to signals with low top squark masses.
On the other hand, events with high \ptmiss, \mTii, and \HT (or \ST) are sensitive to models with high top squark and low LSP masses.
To obtain the highest sensitivity over the entire phase space, the selected events are categorized in 15 bins as a function of the measured \ptmiss, \mTii, and \HT (or \ST), as illustrated in Fig.~\ref{fig:sigBin}.

\section{Background estimation}
\label{sec:BkgEstimation}

The most significant backgrounds contributing to the SR are single top quark and \ttbar processes in which the top quark decays to a lepton and a neutrino. With the \ptmiss from the neutrino, these events have a final state that is very similar to that of the signal model.
Events from \ttbar and single top quark production having two genuine \tauh decays (one lepton and one genuine \tauh decay) account for about 47 (75--78)\% of the total SM background in the \tauhtauh (\ltauh) category.
The contribution from these processes is estimated from simulation. The predicted yield in each SR bin is multiplied by a correction factor derived from CRs in data and simulation.

Events with one or two jets that are misidentified as \tauh candidates arise mostly from single top quark and \ttbar processes.
The contribution from multijet events is significantly diminished because of the requirements $ \ptmiss > 50\GeV $ and $ n_{\PQb} \geq 1 $, the latter of which also reduces the contribution from {\PW}+jets events with a misidentified \tauh candidate.
The contribution from processes with one or more jets misidentified as a \tauh candidate, estimated using data CRs, is about 42 (18)\% of the total background in the \tauhtauh (\ltauh) category.

The background contribution from DY+jets events via $\PZ/\PGg^{*} \to \PGt\PGt$ decays is typically small in the most sensitive bins, amounting to about 9\% in the \tauhtauh category and a few percent in the \ltauh category. This background is estimated using simulation for both the categories.
To account for residual discrepancies between data and the LO DY+jets sample, correction factors for simulated events are derived as functions of the dimuon invariant mass and \pt from a DY-enriched dimuon CR in data and simulation~\cite{CMS:2019eln}.
Other less significant backgrounds, such as {\PW}+jets, $ \PV\PV $, $ \PV\PH $, and $ \ttbar\PV $ can also contribute to the SR via vector bosons decaying to leptons and the Higgs boson decaying to a pair of tau leptons.
The total contribution from these processes, which is estimated from simulation, is below 1\%.

\subsection{Tau leptons from top quark production}
\label{sec:topBkg}

The estimation of the background from \ttbar and single top quark processes (collectively called top quark events) with two genuine \tauh decays (or one genuine \tauh decay and a lepton) is extrapolated from an \emu CR, based on the method described in Refs.~\cite{CMS:2019lrh, Sirunyan:2645851}.
The single top quark events contributing to this final state are mostly from the \tW process.
The contribution from these processes in the SR is obtained by multiplying the predicted yield in each SR bin from simulation by correction factor derived from a CR enriched in top quark events.

\begin{figure}[!htbp]
	\centering
	
	\includegraphics[width=0.495\textwidth]{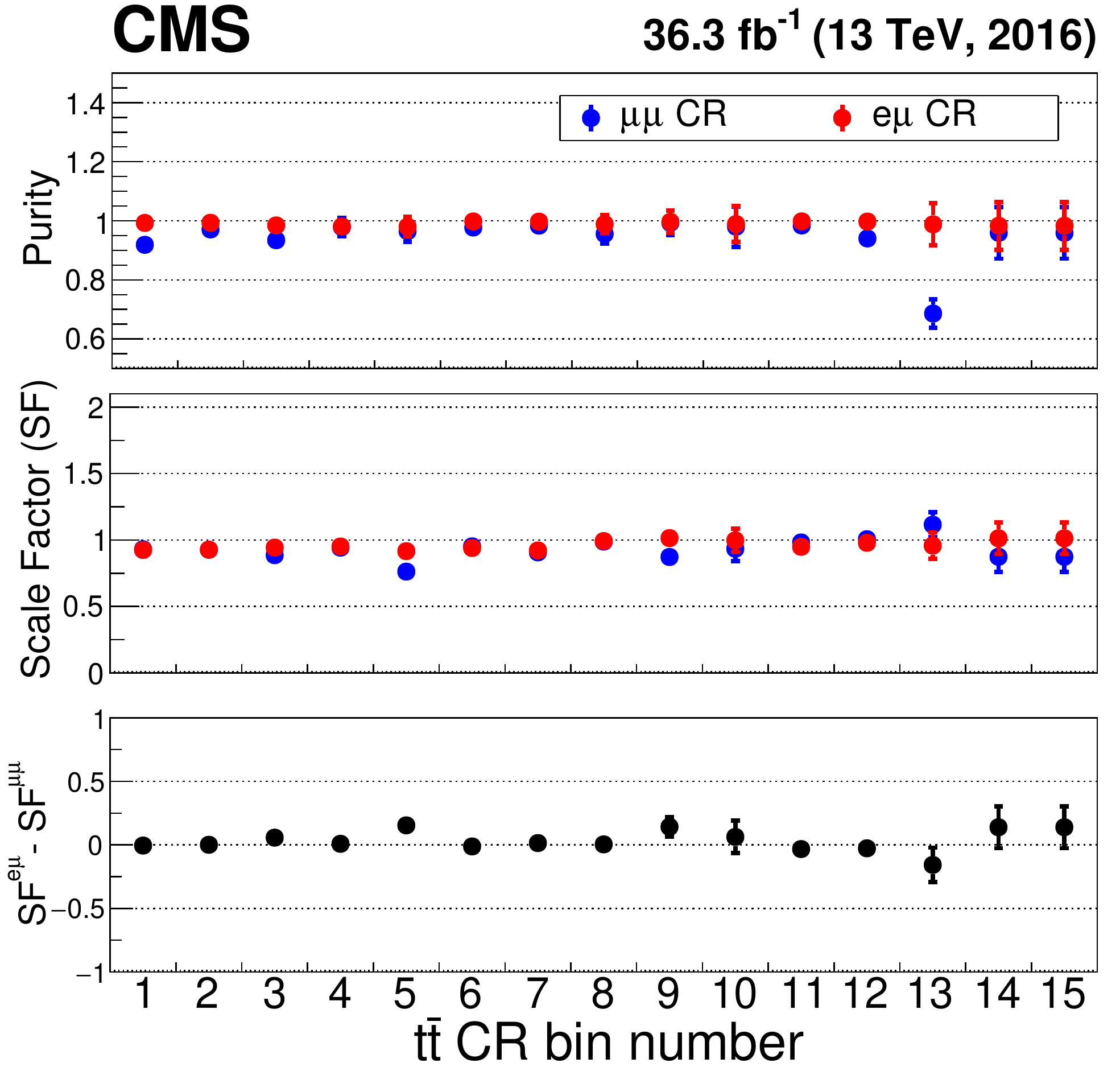}
	\includegraphics[width=0.495\textwidth]{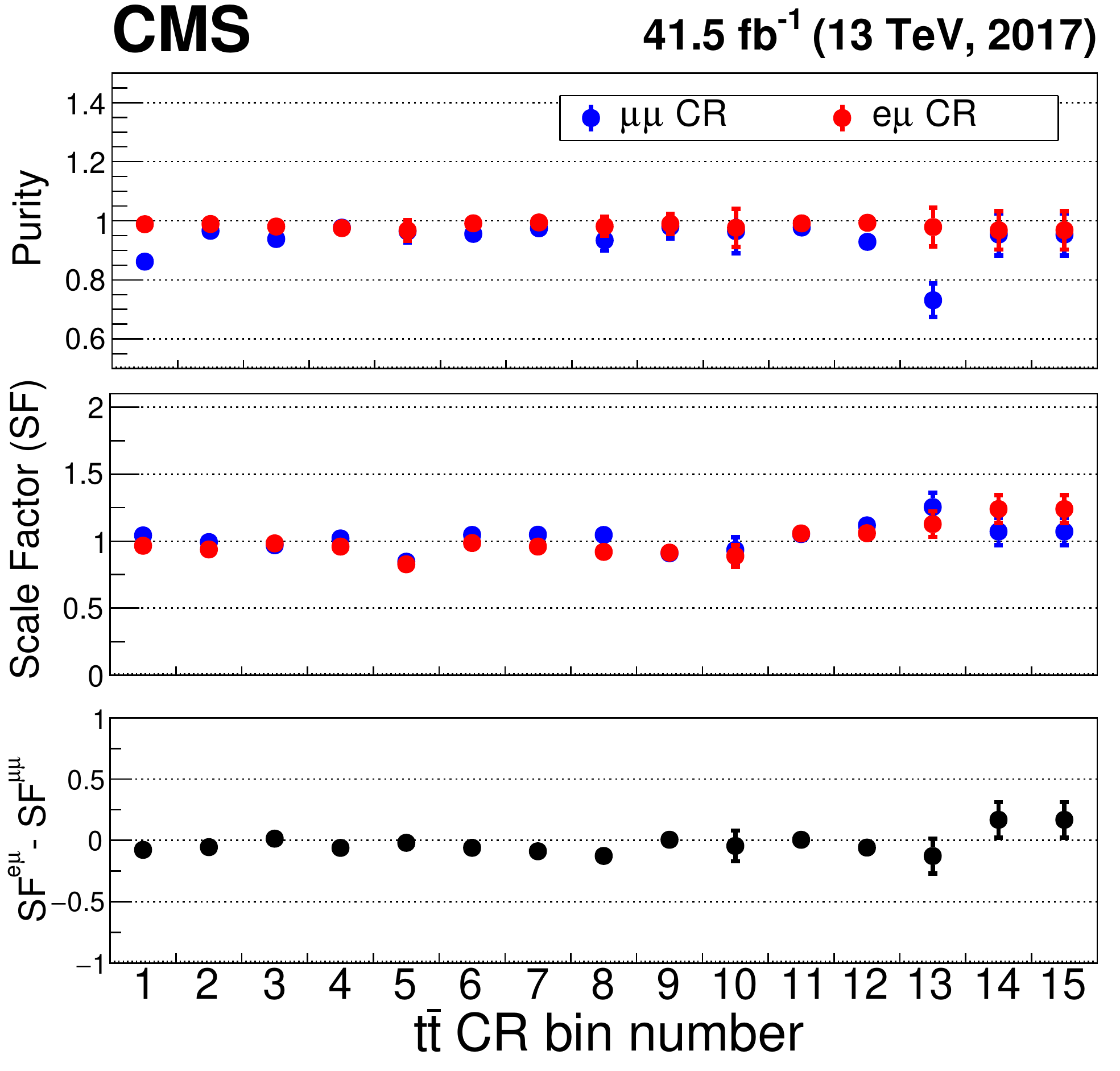} \\
	\includegraphics[width=0.495\textwidth]{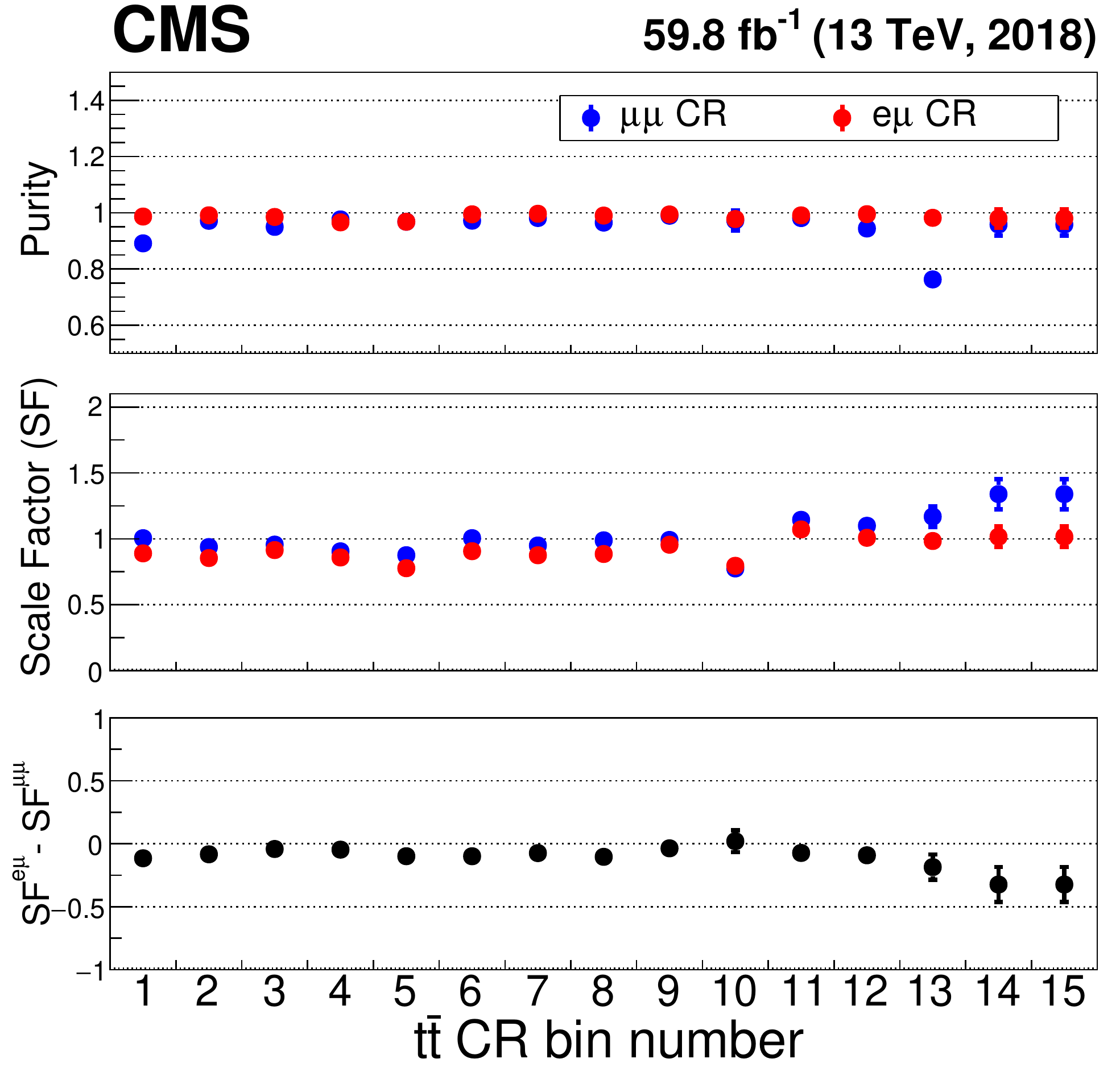}
	
	\caption{
		The purities in top quarks, the scale factors $\text{SF}$ from simulation to data, and the $ \text{SF}^{\emu} - \text{SF}^{\mumu} $ differences in the various bins (as defined in Fig.~\ref{fig:sigBin}) of the top enriched CR, where the purity is estimated from simulation.
		The upper left, upper right, and lower subfigures correspond to 2016, 2017, and 2018 data, respectively.
		To mitigate the effect of statistical fluctuations, bins 14 and 15 are merged to provide the same SF in both bins for subsequent calculations.
	}
	\label{fig:ttbarCRpuritySF}
\end{figure}

The CR enriched in top quark events is identified by selecting events with an \emu pair with opposite-sign charge.
These events are selected with \emu triggers, and are required to satisfy the same requirements on \ptmiss, \ST, and $n_{\PQb}$ as those of the SR.
The \emu triggers are $ {\approx}$95\% efficient for lepton candidates.
To reduce possible DY contamination in this CR coming from the tail of the \emu invariant mass distribution in the process $ \PZ/\Pgg^{*} \to \PGt\PGt \to \emu $, events are vetoed if the invariant mass of the \emu system is $ 60 < m_{\emu} < 120\GeV$.
This selection on the dilepton invariant mass is more useful for reducing the DY contribution in the \mumu CR (discussed later), but is also applied here to be consistent.
The purity of top quark events in the CR, \ie, the fraction of top quark events in most of the bins is $ {\gtrsim} $85\% in simulation, as shown in Fig.~\ref{fig:ttbarCRpuritySF}, in the upper panels of each subfigure.
The small contamination from other processes is found to have no significant effect on the results.

Residual differences between data and simulation are quantified by scale factors (SFs).
For a given SR bin ($ i $) we define
\begin{linenomath}
	\begin{equation}
	\begin{aligned}
	\text{SF}_{i} &= \frac{N^{\emu \ \text{CR}}_{i,\ \text{data}}}{N^{\emu \ \text{CR}}_{i,\ \text{MC}}} ,
	\end{aligned}
	\label{eq:ttbarSF}
	\end{equation}
\end{linenomath}
where the numerator and the denominator represent the yields in the CR in data and simulation, respectively.
The contamination from the signal process in the CR is found to be negligible.
The single top quark and \ttbar backgrounds are treated together when deriving and applying the SFs since it is difficult to find a CR that is highly pure in single top quark events alone, and also has sufficient event count to obtain the SFs bin by bin.
The ratio of single top quark to \ttbar yields in the CR bins is very similar to that in the corresponding SR bins, that is, the relative kinematics of the two processes are similar in the CR and SR.
The corrected \ttbar and single top yield in simulation in each bin of the SR is then obtained as:
\begin{linenomath}
	\begin{equation}
	\begin{aligned}
	N^{\text{SR}}_{i,\ \text{corr. top}} &=
	N^{\text{SR}}_{i,\ \text{top} \ \text{MC}}\, \text{SF}_{i} &=
	\frac{N^{\emu \ \text{CR}}_{i,\ \text{data}}\, \ N^{\text{SR}}_{i,\ \text{top} \ \text{MC}}}{N^{\emu \ \text{CR}}_{i,\ \text{MC}}}
	,
	\end{aligned}
	\label{eq:ttbarSFcorr}
	\end{equation}
\end{linenomath}
where $ N^{\text{SR}}_{i,\ \text{top MC}} $ is the prediction from simulated \ttbar and single top events in the SR.
Only the contribution from events with two genuine \tauh candidates (or one genuine \tauh candidate and a lepton) is corrected using the procedure described above.
The SFs in different bins, shown in Fig.~\ref{fig:ttbarCRpuritySF} (middle row) for 2016, 2017, and 2018 data, are mostly found to be within ${\approx}$10\% of unity.
We note that bins 14 and 15 in the CR are merged and a single SF is used for both bins in subsequent calculations to reduce the statistical uncertainty.

To cross check the validity of this method, the same technique is applied to an independent top-quark-enriched CR with an oppositely charged \mumu pair in the final state.
These events are selected with single-muon triggers that reach ${\approx}$95\% efficiency.
The event selection for the \mumu CR is the same as that for the \emu CR.
This cross check evaluates the effect of possible contamination from DY events, since the branching fraction of $ \PZ/\PGg^{*} \to \PGm\PGm $ is much higher than that of $ \PZ/\PGg^{*} \to \PGt\PGt \to \emu $. It is also useful for checking any dependence of SFs on lepton reconstruction.
The differences between the SFs calculated in the main and cross check CRs, shown in Fig.~\ref{fig:ttbarCRpuritySF} (lower row) are small (within $ {\approx}$10\% in most cases), and are taken as an uncertainty in SFs.
These are added in quadrature to the statistical uncertainty in SFs, and propagated to the uncertainty in the final top quark background prediction.
The different sources of systematic uncertainties in the terms estimated from simulation in the numerator and denominator of Eq.~\ref{eq:ttbarSFcorr}, are included in the final prediction. These uncertainties are described in Sec.~\ref{sec:syst_unc}.

\subsection{Misidentified hadronically decaying tau lepton candidates}
\label{sec:fakeBkg}

A major component of the total background originates from processes with a quark or gluon jet that is misidentified as a \tauh candidate.
The largest sources of such events in the SR are semileptonic \ttbar and single top quark decays.
Events with one genuine electron (muon) and a jet misidentified as a \tauh candidate contribute to the \etauh (\mutauh) category whereas those with one or two misidentified \tauh candidates contribute to the \tauhtauh category.
These background contributions are estimated from CRs in data and simulation, which are obtained by requiring \tauh candidates to pass a looser WP but fail the tight requirements.
The yields observed in the CRs are extrapolated to the SR in a way described next for the \tauhtauh and \ltauh categories.
The \tauhtauh category receives contributions from events with one or two misidentified \tauh candidates, whereas the \ltauh category contains events with only one misidentified \tauh candidate.
Hence different extrapolation methods are used for the two categories.

\subsubsection{Estimation for the \texorpdfstring{$\tauhtauh$}{tauhtauh} category}
\label{sec:fakeBkg_TauhTauh}

The misidentified \tauh background in the \tauhtauh category is estimated following the strategy described in Refs.~\cite{CMS:2019lrh, Sirunyan:2628769}.
For a genuine (misidentified) \tauh passing the loose identification requirements, we define $g$ ($f$) as the probability that it also passes the tight identification requirements.
The number of \tauhtauh events where the \tauh candidate with the highest \pt is genuine and that with the second-highest \pt is misidentified, is denoted as $ N_\text{gf}$.
Other terms, $ N_\text{fg}$, $ N_\text{gg}$, and $ N_\text{ff}$ are defined similarly.
The number of \tauhtauh events where the candidate with the highest \pt passes the tight identification criteria and that with the second-highest \pt fails, but passes the loose criteria, is denoted as $N_\text{TL}$.
Other terms, $N_\text{LT}$, $N_\text{LL}$, and $N_\text{TT}$ are defined similarly.
The events are required to satisfy the selections described in Sec.~\ref{sec:evt_sel}.
If $ N $ is the total number of events, the following set of equations can be established:

{\small
	\begin{linenomath}
		\begin{equation}
		\begin{aligned}
		&N = N_\text{gg} + N_\text{fg} + N_\text{gf} + N_\text{ff} = N_\text{TT} + N_\text{LT} + N_\text{TL} + N_\text{LL}, \\
		&N_\text{LL} = (1-g_{1})(1-g_{2}) N_\text{gg} + (1-f_{1})(1-g_{2}) N_\text{fg} + (1-g_{1})(1-f_{2}) N_\text{gf} + (1-f_{1})(1-f_{2}) N_\text{ff}, \\
		&N_\text{LT} = (1-g_{1}) g_{2} N_\text{gg} + (1-f_{1}) g_{2} N_\text{fg} + (1-g_{1}) f_{2} N_\text{gf} + (1-f_{1}) f_{2} N_\text{ff}, \\
		&N_\text{TL} = g_{1} (1-g_{2}) N_\text{gg} + f_{1} (1-g_{2}) N_\text{fg} + g_{1} (1-f_{2}) N_\text{gf} + f_{1} (1-f_{2}) N_\text{ff}, \\
		&N_\text{TT} = g_{1} g_{2} N_\text{gg} + f_{1} g_{2} N_\text{fg} + g_{1} f_{2} N_\text{gf} + f_{1} f_{2} N_\text{ff} ,
		\end{aligned}
		\label{FakeEquation_1}
		\end{equation}
	\end{linenomath}
}

where the subscripts 1 and 2 on $g$ and $f$ refer to the \tauh candidates with the highest and second-highest \pt, respectively.
The above equations can be inverted to give the numbers of genuine and misidentified {\tauh}\tauh candidate events in the SR:
\begin{linenomath}
	\begin{equation}
	N_\text{TT} = N^{\text{gen}}_\text{TT} + N^{\text{misid}}_\text{TT},
	\end{equation}
	\text{where}
	\begin{equation*}
	\begin{aligned}
	N^{\text{gen}}_\text{TT} &= g_{1} g_{2} N_\text{gg}, \\
	N^{\text{misid}}_\text{TT} &= f_{1} g_{2} N_\text{fg} + g_{1} f_{2} N_\text{gf} + f_{1} f_{2} N_\text{ff} .	
	\end{aligned}
	\label{FakeEquation_3}
	\end{equation*}
\end{linenomath}
Here $ N^{\text{gen}}_\text{TT} $ represents the number of events in the SR with two genuine \tauh candidates in the final state and $N^{\text{misid}}_\text{TT} $ stands for the number of events in the SR with one or two misidentified \tauh candidates.

The probability $g$ is evaluated using \ttbar simulation for different decay modes of the reconstructed \tauh candidate, as a function of its \pt.
It is observed to be about 80\% with a mild dependence on the \pt of the \tauh for the decay modes containing either one charged hadron and up to two neutral pions, or three charged hadrons and no neutral pions.
For the decay mode with three charged hadrons and one neutral pion, $g$ is found to vary between 50 and 70\% depending on the \pt of the \tauh candidate.

The misidentification rate $ f $ is estimated from data using a multijet-enriched CR.
This CR is defined by requiring a same-sign \tauh pair (satisfying the \tauh selection criteria used in the SR), and by requiring $ \ptmiss < 50\GeV$.
There can be small correlations between the probabilities of the two \tauh candidates to pass the tight criteria.
This causes the value of $f$ to differ by a few percent depending on which of the two \tauh candidates is required to pass the tight criteria.
This difference is included as an uncertainty in $f$.
The misidentification rate is measured as a function of its \pt for different decay modes of the reconstructed \tauh candidate; it varies between 25 and 45\%.
In simulation studies~\cite{Sirunyan:2628769} we find that the misidentification rate also depends on the flavor of the parton corresponding to the jet that is misidentified as a \tauh candidate.
Since the jet flavor cannot be reliably determined in data, an additional 30\% uncertainty in $f$ is included~\cite{CMS-PAS-SUS-21-001}.
This uncertainty is evaluated as the relative difference between the average and the maximum of the misidentification rates corresponding to the different jet flavors being up, down, strange, charm, and bottom quarks, and gluons, estimated using simulated {\PW}+jets events.

\subsubsection{Estimation for the \texorpdfstring{$\Pell\tauh$}{Pelltauh} category}
\label{sec:fakeBkg_LepTauh}

The misidentified \tauh background in the $\ltauh$ category is estimated by selecting a sideband region (SbR) where all the SR selections are applied, except that the \tauh candidate is required to pass the very loose (VL) WP but fail the tight (T) identification criteria.
This identification requirement is indicated as ``VL \& !T" in the following discussion.
The yields from this SbR are extrapolated to obtain the contribution from misidentified \tauh candidates to the $\ltauh$ SR.
The extrapolation factor is determined in a CR~\cite{CMS:2019eln} enriched in {\PW}+jets events containing a misidentified \tauh candidate.
This CR is obtained from data by requiring exactly one \PGm, exactly one \tauh, $\ptmiss>50\GeV$ and $60<\mT<130\GeV$ where \mT is the transverse mass computed using \ptvecmiss and the transverse momentum of the \Pgm.
Events with any \PQb-tagged jet passing the loose WP are vetoed to remove any overlap between this CR and both the SR and SbR.
The fraction of {\PW}+jets events in this CR is calculated to be ${\approx}83\%$ using simulation.
The remaining contribution from non-{\PW}+jets events is estimated from simulation and subtracted from the data.
The ratio, $R$, of the number of misidentified \tauh events in SR to that in SbR is defined in the {\PW}+jets CR as:
\begin{linenomath}
	\begin{equation}
	R =
	\frac{N_\text{data}^\text{CR} (\tauh^{\text{T}}) - N_\text{non-{\PW}+jets MC}^\text{CR} (\tauh^{\text{T}})}
	{N_\text{data}^\text{CR} (\tauh^{\text{VL \& !T}})-N_\text{non-{\PW}+jets MC}^{\text{CR}}(\tauh^{\text{VL \& !T}})} .
	\label{eq:transferFactor}
	\end{equation}
\end{linenomath}
Here $N_{\text{data}}^{\text{CR}}$ is the number of events in the {\PW}+jets CR obtained from data and $N_\text{MC, non-{\PW}+jets}^\text{CR}$ is the number of simulated events except {\PW}+jets in the same CR.
The value of $R$ is calculated as a function of the \tauh candidate's \pt and $\eta$, and it varies between 15 and 30\%.
A variation of 50\% in the non-{\PW}+jets contribution is found to change the value of $R$ by up to 10\%, which is included as an uncertainty in $R$.
Similar to $f$ in Sec.~\ref{sec:fakeBkg_TauhTauh}, $R$ is also found to vary by ${\approx}30\%$ in simulation depending on the flavor of the parton corresponding to the jet that is misidentified as a \tauh.
Hence an uncertainty of 30\% in $R$ is included.
The contribution from misidentified \tauh candidates to each of the \ltauh SR bins is then evaluated as
\begin{linenomath}
	\begin{equation}
	N^{\text{misid, SR}} = R \, N^{\text{misid, SbR}} = R \, [N_{\text{data}}^{\text{SbR}} - N_{\text{MC, genuine \tauh}}^{\text{SbR}}],
	\end{equation}
\end{linenomath}
where $N_{\text{data}}^{\text{SbR}}$ is the number of events obtained in the sideband region from data, and $N_{\text{MC, genuine \tauh}}^{\text{SbR}}$ represents the contribution to the sideband region from simulated events where the $\tauh$ candidate is genuine.
The contribution to $N^{\text{misid, SbR}}$ from events with \tauh candidates in a particular \pt and $\eta$ range is multiplied by $R$ measured in the same range.

\section{Systematic uncertainties}
\label{sec:syst_unc}

There are several sources of systematic uncertainties that are propagated to the prediction of the final signal and background yields.
For the \tauhtauh category, the most significant is the uncertainty in the modeling of the \tauh trigger (8--12\%). The uncertainty due to \tauh identification and isolation (ID-iso) requirements~\cite{Sirunyan:CMS-TAU-16-003} is 6--8\%.
In the $\ltauh$ category, the major uncertainty arises from \tauh ID-iso requirements (3--4\%), followed by the uncertainty in the SF for top quark events (${\approx}4\%$).
The other sources of uncertainty affecting all processes include the jet energy scale (JES) and jet energy resolution (JER), the \tauh energy scale, the effect of unclustered components in calculating \ptmiss, pileup reweighting, and the \PQb tagging efficiency.
The simulation is reweighted to make its pileup vertex distribution identical to that of the data.
The uncertainty in estimating the number of pileup interactions is estimated by varying the total inelastic cross section by $\pm$4.6\%~\cite{Sirunyan:2018nqx}. This is propagated as an uncertainty in the pileup reweighting factor applied to the simulation.

Since the \ttbar and single top quark contribution in the SR is obtained by multiplying the simulated yield by SFs (Eq.~\ref{eq:ttbarSFcorr}), several MC uncertainties cancel in the ratio.
However, some small residual effects arising from JER, JES, and unclustered energies may remain after the first order cancellation.
This is because the \ptmiss spectrum in \emu and \mumu CRs is different from that in \tauhtauh SR where extra neutrinos exist.
These small uncertainties are also included in the estimation of the \ttbar and single top quark contributions.
As mentioned earlier, the difference between the SFs obtained in the \emu and \mumu CRs, added in quadrature with the statistical uncertainty, is assumed to be the uncertainty in this method.
The flavor dependence of the \tauh misidentification rates $f$ and $R$ is accounted for by including an uncertainty of 30\% in the rates.
The \tauh misidentification rate in the $\tauhtauh$ category has an additional uncertainty of about 4\%, which arises from small correlations between the probabilities of the two \tauh candidates to pass the tight identification criteria.

The factorization ($ \mu_{\text{F}} $) and renormalization ($ \mu_{\text{R}} $) scales used in the simulation are varied up and down by a factor of two to account for missing higher order corrections, while avoiding the cases in which one is doubled and the other is halved. The \textsc{SysCalc} package~\cite{Kalogeropoulos:2018cke} is used for this purpose. The resulting uncertainty is estimated to be less than 6\% for both signal and background processes estimated from simulation.
The uncertainty in the measured integrated luminosity amounts to 1.2, 2.3, and 2.5\% in 2016, 2017, and 2018~\cite{CMS-LUM-17-003,CMS-PAS-LUM-17-004, CMS-PAS-LUM-18-002}, respectively.
The uncertainty in the \PZ boson \pt correction applied to DY+jets events is assumed to be equal to the deviation of the correction factor from unity.
This correction is derived as a function of the \PZ boson \pt.
A normalization uncertainty of 15\% is assigned to the production cross sections of the background processes that are evaluated directly from simulation \cite{Sirunyan:2018owv, Sirunyan:2018ucr, Aaboud:2017qkn, CMS:2019too, Sirunyan:2019bez, Sirunyan:2017wgx}.

Since the simulation of the detector for signal events is performed using \textsc{FastSim}, the signal yields are corrected to account for the differences in the \Pe, \PGm, and \tauh identification efficiencies with respect to the \GEANTfour simulation used for the backgrounds.
The statistical uncertainty associated with this correction is propagated to the final results as a part of the systematic uncertainties.
The \textsc{FastSim} package has a worse \ptmiss resolution than the full \GEANTfour simulation that can potentially result in an artificial enhancement of the signal yields.
Therefore the signal yields are corrected, and the uncertainty in the resulting correction to the yield is estimated to be less than ${\approx}8\%$.

The region with $\ptmiss>400\GeV$ ($\mTii>110\GeV$) in the background MC simulation was not adequately modeled in 2017 (2018).
To account for this, an additional uncertainty of 40 (38)\% is applied to background MC events with $\ptmiss>400\GeV$ ($\mTii>110\GeV$) in 2017 (2018) samples. These uncertainty values correspond to the sizes of the discrepancies observed in those regions, in the \emu and \mumu CRs.

The uncertainties in the signal and background from all sources are presented in Tables~\ref{tab:unc_TauTauFullRun2},~\ref{tab:unc_ETauFullRun2} and~\ref{tab:unc_MuTauFullRun2} for the $\tauhtauh$, $\etauh$ and $\mutauh$ categories respectively.
Upper and lower numbers correspond to the relative uncertainties due to the upward and downward variations of the signal or background yields due to the variations of the respective source within uncertainties.
These values are the weighted averages of the relative uncertainties in the various search bins with the weights being the predicted yields in the respective bins.
The uncertainty from a given source is considered to be correlated across the 15 search bins, whereas the different sources are treated as uncorrelated with each other.
In addition, the statistical uncertainties are also included and are considered to be uncorrelated across the bins.

\begin{table}[!htbp]
	\topcaption{Relative systematic uncertainties for the \tauhtauh category from various sources in signal and background
		yields.
		These values are averages of the relative uncertainties in the different search regions, weighted by the yields in the respective bins.
		For the asymmetric uncertainties, the upper (lower) entry is the uncertainty due to the upward (downward) variation, which can be in the same direction as a result of taking the weighted average.
		In the header row, the top squark and LSP masses in \GeV are indicated in parentheses.
		The uncertainty values shown here are prior to the maximum likelihood fit described in Sec.~\ref{sec:results}.
	}
	\cmsTable{
		\begin{tabular}{ l  c  c  c  c  c  c  c  c }
			\hline
			\\[\cmsTabSkipSmall]
			Uncertainty source & 
			\vtop{\hbox{\strut $ x = 0.5 $}\hbox{\strut $ \PSQtDo(300)$}\hbox{\strut $ \PSGczDo(100) $}} & 
			\vtop{\hbox{\strut $ x = 0.5 $}\hbox{\strut $ \PSQtDo(500)$}\hbox{\strut $ \PSGczDo(350) $}} & 
			\vtop{\hbox{\strut $ x = 0.5 $}\hbox{\strut $ \PSQtDo(800)$}\hbox{\strut $ \PSGczDo(300) $}} & 
			\vtop{\hbox{\strut $ x = 0.5 $}\hbox{\strut $ \PSQtDo(1000)$}\hbox{\strut $ \PSGczDo(1) $}} & 
			$\ttbar$ + single t & 
			DY+jets & 
			Other SM & 
			Misid. \tauh \\[\cmsTabSkipLarge]
			~ & ~ & ~ & ~ & ~ & ~ & ~ & ~ & ~\\
			\hline
			\\[\cmsTabSkipSmall]
			Signal cross section & 
			$\pm$6.7\%  & 
			$\pm$7.5\%  & 
			$\pm$9.5\% & 
			$\pm$11\%  & 
			\NA & 
			\NA & 
			\NA & 
			\NA \\[\cmsTabSkip]
			
			\textsc{FastSim} $ \ptmiss $ resolution & 
			$\pm$7.8\% & 
			$\pm$6\%  & 
			$\pm$4.5\% & 
			$\pm$2.3\% & 
			\NA & 
			\NA & 
			\NA & 
			\NA \\[\cmsTabSkip]
			
			$ \tauh $ \textsc{FastSim}/{\GEANTfour} & 
			\vtop{\hbox{\strut $+$4.0\%}\hbox{\strut $-$3.9\%}} & 
			\vtop{\hbox{\strut $+$3.1\%}\hbox{\strut $-$3.0\%}} & 
			\vtop{\hbox{\strut $+$6.3\%}\hbox{\strut $-$6.1\%}} & 
			\vtop{\hbox{\strut $+$15.9\%}\hbox{\strut $-$14.5\%}} & 
			\NA & 
			\NA & 
			\NA & 
			\NA \\[\cmsTabSkipLarge]
			
			JER & 
			\vtop{\hbox{\strut $<$0.1\%}\hbox{\strut $<$0.1\%}} & 
			\vtop{\hbox{\strut $-$1.1\%}\hbox{\strut $+$0.8\%}} & 
			\vtop{\hbox{\strut $<$0.1\%}\hbox{\strut $+$0.11\%}} & 
			\vtop{\hbox{\strut $+$0.23\%}\hbox{\strut $<$0.1\%}} & 
			\vtop{\hbox{\strut $+$1.3\%}\hbox{\strut $-$1.3\%}} & 
			\vtop{\hbox{\strut $+$8.4\%}\hbox{\strut $-$3.6\%}} & 
			\vtop{\hbox{\strut $+$1.7\%}\hbox{\strut $-$3.0\%}} & 
			\NA \\[\cmsTabSkipLarge]
			
			2018 \mTii uncertainty & 
			\NA & 
			\NA & 
			\NA & 
			\NA & 
			\vtop{\hbox{\strut $<$0.1\%}\hbox{\strut $<$ 0.1\%}} & 
			\vtop{\hbox{\strut $+$0.29\%}\hbox{\strut $-$0.29\%}} & 
			\vtop{\hbox{\strut $+$0.38\%}\hbox{\strut $-$0.38\%}} & 
			\NA \\[\cmsTabSkipLarge]
			
			Pileup & 
			\vtop{\hbox{\strut $-$0.3\%}\hbox{\strut $+$0.3\%}} & 
			\vtop{\hbox{\strut $+$0.57\%}\hbox{\strut $-$0.62\%}} & 
			\vtop{\hbox{\strut $+$0.2\%}\hbox{\strut $-$0.2\%}} & 
			\vtop{\hbox{\strut $<$ 0.1\%}\hbox{\strut $<$0.1\%}} & 
			\NA & 
			\vtop{\hbox{\strut $+$1.7\%}\hbox{\strut $-$1.8\%}} & 
			\vtop{\hbox{\strut $-$1.8\%}\hbox{\strut $+$1.9\%}} & 
			\NA \\[\cmsTabSkipLarge]
			
			JES & 
			\vtop{\hbox{\strut $+$1.4\%}\hbox{\strut $-$0.5\%}} & 
			\vtop{\hbox{\strut $+$0.54\%}\hbox{\strut $-$0.12\%}} & 
			\vtop{\hbox{\strut $<$0.1\%}\hbox{\strut $<$0.1\%}} & 
			\vtop{\hbox{\strut $<$0.1\%}\hbox{\strut $<$0.1\%}} & 
			\vtop{\hbox{\strut $+$2.6\%}\hbox{\strut $-$2.6\%}} & 
			\vtop{\hbox{\strut $+$8.5\%}\hbox{\strut $-$6.0\%}} & 
			\vtop{\hbox{\strut $+$2.4\%}\hbox{\strut $-$1.9\%}} & 
			\NA \\[\cmsTabSkipLarge]
			
			$ \tauh $ ID-iso & 
			\vtop{\hbox{\strut $+$6.5\%}\hbox{\strut $-$8.1\%}} & 
			\vtop{\hbox{\strut $+$6.4\%}\hbox{\strut $-$8.1\%}} & 
			\vtop{\hbox{\strut $+$6.6\%}\hbox{\strut $-$8.1\%}} & 
			\vtop{\hbox{\strut $+$6.6\%}\hbox{\strut $-$8.2\%}} & 
			\vtop{\hbox{\strut $+$6.6\%}\hbox{\strut $-$8.1\%}} & 
			\vtop{\hbox{\strut $+$6.5\%}\hbox{\strut $-$8.1\%}} & 
			\vtop{\hbox{\strut $+$6.8\%}\hbox{\strut $-$8.1\%}} & 
			\NA \\[\cmsTabSkipLarge]
			
			\ptmiss unclustered energy & 
			\vtop{\hbox{\strut $+$0.47\%}\hbox{\strut $+$0.33\%}} & 
			\vtop{\hbox{\strut $-$0.46\%}\hbox{\strut $-$0.26\%}} & 
			\vtop{\hbox{\strut $+$0.13\%}\hbox{\strut $<$0.1\%}} & 
			\vtop{\hbox{\strut $<$0.1\%}\hbox{\strut $<$0.1\%}} & 
			\vtop{\hbox{\strut $+$1.2\%}\hbox{\strut $-$1.2\%}} & 
			\vtop{\hbox{\strut $+$4.9\%}\hbox{\strut $-$4.6\%}} & 
			\vtop{\hbox{\strut $+$1.7\%}\hbox{\strut $-$0.2\%}} & 
			\NA \\[\cmsTabSkipLarge]
			
			Background normalization & 
			\NA & 
			\NA & 
			\NA & 
			\NA & 
			\NA & 
			$ \pm 15 \% $ & 
			$ \pm 15 \% $ & 
			\NA \\[\cmsTabSkip]
			
			$ \tauh $ energy scale & 
			\vtop{\hbox{\strut $+$2.5\%}\hbox{\strut $-$2.7\%}} & 
			\vtop{\hbox{\strut $+$2.4\%}\hbox{\strut $-$3.5\%}} & 
			\vtop{\hbox{\strut $+$1.1\%}\hbox{\strut $-$1.3\%}} & 
			\vtop{\hbox{\strut $+$1.1\%}\hbox{\strut $-$1.1\%}} & 
			\vtop{\hbox{\strut $+$1.7\%}\hbox{\strut $-$1.8\%}} & 
			\vtop{\hbox{\strut $+$3.6\%}\hbox{\strut $-$3.4\%}} & 
			\vtop{\hbox{\strut $+$1.7\%}\hbox{\strut $-$4.6\%}} & 
			\NA \\[\cmsTabSkipLarge]
			
			$ \mu_{\text{R}} $ and $ \mu_{\text{F}} $ scales & 
			\vtop{\hbox{\strut $+$0.8\%}\hbox{\strut $-$0.8\%}} & 
			\vtop{\hbox{\strut $+$1.7\%}\hbox{\strut $-$1.8\%}} & 
			\vtop{\hbox{\strut $+$0.57\%}\hbox{\strut $-$0.64\%}} & 
			\vtop{\hbox{\strut $+$0.41\%}\hbox{\strut $-$0.46\%}} & 
			\NA & 
			\vtop{\hbox{\strut $+$2.1\%}\hbox{\strut $-$2.9\%}} & 
			\vtop{\hbox{\strut $+$4.1\%}\hbox{\strut $-$3.4\%}} & 
			\NA \\[\cmsTabSkipLarge]
			
			Luminosity & 
			$ \pm$2.1\% & 
			$ \pm$2.1\% & 
			$ \pm$2.1\% & 
			$ \pm$2.1\% & 
			\NA & 
			$ \pm$2.1\% & 
			$ \pm$2.1\% & 
			\NA \\[\cmsTabSkip]
			
			\PQb tagging & 
			\vtop{\hbox{\strut $<$0.1\%}\hbox{\strut $<$0.1\%}} & 
			\vtop{\hbox{\strut $<$0.1\%}\hbox{\strut $<$0.1\%}} & 
			\vtop{\hbox{\strut $<$0.1\%}\hbox{\strut $<$0.1\%}} & 
			\vtop{\hbox{\strut $+$0.14\%}\hbox{\strut $-$0.15\%}} & 
			\NA & 
			\vtop{\hbox{\strut $+$7.7\%}\hbox{\strut $-$7.8\%}} & 
			\vtop{\hbox{\strut $+$7.9\%}\hbox{\strut $-$8.0\%}} & 
			\NA \\[\cmsTabSkipLarge]
			
			2017 \ptmiss uncertainty & 
			\NA & 
			\NA & 
			\NA & 
			\NA & 
			\vtop{\hbox{\strut $<$0.1\%}\hbox{\strut $<$0.1\%}} & 
			\vtop{\hbox{\strut $<$0.1\%}\hbox{\strut $<$0.1\%}} & 
			\vtop{\hbox{\strut $+$1.2\%}\hbox{\strut $-$1.2\%}} & 
			\NA \\[\cmsTabSkipLarge]
			
			Trigger & 
			\vtop{\hbox{\strut $+$7.9\%}\hbox{\strut $-$7.5\%}} & 
			\vtop{\hbox{\strut $+$7.8\%}\hbox{\strut $-$7.5\%}} & 
			\vtop{\hbox{\strut $+$8.0\%}\hbox{\strut $-$7.7\%}} & 
			\vtop{\hbox{\strut $+$8.1\%}\hbox{\strut $-$7.8\%}} & 
			\vtop{\hbox{\strut $+$11.8\%}\hbox{\strut $-$11.2\%}} & 
			\vtop{\hbox{\strut $+$11.6\%}\hbox{\strut $-$10.9\%}} & 
			\vtop{\hbox{\strut $+$11.6\%}\hbox{\strut $-$10.9\%}} & 
			\NA \\[\cmsTabSkipLarge]
			
			$ \ttbar $ + single t SF & 
			\NA & 
			\NA & 
			\NA & 
			\NA & 
			$\pm$3.4\% & 
			\NA & 
			\NA & 
			\NA \\[\cmsTabSkip]
			
			\PZ $ \pt $ reweighting & 
			\NA & 
			\NA & 
			\NA & 
			\NA & 
			\NA & 
			\vtop{\hbox{\strut $+$2.0\%}\hbox{\strut $-$2.0\%}} & 
			\NA & 
			\NA \\[\cmsTabSkipLarge]
			
			$ \tauh $ misid. rate (parton flavor) & 
			\NA & 
			\NA & 
			\NA & 
			\NA & 
			\NA & 
			\NA & 
			\NA & 
			\vtop{\hbox{\strut $+$36.5\%}\hbox{\strut $-$31.8\%}} \\[\cmsTabSkipLarge]
			
			$ \tauh $ misid. rate (correlations) & 
			\NA & 
			\NA & 
			\NA & 
			\NA & 
			\NA & 
			\NA & 
			\NA & 
			\vtop{\hbox{\strut $+$3.7\%}\hbox{\strut $-$3.8\%}} \\[\cmsTabSkipLarge]
			
			\hline
		\end{tabular}
	}
	\label{tab:unc_TauTauFullRun2}
\end{table}

\begin{table}[!htbp]
	\topcaption{Relative systematic uncertainties for the \etauh category from various sources in signal and background
		yields.
		These values are averages of the relative uncertainties in the different search regions, weighted by the yields in the respective bins.
		For the asymmetric uncertainties, the upper (lower) entry is the uncertainty due to the upward (downward) variation, which can be in the same direction as a result of taking the weighted average.
		In the header row, the top squark and LSP masses in \GeV are indicated in parentheses.
		The uncertainty values shown here are prior to the maximum likelihood fit described in Sec.~\ref{sec:results}.
	}
	\cmsTable{
		\begin{tabular}{lcccccccc}
			\hline
			\\[\cmsTabSkipSmall]
			~Uncertainty source & 
			\vtop{\hbox{\strut $ x = 0.5 $}\hbox{\strut $ \PSQtDo(300)$}\hbox{\strut $ \PSGczDo(100) $}} &
			\vtop{\hbox{\strut $ x = 0.5 $}\hbox{\strut $ \PSQtDo(500)$}\hbox{\strut $ \PSGczDo(350) $}} & 
			\vtop{\hbox{\strut $ x = 0.5 $}\hbox{\strut $ \PSQtDo(800)$}\hbox{\strut $ \PSGczDo(300) $}} & 
			\vtop{\hbox{\strut $ x = 0.5 $}\hbox{\strut $ \PSQtDo(1000)$}\hbox{\strut $ \PSGczDo(1) $}} &  
			$\ttbar$ & Single \PQt & 
			\vtop{\hbox{\strut (DY+jets)}\hbox{\strut+ Other SM}} & Misid. \tauh\\
			~ & ~ & ~ & ~ & ~ & ~ & ~ & ~ & ~\\
			\hline
			\\[\cmsTabSkipSmall]
			Signal cross-section & $\pm$6.9\% & $\pm$7.5\% &$\pm$9.5\% &$\pm$11\% &\NA & \NA & \NA & \NA \\
			[\cmsTabSkip]
			\textsc{FastSim} \ptmiss resolution & $\pm$0.6\% & $\pm$0.5\% & $<$0.1\% & $<$0.1\% & \NA & \NA & \NA & \NA\\
			[\cmsTabSkip]
			\tauh \textsc{FastSim}/\GEANTfour & $\pm$0.9\% & $\pm$0.8\% &$\pm$1.1\% & $\pm$1.6\% &\NA & \NA & \NA & \NA\\
			[\cmsTabSkip]
			$\Pe$ \textsc{FastSim}/\GEANTfour & $\pm$1.7\% & $\pm$1.4\% &$\pm$3.1\%& $\pm$3.1\% &\NA & \NA & \NA & \NA\\
			[\cmsTabSkip]
			JER & $+$0.1\% & $+$0.2\% &$<$0.1\% & $+$0.1\%& \NA & \NA & $+$2.5\% & $+$0.1\%\\
			~& $-$0.4\% & $-$1.5\% &$-$0.1\%& $+$0.1\%& \NA & \NA & $+$0.3\% & $-$0.4\%\\
			[\cmsTabSkip]
			2018 $\mTii$ uncertainty & \NA & \NA & \NA & \NA & $<$0.1\% & $<$0.1\% & $<$0.1\% & $<$0.1\%\\
			[\cmsTabSkip]
			JES & $+$0.2\% & $-$0.2\% &$+$0.1\% & $+$0.1\% & \NA & \NA & $+$3.2\% & $+$0.4\%\\
			~& $-$0.2\% & $-$0.3\% & $-$0.1 & $-$0.1\% & \NA  & \NA & $-$2.0\% & $-$0.4\%\\
			[\cmsTabSkip]
			$\mu_{\text{R}}$ and $\mu_{\text{F}}$ scale & $+$0.5\% & $+$1.02\% & $+$0.5\% & $+$0.3\%& \NA & \NA & $+$3.2\% & $+$5.5\%\\
			~ & $-$0.4\% & $-$1.1\% & $-$0.5\% & $-$0.4\% & \NA & \NA & $-$4.6\% & $-$5.5\%\\
			[\cmsTabSkip]
			\tauh Id-iso & $+$3.2\% & $+$3.2\% & $+$3.2\% & $+$3.2\% & $+$3.1\% & $+$3.1\% & $+$3.1\% & $+$1.7\%\\ 
			~& $-$3.9\% & $-$4.3\% & $-$4.1\% & $-$4.1\% & $-$3.7\% & $-$3.9\% & $-$3.7\% & $-$1.4\%\\
			[\cmsTabSkip]
			Pileup & $+$0.3\% & $+$1.3\% & $+$0.7\% & $+$0.7\% & \NA & \NA & $+$0.2\% & $+$0.5\%\\
			~ & $-$0.3\% & $-$1.3\% & $-$0.7\% & $-$0.7\% & \NA & \NA & $-$0.2\% & $-$0.5\%\\
			[\cmsTabSkip]
			\ptmiss unclustered energy & $+$0.6\% & $+$0.8\% & $+$0.2\% & $<$0.1\% & \NA & \NA & $+$3.6\% & $+$0.2\%\\
			~ & $-$0.4\% & $-$0.7\% & $-$0.2\% & $-$0.1\% & \NA & \NA & $-$1.9\% & $-$0.4\%\\
			[\cmsTabSkip]
			Background normalization & \NA & \NA & \NA & \NA & \NA & \NA & $\pm$15\%  & \NA\\
			[\cmsTabSkip]
			Luminosity & $\pm$2.1\% & $\pm$2.1\% & $\pm$2.1\% & $\pm$2.1\% & \NA & \NA & $\pm$2.1\% & \NA\\
			[\cmsTabSkip]
			b tagging & $\pm$0.1\% & $<$0.1\% &$\pm$0.2\% & $\pm$0.5\%& \NA & \NA & $\pm$4.9\% & $\pm$0.8\%\\
			[\cmsTabSkip]
			2017 $\ptmiss$ uncertainty & \NA & \NA & \NA & \NA & $<$0.1\% & $<$0.1\% & $<$0.1\% & $<$0.1\% \\
			[\cmsTabSkip]
			Trigger & $<$0.1\% & $<$0.1\% &$<$0.1\% &$<$0.1\% &$<$0.1\% & $<$0.1\% & $<$0.1\% & $<$0.1\%\\
			[\cmsTabSkip]
			\tauh energy scale & $-$0.6\% & $-$0.1\% & $-$0.1\% & $<$0.1\% & $<$0.1\%  & $+$0.1\% & $+$1.5\% & $<$0.1\%\\
			~ & $-$0.7\% & $-$0.4\% & $-$0.1\% & $<$0.1\% & $<$0.1\% & $-$0.1\% & $-$3.4\%\\
			[\cmsTabSkip]
			$\ttbar$ + single \PQt SF & \NA & \NA & \NA & \NA & $\pm$3.8\% & $\pm$4.0\% & \NA & \NA\\
			[\cmsTabSkip]
			\tauh misid. rate (parton flavor) & \NA & \NA & \NA & \NA & \NA & \NA & \NA & $\pm$ 30\%\\
			[\cmsTabSkip]
			Non-{\PW}+jets background & & & & & & & &\\
			modeling in $R$ & \NA & \NA &\NA& \NA & \NA & \NA & \NA & $\pm$10\%\\
			\hline
		\end{tabular}
	}
	\label{tab:unc_ETauFullRun2}
\end{table}

\begin{table}[!htbp]
	\topcaption{Relative systematic uncertainties for the \mutauh category from various sources in signal and background
		yields.
		These values are averages of the relative uncertainties in the different search regions, weighted by the yields in the respective bins.
		For the asymmetric uncertainties, the upper (lower) entry is the uncertainty due to the upward (downward) variation, which can be in the same direction as a result of taking the weighted average.
		In the header row, the top squark and LSP masses in \GeV are indicated in parentheses.
		The uncertainty values shown here are prior to the maximum likelihood fit described in Sec.~\ref{sec:results}.
	}
	\cmsTable{
		\begin{tabular}{lcccccccc}
			\hline
			\\[\cmsTabSkipSmall]
			~Uncertainty source & 
			\vtop{\hbox{\strut $ x = 0.5 $}\hbox{\strut $ \PSQtDo(300)$}\hbox{\strut $ \PSGczDo(100) $}} & 
			\vtop{\hbox{\strut $ x = 0.5 $}\hbox{\strut $ \PSQtDo(500)$}\hbox{\strut $ \PSGczDo(350) $}} & 
			\vtop{\hbox{\strut $ x = 0.5 $}\hbox{\strut $ \PSQtDo(800)$}\hbox{\strut $ \PSGczDo(300) $}} & 
			\vtop{\hbox{\strut $ x = 0.5 $}\hbox{\strut $ \PSQtDo(1000)$}\hbox{\strut $ \PSGczDo(1) $}} &  
			$\ttbar$ & Single \PQt & 
			\vtop{\hbox{\strut (DY+jets)}\hbox{\strut+ Other SM}} & Misid. \tauh\\
			~ & ~ & ~ & ~ & ~ & ~ & ~ & ~ & ~\\
			\hline
			\\[\cmsTabSkipSmall]
			Signal cross-section & $\pm$6.9\% & $\pm$7.5\% &$\pm$9.5\%& $\pm$11\% & \NA & \NA & \NA & \NA \\
			[\cmsTabSkip]
			\textsc{FastSim} \ptmiss resolution & $\pm$1.6\% & $\pm$1.6\% & $\pm$0.3 & $\pm$0.1\% & \NA & \NA & \NA & \NA\\
			[\cmsTabSkip]
			\tauh \textsc{FastSim}/\GEANTfour & $\pm$0.7\% & $\pm$0.7\% & $\pm$0.9\% & $\pm$1.3\%& \NA & \NA & \NA & \NA\\
			[\cmsTabSkip]
			$\PGm$ \textsc{FastSim}/\GEANTfour & $\pm$1.7\% & $\pm$1.4\% & $\pm$2.9\% & $\pm$3.1\%& \NA & \NA & \NA & \NA\\
			[\cmsTabSkip]
			JER & +0.6\% & +0.3\%  &$<$0.1\% & +0.1\%& \NA & \NA & +4.2\% & +0.1\%\\
			~& -0.1\% & -0.5\% &$<$0.1\% & $<$0.1\% & \NA & \NA & -1.5\% & -0.4\%\\
			[\cmsTabSkip]
			2018 $\mTii$ uncertainty & \NA & \NA & \NA & \NA & $<$0.1\% & $<$0.1\% & $<$0.1\% & $<$0.1\%\\
			[\cmsTabSkip]
			JES & +0.1\% & +0.2\% &$<$0.1\% & +0.1\%& \NA & \NA & +4.7\% & +0.4\%\\
			~& -0.3\% & -0.5\% &$<$0.1\% & -0.1\%& \NA & \NA & -3.0\% & -0.4\%\\
			[\cmsTabSkip]
			$\mu_{\text{R}}$ and $\mu_{\text{F}}$ scales & 0.5\% & +0.8\% &+0.2\%& +0.2\%&\NA & \NA & +4.0\% & +4.9\%\\
			~ & -0.5\% & -0.8\% &-0.3\%& -0.3\%& \NA & \NA & -5.1\% & -5.1\%\\
			[\cmsTabSkip]
			\tauh Id-iso & +3.2\% & +3.2\% &+3.2\% & +3.2\%& +3.1\% & +3.1\% & 3.1\% & +1.6\%\\ 
			~& -3.9\% & -3.8\% &-4.1\% & -4.1\%& -3.8\% & -3.9\% & -3.6\% & -1.3\%\\
			[\cmsTabSkip]
			Pileup & +1.1\% & +0.2\% &+0.5& +0.7\%& \NA & \NA & +0.7\% & +0.3\%\\
			~ & -1.1\% & -0.2\% &-0.5& -0.7\%& \NA & \NA & -0.7\% & -0.3\%\\
			[\cmsTabSkip]
			\ptmiss unclustered energy & $<$0.1\% & $<$0.1\% &+0.1\% & $<$0.1\% & \NA & \NA & +5.0\% & 0.2\%\\
			~ & $<$0.1\% & 0.1\% &$<$0.1\% & -0.1\%& \NA & \NA & -3.2\% & -0.3\%\\
			[\cmsTabSkip]
			Background normalization & \NA & \NA & \NA & \NA & \NA & \NA & $\pm$15\%  & \NA\\
			[\cmsTabSkip]
			b tagging & $<$0.1\% & $\pm$0.1\% &$\pm$0.14\% & $\pm$0.4\% & \NA & \NA & $\pm$5.3\% & $\pm$0.7\%\\
			[\cmsTabSkip]
			Luminosity & $\pm$2.1\% & $\pm$2.1\% & $\pm$2.1\% & $\pm$2.1\% & \NA & \NA & $\pm$2.1\% & \NA\\
			[\cmsTabSkip]
			2017 $\ptmiss$ uncertainty & \NA & \NA & \NA & \NA & $<$0.1\% & $<$0.1\% & $<$0.1\% & $<$0.1\% \\
			[\cmsTabSkip]
			\tauh energy scale & -0.6\% & -0.05\% & -0.3\% & $<$0.1\% & +0.1\% & +0.1\% & +2.5\% & +0.1\%\\
			~ & -0.1\% & -0.6\% &-0.1\% & $<$0.1\% & -0.1\% & -0.1\% & -3.8\% & -0.1\%\\
			[\cmsTabSkip]
			Trigger & $<$0.1\% & $<$0.1\% &$<$0.1\% &$<$0.1\% &$<$0.1\% & $<$0.1\% & $<$0.1\% & $<$0.1\%\\
			[\cmsTabSkip]
			$\ttbar$ + single \PQt SF & \NA & \NA & \NA &\NA& $\pm$3.8\% & $\pm$3.9\% & \NA & \NA\\
			[\cmsTabSkip]
			\tauh misid. rate (parton flavor) & \NA & \NA &\NA& \NA & \NA & \NA & \NA & $\pm$30\%\\
			[\cmsTabSkip]
			Non-{\PW}+jets background & & & & & & & &\\
			modeling in $R$ & \NA & \NA &\NA& \NA & \NA & \NA & \NA & $\pm$10\%\\
			\hline
		\end{tabular}
	}
	\label{tab:unc_MuTauFullRun2}
\end{table}

\section{Results}
\label{sec:results}

\begin{figure}[!h]
	\centering
	\includegraphics[width=0.495\textwidth]{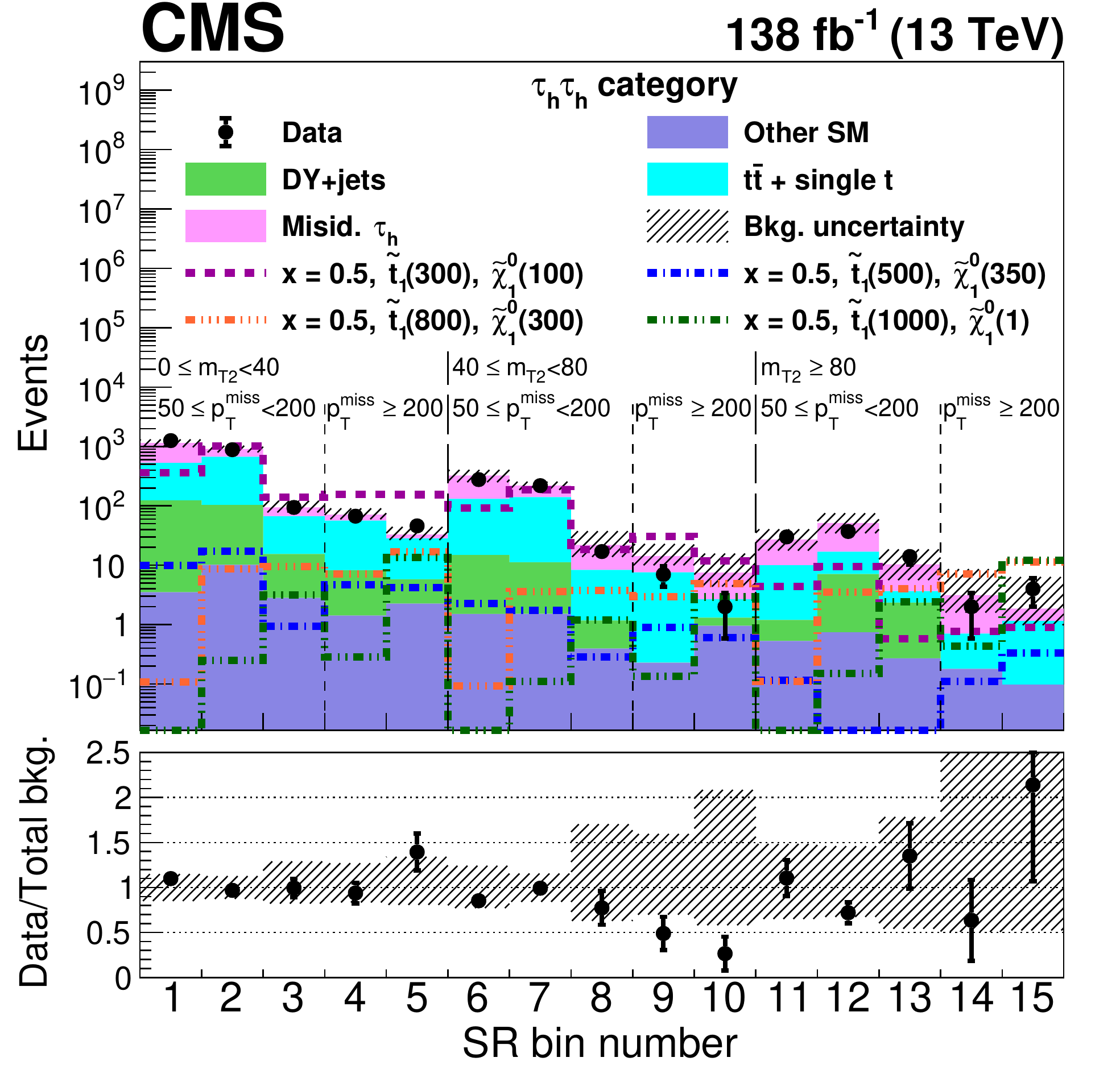} \\
	\includegraphics[width=0.495\textwidth]{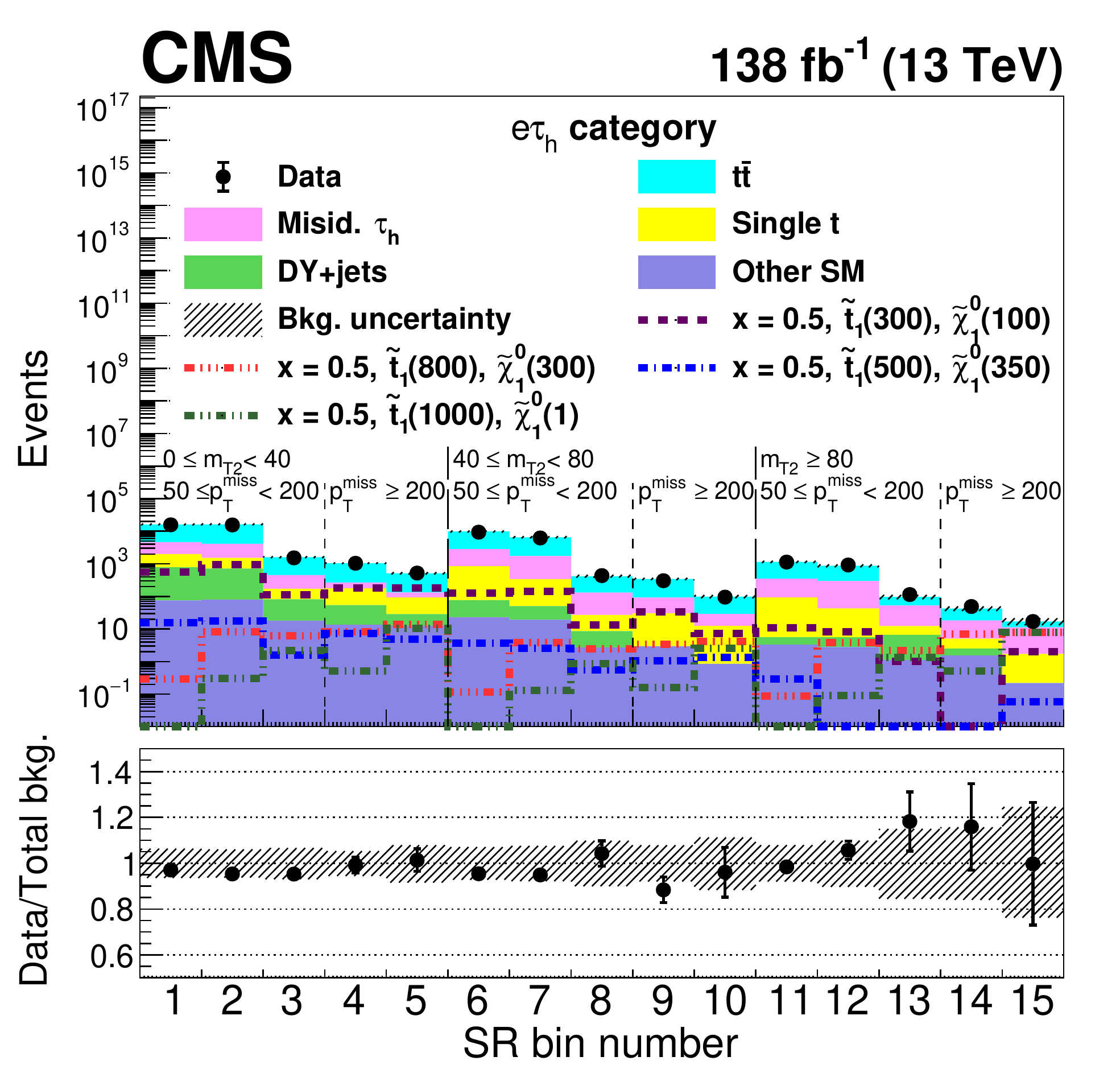}
	\includegraphics[width=0.495\textwidth]{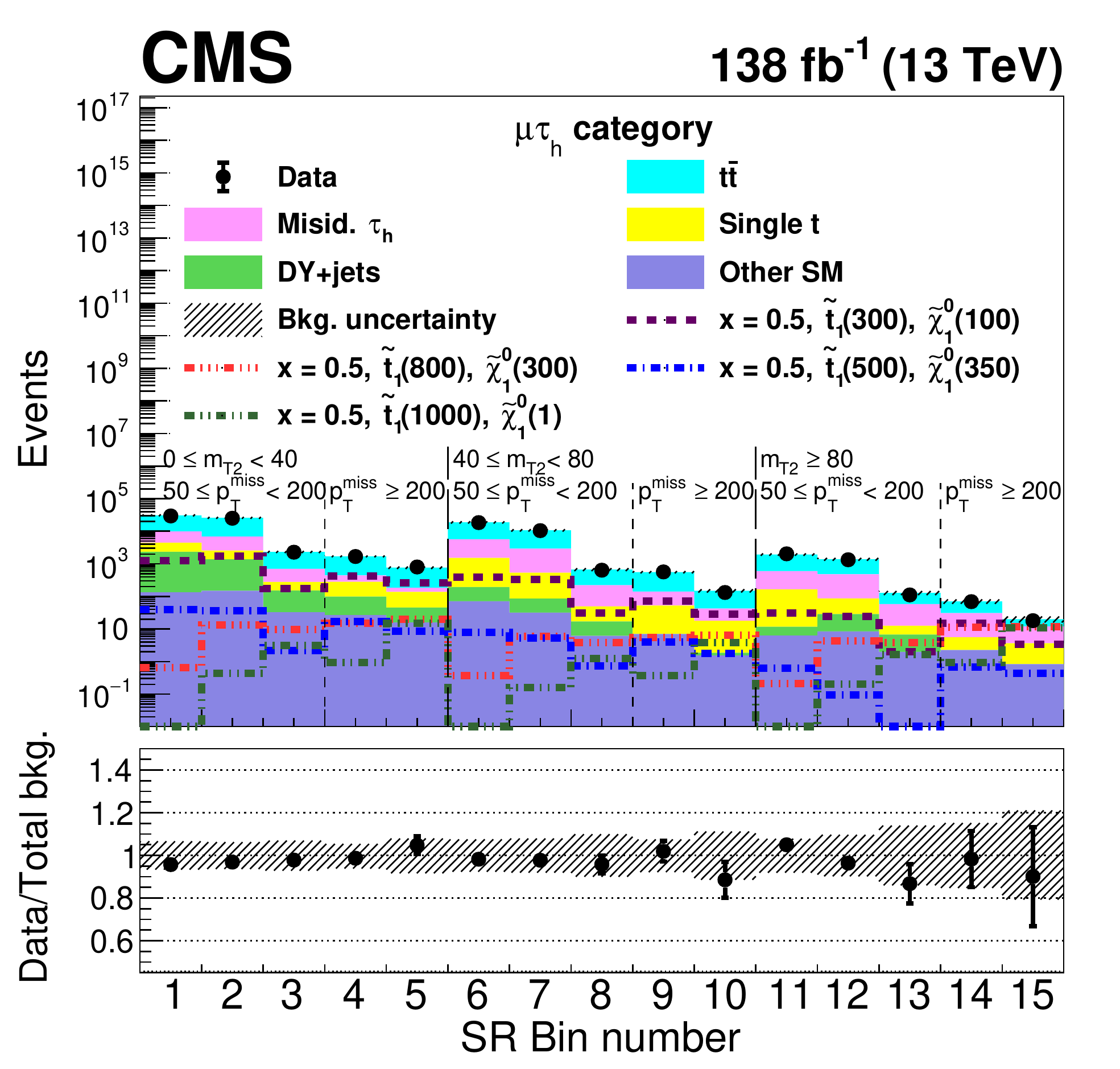}
	
	\caption{Event yields in the 15 search bins as defined in Fig. \ref{fig:sigBin}, for the \tauhtauh (upper), $\etauh$ (lower left), and $\mutauh$ (lower right) categories . The yields for the background processes are stacked, and those expected for a few representative sets of model parameter values are overlaid: $x = 0.5$ and [$m_{\PSQtDo}$, $m_{\PSGczDo}$] = [300, 100], [500, 350], [800, 300], and [1000, 1] \GeV. The \ptmiss and \mTii bin definitions are shown in \GeV. The lower panel indicates the ratio of the observed number of events to the total predicted number of background events in each bin. The shaded bands indicate the statistical and systematic uncertainties in the background, added in quadrature. The predicted yields and uncertainties shown here are prior to the maximum likelihood fit described in Sec.~\ref{sec:results}.}
	\label{fig:BGyield}
\end{figure}

We present the observed and expected yields along with their uncertainties in all 15 search bins in Tables \ref{tab:TauTau_fullRun2}, \ref{tab:EleTau_fullRun2}, and \ref{tab:MuTau_fullRun2} for the \tauhtauh, $\etauh$, and $\mutauh$ categories, respectively.
Figure \ref{fig:BGyield} shows the observed data in all search bins compared with the signal and background predictions.
As expected, the dominant contributions in the sensitive signal bins are from \ttbar and misidentified  \tauh backgrounds.
In cases where the background prediction of a process in a given bin is negligible, the statistical uncertainty is modeled by a gamma distribution~\cite{CMS-NOTE-2011-005} in the likelihood function used for the statistical interpretation, and the Poissonian upper limit at 68\% confidence level (\CL) is shown as a positive uncertainty in the Tables \ref{tab:TauTau_fullRun2}, \ref{tab:EleTau_fullRun2} and \ref{tab:MuTau_fullRun2}.
The number of events observed in data is found to be consistent with the SM background prediction.

The test statistic used for the interpretation of the result is the profile likelihood ratio $q_{\mu} = -2 \ln{(\mathcal{L}_{\mu} / \mathcal{L}_{\text{max}})}$, where $ \mathcal{L}_{\mu} $ is the maximum likelihood for a fixed signal strength modifier $ \mu $, and $ \mathcal{L}_{\text{max}} $ is the global maximum of the likelihood~\cite{CMS-NOTE-2011-005}.
The systematic uncertainties discussed in Section~\ref{sec:syst_unc} are modeled by log-normal distributions~\cite{CMS-NOTE-2011-005} in the likelihood function.
We set upper limits on signal production at 95\% \CL using a modified frequentist approach with a  \CLs criterion~\cite{JUNK1999435, Read_2002} that is implemented through an asymptotic approximation of the test statistic \cite{Cowan2011}.
In this calculation all the background and signal uncertainties are incorporated as nuisance parameters and profiled in the maximum likelihood fit~\cite{CMS-NOTE-2011-005}.

Final results are obtained by simultaneously fitting all the SR bins in the \tauhtauh, $\etauh$, and $\mutauh$ categories from the 2016, 2017, and 2018 data sets.
The contributions from \ttbar, single top quark, DY+jets, and misidentified \tauh candidates are modeled separately in the fit, whereas the rest of the minor SM backgrounds are treated as a single component.
The uncertainty in the integrated luminosity is treated as partially correlated between the three data sets.
The systematic uncertainties due to JES, factorization and renormalization scales, misidentification rate measurement, and \textsc{FastSim} \ptmiss correction are assumed to be  correlated, and the rest of the uncertainties are treated as uncorrelated among the three data sets.
All sources of systematic uncertainties that are common to the \tauhtauh and $\ltauh$ categories, are assumed to be correlated among the categories.

\begin{figure}[!htbp]
	\centering
	
	\includegraphics[width=0.495\textwidth]{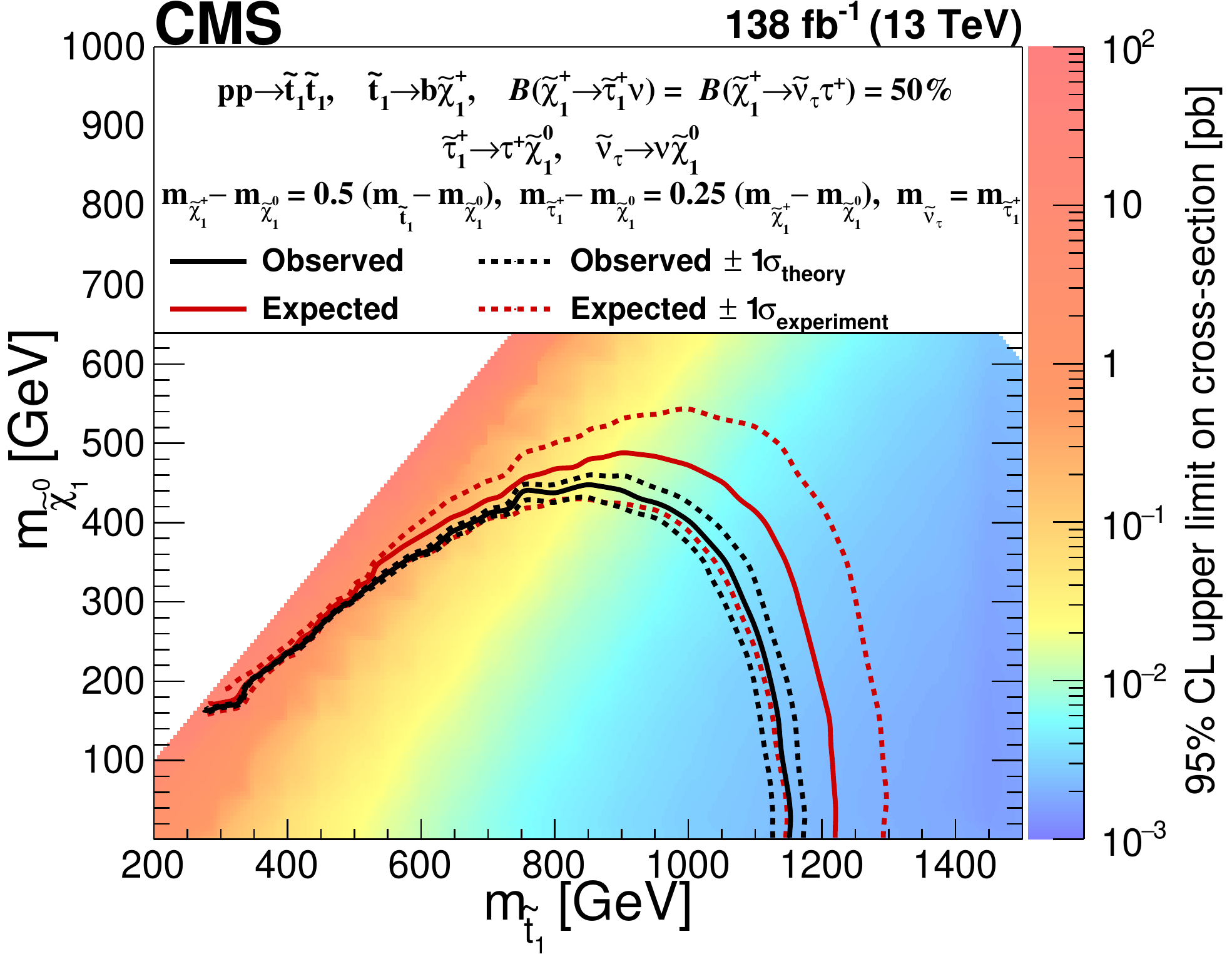}
	\includegraphics[width=0.495\textwidth]{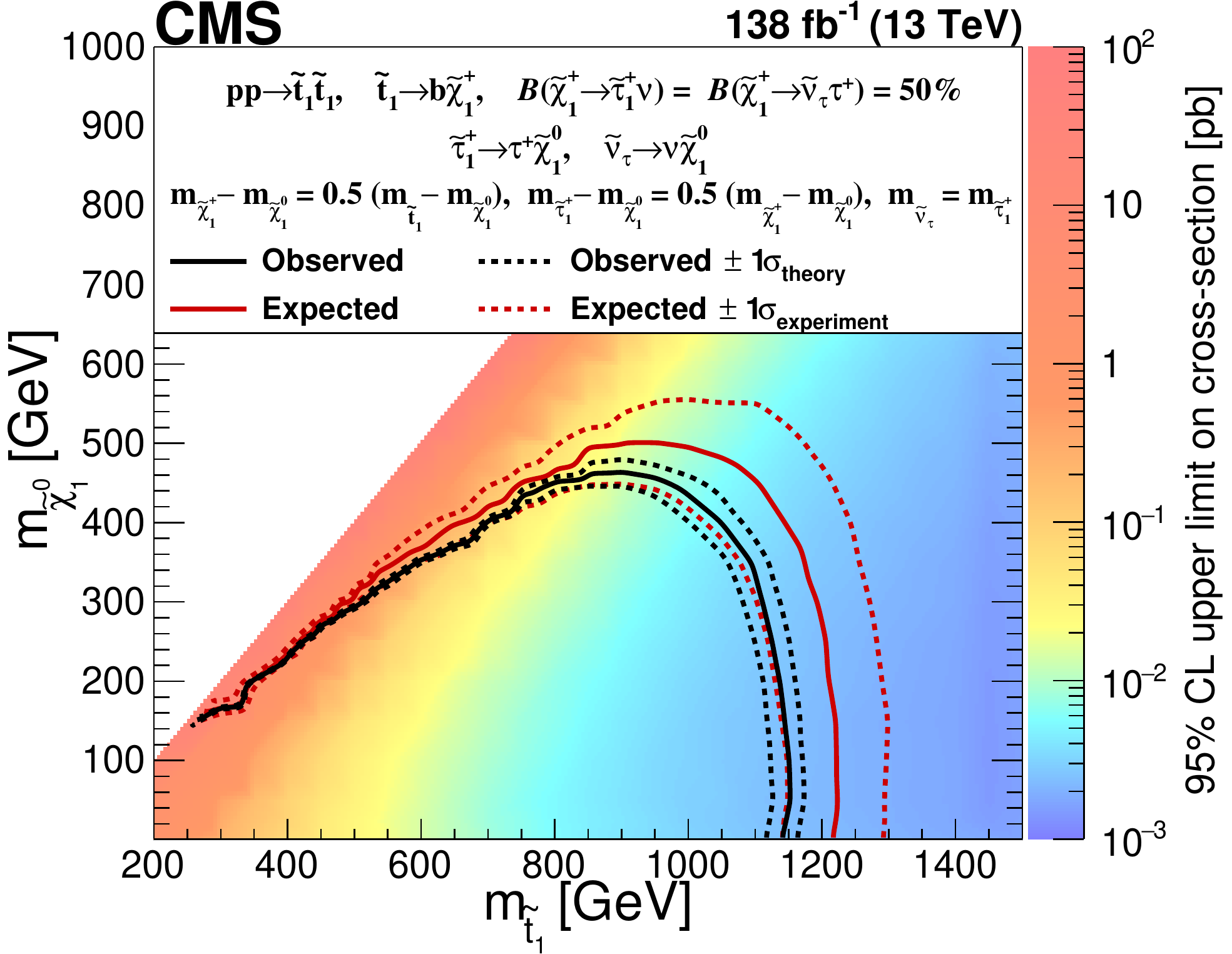} \\
	\includegraphics[width=0.495\textwidth]{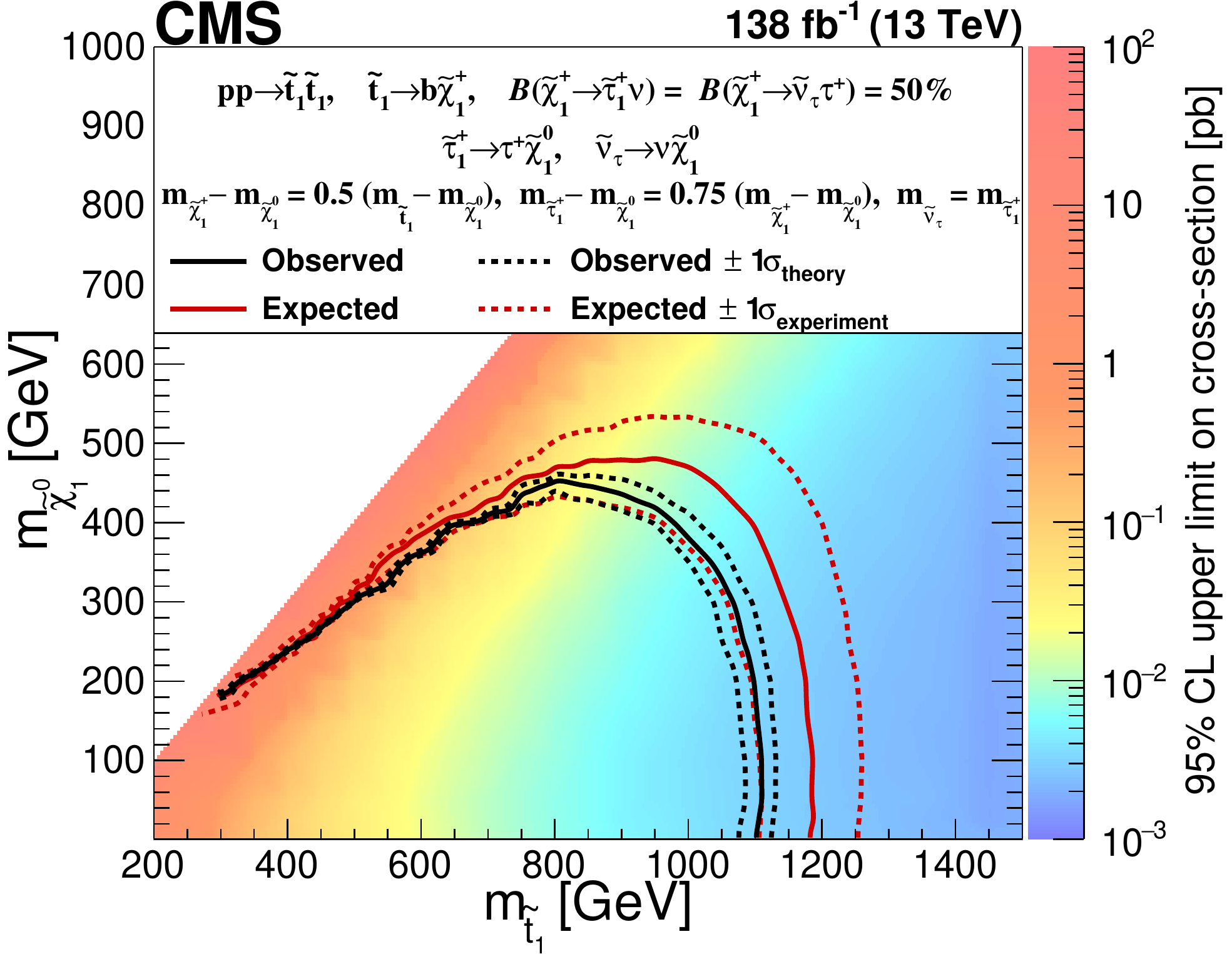}
	
	\caption{
		Exclusion limits at 95\% \CL for the pair production of top squarks decaying to \taultauh or \tauhtauh final states, displayed in the $ {m_{\PSQtDo}}$-$m_{\PSGczDo} $ plane for $ x =  0.25$ (upper left), 0.5 (upper right) and 0.75 (lower), as described in Eq.~(\ref{eq:mass}).
		Branching fractions are denoted by $\mathit{B}$.
		The color axis represents the observed upper limit in the cross section, while the black (red) lines represent the observed (expected) upper mass limits.
		The signal cross sections are evaluated using NNLO$+$NLL calculations.
		The solid lines represent the central values.
		The dashed red lines indicate the region containing 68\% of the distribution of limits expected under the background-only hypothesis. The dashed black lines show the change in the observed limit due to variation of the signal cross sections within their theoretical uncertainties.
	}
	\label{fig:2Dlimit}
\end{figure}

The observed and expected exclusion limits are presented in the plane of the top squark and LSP masses, in Fig.~\ref{fig:2Dlimit}.
Top squark masses up to $ 1150\GeV$ are excluded for a nearly massless LSP, and LSP masses up to $ 450\GeV$ are excluded for a top squark mass of $900\GeV$.
The exclusion limits are not very sensitive to the choice of the $\PSGt_{1}$ mass parameter $x$ because of the complementary nature of the signal diagrams, as discussed in Section \ref{sec:mc_sim}.

The final limits are generally driven by the yields in the \tauhtauh category because of its higher signal-to-background ratio compared with the $\ltauh$ category.
The most sensitive search bin for the higher top squark masses (${\approx}$\TeV) is bin 15, which is the highest \ptmiss, \mTii and \HT bin.
The observed $\tauhtauh$ yield in this bin is greater than the total background prediction, resulting in the observed limit being lower than the expected one by approximately one standard deviation in that region of the $ {m_{\PSQtDo}}$-$m_{\PSGczDo} $ plane.
The total background predictions in bins 9 and 10 of the $\tauhtauh$ category and bin 9 of the $\etauh$ category are greater than the observed yields by about 1--2 standard deviations. These correspond to the highest \ptmiss and intermediate \mTii bins. However, these bins are not among the most sensitive ones and hence do not affect the final limits to any appreciable degree.
The limits become weaker with decreasing $ \Delta m = m_{\PSQtDo} - m_{\PSGczDo} $, corresponding to a parameter space with final-state particles having lower momentum.

\begin{table}[!htbp]
	\centering
	\topcaption{Predicted background yields along with uncertainties for the \tauhtauh category in the 15 search bins, as defined in Fig.~\ref{fig:sigBin}. The number of events observed in data is also shown. The first uncertainty value listed is statistical and the second is systematic. The uncertainties smaller than 0.05 are listed as 0.0. The background yields and uncertainties shown here are prior to the maximum likelihood fit described in Sec.~\ref{sec:results}.}
		\begin{tabular}{ l  c  c  c  c  c  c }
			\hline
			\\[\cmsTabSkipSmall]
			SR & 
			$\ttbar$ + single t & 
			DY+jets & 
			Other SM & 
			Misid. \tauh & 
			Total bkg. & 
			Data \\[\cmsTabSkip]
			
			\hline
			\\[\cmsTabSkipSmall]
			1 & 
			$ 407 ^{+9 +36} _{-9 -36} $ & 
			$ 120 ^{+14 +29} _{-14 -30} $ & 
			$ 3.5 ^{+1.4 +1.0} _{-1.4 -1.1} $ & 
			$ 612 ^{+44 +154} _{-44 -155} $ & 
			$ 1142 ^{+47 +164} _{-47 -165} $ & 
			$ 1255 $ \\[\cmsTabSkip]
			
			2 & 
			$ 568 ^{+11 +49} _{-11 -51} $ & 
			$ 94 ^{+10 +24} _{-10 -18} $ & 
			$ 10 ^{+3 +2} _{-3 -2} $ & 
			$ 239 ^{+27 +90} _{-27 -92} $ & 
			$ 911 ^{+31 +111} _{-31 -111} $ & 
			$ 882 $ \\[\cmsTabSkip]
			
			3 & 
			$ 51 ^{+3 +4} _{-3 -4} $ & 
			$ 13 ^{+3 +3} _{-3 -2} $ & 
			$ 2.8 ^{+0.6 +0.6} _{-0.6 -0.8} $ & 
			$ 28 ^{+8 +25} _{-8 -13} $ & 
			$ 95 ^{+9 +26} _{-9 -14} $ & 
			$ 94 $ \\[\cmsTabSkip]
			
			4 & 
			$ 48 ^{+3 +4} _{-3 -4} $ & 
			$ 6.8 ^{+2.1 +2.4} _{-2.1 -2.1} $ & 
			$ 1.4 ^{+0.7 +0.4} _{-0.7 -0.4} $ & 
			$ 15 ^{+6 +17} _{-6 -8} $ & 
			$ 71 ^{+7 +18} _{-7 -9} $ & 
			$ 67 $ \\[\cmsTabSkip]
			
			5 & 
			$ 23 ^{+2 +2} _{-2 -2} $ & 
			$ 3.5 ^{+5.1 +0.9} _{-1.7 -0.7} $ & 
			$ 2.3 ^{+1.1 +0.5} _{-1.1 -0.5} $ & 
			$ 4.6 ^{+4.2 +8.3} _{-4.2 -2.6} $ & 
			$ 33 ^{+7 +9} _{-5 -4} $ & 
			$ 46 $ \\[\cmsTabSkip]
			
			6 & 
			$ 116 ^{+5 +12} _{-5 -12} $ & 
			$ 13 ^{+3 +3} _{-3 -6} $ & 
			$ 1.5 ^{+1.2 +0.4} _{-1.2 -0.7} $ & 
			$ 194 ^{+21 +74} _{-21 -69} $ & 
			$ 324 ^{+22 +75} _{-22 -71} $ & 
			$ 277 $ \\[\cmsTabSkip]
			
			7 & 
			$ 129 ^{+5 +13} _{-5 -14} $ & 
			$ 9.7 ^{+2.9 +5.0} _{-2.9 -5.2} $ & 
			$ 1.5 ^{+1.4 +0.4} _{-1.4 -0.4} $ & 
			$ 81 ^{+15 +25} _{-15 -27} $ & 
			$ 221 ^{+16 +30} _{-16 -31} $ & 
			$ 219 $ \\[\cmsTabSkip]
			
			8 & 
			$ 7.2 ^{+1.2 +0.8} _{-1.2 -0.7} $ & 
			$ 0.8 ^{+4.1 +0.2} _{-0.4 -0.4} $ & 
			$ 0.4 ^{+0.1 +0.1} _{-0.1 -0.1} $ & 
			$ 14 ^{+4 +14} _{-4 -7} $ & 
			$ 22 ^{+6 +14} _{-4 -7} $ & 
			$ 17 $ \\[\cmsTabSkip]
			
			9 & 
			$ 7.4 ^{+1.2 +0.7} _{-1.2 -0.7} $ & 
			$ 0.0 ^{+3.5 +0.0} _{-0.0 -0.0} $ & 
			$ 0.2 ^{+0.2 +0.1} _{-0.2 -0.1} $ & 
			$ 6.7 ^{+2.3 +7.3} _{-2.3 -3.4} $ & 
			$ 14 ^{+4 +7} _{-3 -3} $ & 
			$ 7 $ \\[\cmsTabSkip]
			
			10 & 
			$ 1.4 ^{+0.6 +0.1} _{-0.6 -0.4} $ & 
			$ 0.4 ^{+6.0 +0.1} _{-0.4 -0.1} $ & 
			$ 1.0 ^{+0.8 +0.5} _{-0.8 -0.4} $ & 
			$ 4.9 ^{+1.3 +5.3} _{-1.3 -2.6} $ & 
			$ 7.7 ^{+6.2 +5.4} _{-1.6 -2.7} $ & 
			$ 2 $ \\[\cmsTabSkip]
			
			11 & 
			$ 8.8 ^{+1.4 +1.0} _{-1.4 -1.2} $ & 
			$ 0.7 ^{+5.4 +0.2} _{-0.7 -0.7} $ & 
			$ 0.5 ^{+0.3 +0.2} _{-0.3 -0.2} $ & 
			$ 17 ^{+7 +10} _{-7 -6} $ & 
			$ 27 ^{+9 +10} _{-7 -7} $ & 
			$ 30 $ \\[\cmsTabSkip]
			
			12 & 
			$ 9.8 ^{+1.5 +1.7} _{-1.5 -1.2} $ & 
			$ 6.4 ^{+2.3 +4.8} _{-2.3 -2.0} $ & 
			$ 0.7 ^{+0.2 +0.2} _{-0.2 -0.2} $ & 
			$ 35 ^{+7 +22} _{-7 -15} $ & 
			$ 52 ^{+7 +22} _{-7 -15} $ & 
			$ 37 $ \\[\cmsTabSkip]
			
			13 & 
			$ 1.3 ^{+0.6 +0.2} _{-0.5 -0.4} $ & 
			$ 2.1 ^{+5.0 +1.0} _{-1.3 -0.8} $ & 
			$ 0.3 ^{+0.1 +0.1} _{-0.1 -0.1} $ & 
			$ 6.7 ^{+3.1 +5.5} _{-3.1 -3.2} $ & 
			$ 10 ^{+6 +6} _{-3 -3} $ & 
			$ 14 $ \\[\cmsTabSkip]
			
			14 & 
			$ 0.5 ^{+0.7 +0.1} _{-0.5 -0.1} $ & 
			$ <3.5 $ & 
			$ 0.2 ^{+0.1 +0.0} _{-0.1 -0.0} $ & 
			$ 2.4 ^{+1.9 +3.7} _{-0.6 -1.3} $ & 
			$ 3.1 ^{+4.0 +3.7} _{-0.8 -1.3} $ & 
			$ 2 $ \\[\cmsTabSkip]
			
			15 & 
			$ 1.1 ^{+0.6 +0.2} _{-0.6 -0.2} $ & 
			$ <3.5 $ & 
			$ 0.1 ^{+0.0 +0.0} _{-0.0 -0.0} $ & 
			$ 0.7 ^{+2.6 +0.6} _{-0.5 -0.4} $ & 
			$ 1.9 ^{+4.4 +0.7} _{-0.8 -0.4} $ & 
			$ 4 $ \\[\cmsTabSkip]
			
			Total & 
			$ 1380 ^{+17 +123} _{-17 -126} $ & 
			$ 270 ^{+23 +66} _{-19 -59} $ & 
			$ 26 ^{+4 +6} _{-4 -6} $ & 
			$ 1261 ^{+59 +461} _{-59 -403} $ & 
			$ 2937 ^{+66 +482} _{-65 -427} $ & 
			$ 2953 $ \\[\cmsTabSkip]
			
			\hline
		\end{tabular}
	\label{tab:TauTau_fullRun2}
\end{table}

\begin{table}[!htbp]
	\centering
	\topcaption{Predicted background yields along with uncertainties for the \etauh category in the 15 search bins, as defined in Fig.~\ref{fig:sigBin}. The number of events observed in data is also shown. The first uncertainty value listed is statistical and the second is systematic. The uncertainties smaller than 0.05 are listed as 0.0. The background yields and uncertainties shown here are prior to the maximum likelihood fit described in Sec.~\ref{sec:results}.}
		\begin{tabular}{l c c c c c c}
			\hline
			\\[\cmsTabSkipSmall]
			SR& 
			$\ttbar$ & 
			Single \PQt & 
			(DY+jets) &
			Misid. \tauh &  
			Total bkg. & 
			Data\\
			~ & ~ & ~ & +Other SM & ~ & ~ & ~\\[\cmsTabSkip]
			\hline
			\\[\cmsTabSkipSmall]
			1 & 
			$11574^{+50+634}_{-50-651}$ & 
			$1210^{+15+62}_{-15-66}$ & 
			$793^{+34+74}_{-34-90}$ &  
			$2646^{+29+849}_{-29-848}$ &
			$16222^{+69+1064}_{-68-1075}$ & 
			$15744$\\[\cmsTabSkip]
			2 & 
			$12239^{+50+568}_{-50-630}$ & 
			$799^{+12+50}_{-12-50}$ & 
			$717^{+26+45}_{-26-55}$ &  
			$2619^{+30+846}_{-30-845}$ &
			$16374^{+65+1021}_{-65-1057}$ & 
			$15605$\\[\cmsTabSkip]
			
			3 & 
			$1151^{+15+57}_{-15-63}$ & 
			$90^{+4+11}_{-4-9}$ & 
			$84^{+7+8}_{-7-6}$ &
			$277^{+10+91}_{-10-94}$ &  
			$1601^{+20+108}_{-20-114}$ & 
			$1524$\\[\cmsTabSkip]
			
			4 & 
			$779^{+13+43}_{-13-46}$ & 
			$123^{+5+10}_{-5-9}$ & 
			$55^{+6+4}_{-6-8}$ &
			$92^{+6+31}_{-6-32}$ &  
			$1048^{+16+54}_{-16-57}$ & 
			$1039$\\[\cmsTabSkip]
			
			5 & 
			$381^{+8+34}_{-8-35}$ & 
			$65^{+4+9}_{-4-7}$ & 
			$30^{+5+4}_{-5-3}$ &  
			$39^{+5+18}_{-5-22}$ &
			$514^{+11+40}_{-11-43}$ & 
			$520$\\[\cmsTabSkip]
			
			6 & 
			$6984^{+40+335}_{-40-368}$ & 
			$774^{+12+38}_{-12-43}$ & 
			$78^{+11+35}_{-11-8}$ &  
			$1989^{+24+635}_{-24-635}$ &
			$9825^{+49+720}_{-49-735}$ & 
			$9372$\\[\cmsTabSkip]
			
			7 & 
			$4822^{+32+251}_{-32-285}$ & 
			$290^{+7+18}_{-7-19}$ & 
			$52^{+6+19}_{-6-6}$ &  
			$1395^{+21+447}_{-21-447}$ &
			$6559^{+39+513}_{-38.9-530}$ & 
			$6222$\\[\cmsTabSkip]
			
			8 & 
			$287^{+8+23}_{-8-24}$ & 
			$18^{+2+2}_{-2-2}$ & 
			$9.2^{+1.9+6.5}_{-1.9-1.1}$ &  
			$104^{+6+34}_{-6-34}$ &
			$418^{+10+41}_{-10-42}$ & 
			$435$\\[\cmsTabSkip]
			
			9 & 
			$251^{+7+17}_{-7-18}$ &  
			$27^{+2+2}_{-2-2}$ & 
			$3.2^{+1.3+2.8}_{-1.3-0.6}$ & 
			$62^{+4+20}_{-4-20}$ &
			$343^{+9+26}_{-9-27}$ & 
			$303$\\[\cmsTabSkip]
			
			10 & 
			$70^{+4+8}_{-4-9}$ & 
			$12^{+1+1}_{-1-1}$ & 
			$1.1^{+0.3+0.3}_{-0.3-0.3}$ & 
			$17^{+2.6+5.7}_{-2.6-6.1}$ & 
			$99^{+4.8+10}_{-4.8-11}$ & 
			$95$\\[\cmsTabSkip]
			
			11 & 
			$800^{+14+41}_{-14-44}$ & 
			$87^{+4+5}_{-4-6}$ & 
			$5.9^{+2.1+1.2}_{-2.1-2.0}$ &  
			$257^{+8+82}_{-8-83}$ &
			$1150^{+17+94}_{-17-94}$ & 
			$1131$\\[\cmsTabSkip]
			
			12 & 
			$575^{+11+35}_{-11-43}$ & 
			$37^{+3+3}_{-3-3}$ & 
			$6.4^{+2.1+8.1}_{-2.1-0.8}$ &  
			$254^{+8+81}_{-8-82}$ &
			$873^{+14+89}_{-14-92}$ & 
			$921$\\[\cmsTabSkip]
			
			13 & 
			$44^{+3+6}_{-3-6}$ & 
			$5.7^{+1.1+1.0}_{-1.1-0.7}$ & 
			$6.8^{+2.8+0.9}_{-2.8-3.3}$ & 
			$40^{+3+13}_{-3-13}$ & 
			$97^{+5+14}_{-5-14}$ & 
			$114$\\[\cmsTabSkip]
			
			14 & 
			$24^{+2+4}_{-2-4}$ & 
			$2.6^{+0.7+0.3}_{-0.7-0.3}$ & 
			$2.7^{+1.2+0.6}_{-1.2-0.9}$ & 
			$13^{+2+4.2}_{-2-4.4}$ & 
			$42^{+3+5.9}_{-3-6.1}$ & 
			$49$\\[\cmsTabSkip]
			
			15 & 
			$5.8^{+0.9+1.8}_{-0.9-1.7}$ & 
			$1.5^{+0.6+0.2}_{-0.6-0.2}$ & 
			$0.3^{+0.1+0.1}_{-0.1-0.1}$ & 
			$9.5^{+1.6+3.4}_{-1.6-3.3}$ & 
			$17^{+2+3.9}_{-2-3.7}$ & 
			$17$\\[\cmsTabSkip]
			
			Total & 
			$39985^{+92+2006}_{-92-2176}$ &  
			$3543^{+26+211}_{-26-217}$ & 
			$1844^{+46+171}_{-46-170}$ & 
			$9811^{+56+3152}_{-56-3154}$ &
			$55183^{+120+3745}_{-120-3841}$ & 
			$53122$\\[\cmsTabSkip]
			\hline
		\end{tabular}
	\label{tab:EleTau_fullRun2}
\end{table}

\begin{table}[!htbp]
	\centering
	\topcaption{Predicted background yields along with uncertainties for the \mutauh category in the 15 search bins, as defined in Fig.~\ref{fig:sigBin}. The number of events observed in data is also shown. The first uncertainty value listed is statistical and the second is systematic. The uncertainties smaller than 0.05 are listed as 0.0. The background yields and uncertainties shown here are prior to the maximum likelihood fit described in Sec.~\ref{sec:results}.}
		\begin{tabular}{l c c c c c c }
			\hline
			\\[\cmsTabSkipSmall]
			SR& 
			$\ttbar$ & 
			Single \PQt & 
			(DY+jets) &   
			Misid. \tauh &
			Total bkg. &
			Data\\
			~ & ~ & ~ & +Other SM & ~ & ~ & ~\\ [\cmsTabSkip]
			\hline
			\\[\cmsTabSkipSmall]
			1 & 
			$20947^{+70+1147}_{-70-1178}$ & 
			$2152^{+21+109}_{-21-118}$ & 
			$2340^{+61+338}_{-61-299}$ &  
			$5391^{+41+1726}_{-41-1724}$ &
			$30801^{+104+2102}_{-104-1212}$ & 
			$29475$\\[\cmsTabSkip]
			
			2 & 
			$18973^{+65+876}_{-65-972}$ & 
			$1206^{+16+75}_{-16-76}$ & 
			$1359^{+37+92}_{-37-97}$ & 
			$4340^{+38+1397}_{-38-1398}$ &
			$25861^{+85+1654}_{-85-1707}$ & 
			$25055$\\[\cmsTabSkip]
			
			3 & 
			$1624^{+18+80}_{-18-90}$ & 
			$126^{+5+14}_{-5-12}$ & 
			$151^{+10+14}_{-10-14}$ & 
			$424^{+12+139}_{-12-144}$ & 
			$2323^{+25+162}_{-25-170}$ & 
			$2273$\\[\cmsTabSkip]
			
			4 & 
			$1258^{+17+70}_{-17-75}$ & 
			$ 182^{+6+14}_{-6-13}$ & 
			$98^{+11+8}_{-11-27}$ &  
			$163^{+8+55}_{-8-56}$ &
			$1700^{+23+91}_{-23-99}$ & 
			$1678$\\[\cmsTabSkip]
			
			5 & 
			$579^{+10+52}_{-10-54}$  & 
			$95^{+4+13}_{-4-10}$ & 
			$45^{+6+4}_{-6-5}$ &  
			$47^{+6+24}_{-6-30}$ &
			$764^{+14+59}_{-14-63}$ & 
			$800$\\[\cmsTabSkip]
			
			6 & 
			$13094^{+56+633}_{-56-692}$  & 
			$1358^{+17+68}_{-17-75} $ & 
			$193^{+13+55}_{-13-28}$ &  
			$4132^{+34+1317}_{-34-1317}$ &
			$18752^{+69+1464}_{-69-1490}$ & 
			$18412$\\[\cmsTabSkip]
			
			7 & 
			$7754^{+42+409}_{-42-459}$ & 
			$453^{+10+30}_{-10-32}$ & 
			$85^{+8+29}_{-8-7}$ &  
			$2398^{+27+768}_{-27-768}$ &
			$10685^{+51+871}_{-51-896}$ & 
			$10441$\\[\cmsTabSkip]
			
			8 & 
			$444^{+10+36}_{-10-37}$ & 
			
			$33.4^{+3+4}_{-3-3.2}$ & 
			$17^{+3+8}_{-3-1.8}$ &  
			$172^{+7+55}_{-7-56}$ &
			$666^{+13+66}_{-13-68}$ & 
			$638$\\[\cmsTabSkip]
			
			9 & 
			$414^{+10+29}_{-10-31}$ &  
			$44.5^{+3.0+3.5}_{-3.0-3.5}$ & 
			$7.0^{+3.0+1.5}_{-3.0-2.4}$ & 
			$88^{+6+29}_{-6-29}$ &
			$554^{+12+41}_{-12-43}$ & 
			$565$\\[\cmsTabSkip]
			
			10 & 
			$107^{+5+13}_{-5-13}$ & 
			$16^{+2+2}_{-2-2}$ & 
			$1.9^{+1.0+2.9}_{-1.0-0.5}$ &  
			$24^{+3+8}_{-3-9}$ &
			$149^{+6+16}_{-6-16}$ & 
			$132$\\[\cmsTabSkip]
			
			11 & 
			$1332^{+18+67}_{-18-78}$ &
			$153^{+6+9}_{-6-10}$ & 
			$12^{+4+6}_{-4-1}$ &  
			$435^{+11.1+139}_{-11-140}$ &
			$1931^{+22+155}_{-22-161}$ & 
			$2027$\\[\cmsTabSkip]
			
			12 & 
			$905^{+15+56}_{-15-62}$ & 
			$59^{+4+4}_{-4-5}$ & 
			$29^{+5+8}_{-5-3}$ &  
			$391^{+10+124}_{-10-126}$ &
			$1383^{+19+137}_{-19-140}$ & 
			$1333$\\[\cmsTabSkip]
			
			13 & 
			$70^{+4+9}_{-4-9}$ &
			$6.7^{+2.0+0.6}_{-2.0-3.0}$ &
			$5.9^{+1.1+0.6}_{-1.1-0.6}$ & 
			$46^{+4+15}_{-4-15}$ &
			$128^{+6+17}_{-6-18}$ & 
			$111$\\[\cmsTabSkip]
			
			14 & 
			$39^{+3+6}_{-3-6}$ &
			$3.1^{+0.9+0.2}_{-0.9-0.2}$ & 
			$2.3^{+0.7+0.2}_{-0.7-0.3}$ &  
			$25^{+3+8}_{-3-8}$ &
			$70^{+4+10}_{-4-10}$ & 
			$69$\\[\cmsTabSkip]
			
			15 & 
			$8.1^{+1.2+2.5}_{-1.2-2.5}$ & 
			$2.7^{+0.8+0.4}_{-0.8-0.3}$ & 
			$0.8^{+0.2+0.2}_{-0.2-0.2}$ & 
			$8.3^{+1.5+2.6}_{-1.5-2.6}$ & 
			$20^{+2+4}_{-2-4}$ & 
			$18$\\[\cmsTabSkip]
			
			Total & 
			$67548^{+125+3395}_{-125-3676}$ &  
			$5890^{+35+350}_{-35-360}$ & 
			$4348^{+75+510}_{-75-449}$ & 
			$18083^{+75+5794}_{-75-5798}$ &
			$95870^{+167+6743}_{-167-6889}$ & 
			$93072$\\[\cmsTabSkip]
			\hline
		\end{tabular}
	\label{tab:MuTau_fullRun2}
\end{table}

\clearpage
\section{Summary}
\label{sec:Summary}

Top squark pair production in final states with two tau leptons has been explored in data collected by the CMS detector during 2016, 2017, and 2018, corresponding to an integrated luminosity of 138\fbinv.
This search improves upon the previous publication~\cite{CMS:2019lrh} by analyzing the entirety of the Run 2 data, adding the \etauh and \mutauh final states, and utilizing improved algorithms for identifying hadronically decaying tau leptons and \PQb quark jets. The dominant standard model backgrounds originate from top quark pair and single top quark production and processes where jets were misidentified as \tauh decays.
Control regions in data are used to estimate these backgrounds, whereas other backgrounds are estimated using simulation.
The simulated objects (leptons, jets, etc.) are corrected using scale factors to account for differences between their performance in simulation and collision data.
No significant excess is observed, and exclusion limits on the top squark and lightest neutralino masses are set at 95\% confidence level within the framework of simplified models where the top squark decays via a chargino to final states including tau leptons.
A branching fraction of 50\% is assumed for each of the two considered decay modes of the chargino, $ \PSGcpDo \to \PSGtpDo\PGnGt$ and $ \PSGcpDo \to \PGtp\PSGnGt$.
These decay modes are motivated by high-\tanb and higgsino-like scenarios where decays to tau leptons are enhanced.
In such models, top squark masses are excluded up to about 1150\GeV for a lightest supersymmetric particle (LSP) of mass 1\GeV, while LSP masses up to 450\GeV are excluded for a top squark mass of 900\GeV.
These are the most stringent exclusion limits to date for the signal models considered in this study.

\clearpage

\begin{acknowledgments}
We congratulate our colleagues in the CERN accelerator departments for the excellent performance of the LHC and thank the technical and administrative staffs at CERN and at other CMS institutes for their contributions to the success of the CMS effort. In addition, we gratefully acknowledge the computing c enters and personnel of the Worldwide LHC Computing Grid and other c enters for delivering so effectively the computing infrastructure essential to our analyses. Finally, we acknowledge the enduring support for the construction and operation of the LHC, the CMS detector, and the supporting computing infrastructure provided by the following funding agencies: BMBWF and FWF (Austria); FNRS and FWO (Belgium); CNPq, CAPES, FAPERJ, FAPERGS, and FAPESP (Brazil); MES and BNSF (Bulgaria); CERN; CAS, MoST, and NSFC (China); MINCIENCIAS (Colombia); MSES and CSF (Croatia); RIF (Cyprus); SENESCYT (Ecuador); MoER, ERC PUT and ERDF (Estonia); Academy of Finland, MEC, and HIP (Finland); CEA and CNRS/IN2P3 (France); BMBF, DFG, and HGF (Germany); GSRI (Greece); NKFIH (Hungary); DAE and DST (India); IPM (Iran); SFI (Ireland); INFN (Italy); MSIP and NRF (Republic of Korea); MES (Latvia); LAS (Lithuania); MOE and UM (Malaysia); BUAP, CINVESTAV, CONACYT, LNS, SEP, and UASLP-FAI (Mexico); MOS (Montenegro); MBIE (New Zealand); PAEC (Pakistan); MES and NSC (Poland); FCT (Portugal); MESTD (Serbia); MCIN/AEI and PCTI (Spain); MOSTR (Sri Lanka); Swiss Funding Agencies (Switzerland); MST (Taipei); MHESI and NSTDA (Thailand); TUBITAK and TENMAK (Turkey); NASU (Ukraine); STFC (United Kingdom); DOE and NSF (USA).

\hyphenation{Rachada-pisek} Individuals have received support from the Marie-Curie program and the European Research Council and Horizon 2020 Grant, contract Nos.\ 675440, 724704, 752730, 758316, 765710, 824093, 884104, and COST Action CA16108 (European Union); the Leventis Foundation; the Alfred P.\ Sloan Foundation; the Alexander von Humboldt Foundation; the Belgian Federal Science Policy Office; the Fonds pour la Formation \`a la Recherche dans l'Industrie et dans l'Agriculture (FRIA-Belgium); the Agentschap voor Innovatie door Wetenschap en Technologie (IWT-Belgium); the F.R.S.-FNRS and FWO (Belgium) under the ``Excellence of Science -- EOS" -- be.h project n.\ 30820817; the Beijing Municipal Science \& Technology Commission, No. Z191100007219010; the Ministry of Education, Youth and Sports (MEYS) of the Czech Republic; the Hellenic Foundation for Research and Innovation (HFRI), Project Number 2288 (Greece); the Deutsche Forschungsgemeinschaft (DFG), under Germany's Excellence Strategy -- EXC 2121 ``Quantum Universe" -- 390833306, and under project number 400140256 - GRK2497; the Hungarian Academy of Sciences, the New National Excellence Program - \'UNKP, the NKFIH research grants K 124845, K 124850, K 128713, K 128786, K 129058, K 131991, K 133046, K 138136, K 143460, K 143477, 2020-2.2.1-ED-2021-00181, and TKP2021-NKTA-64 (Hungary); the Council of Science and Industrial Research, India; the Latvian Council of Science; the Ministry of Education and Science, project no. 2022/WK/14, and the National Science Center, contracts Opus 2021/41/B/ST2/01369 and 2021/43/B/ST2/01552 (Poland); the Funda\c{c}\~ao para a Ci\^encia e a Tecnologia, grant CEECIND/01334/2018 (Portugal); the National Priorities Research Program by Qatar National Research Fund; MCIN/AEI/10.13039/501100011033, ERDF ``a way of making Europe", and the Programa Estatal de Fomento de la Investigaci{\'o}n Cient{\'i}fica y T{\'e}cnica de Excelencia Mar\'{\i}a de Maeztu, grant MDM-2017-0765 and Programa Severo Ochoa del Principado de Asturias (Spain); the Chulalongkorn Academic into Its 2nd Century Project Advancement Project, and the National Science, Research and Innovation Fund via the Program Management Unit for Human Resources \& Institutional Development, Research and Innovation, grant B05F650021 (Thailand); the Kavli Foundation; the Nvidia Corporation; the SuperMicro Corporation; the Welch Foundation, contract C-1845; and the Weston Havens Foundation (USA).
\end{acknowledgments}

\bibliography{auto_generated} 

\providecommand{\href}[2]{#2}\begingroup\raggedright\begin{thebibliography}{10}%
\makeatletter
\providecommand{\hrefCMSnoop }[0]{\@secondoftwo}%
\makeatother
\providecommand{\doi}{\texttt{doi:}\begingroup \urlstyle{tt}\Url}

\bibitem{Ramond:1971gb}
\hrefCMSnoop {}{P.~Ramond, ``Dual theory for free fermions'',} \textit{ Phys.
  Rev. D} \textbf{ 3} (1971) 2415,
\href{http://dx.doi.org/10.1103/PhysRevD.3.2415}{\doi{10.1103/PhysRevD.3.2415}}.

\bibitem{Golfand:1971iw}
\hrefCMSnoop {}{{\relax Yu}.~A. Golfand and E.~P. Likhtman, ``Extension of the
  algebra of {Poincar{\'e}} group generators and violation of p invariance'',}
  \textit{ JETP Lett.} \textbf{ 13} (1971) 323,
\href{http://dx.doi.org/10.1142/9789814542340_0001}{\doi{10.1142/9789814542340_0001}}.

\bibitem{Neveu:1971rx}
\hrefCMSnoop {}{A.~Neveu and J.~H. Schwarz, ``Factorizable dual model of
  pions'',} \textit{ Nucl. Phys. B} \textbf{ 31} (1971) 86,
\href{http://dx.doi.org/10.1016/0550-3213(71)90448-2}{\doi{10.1016/0550-3213(71)90448-2}}.

\bibitem{Wess:1973kz}
\hrefCMSnoop {}{J.~Wess and B.~Zumino, ``A {Lagrangian} model invariant under
  supergauge transformations'',} \textit{ Phys. Lett. B} \textbf{ 49} (1974)
  52,
\href{http://dx.doi.org/10.1016/0370-2693(74)90578-4}{\doi{10.1016/0370-2693(74)90578-4}}.

\bibitem{Fayet:1974pd}
\hrefCMSnoop {}{P.~Fayet, ``Supergauge invariant extension of the {Higgs}
  mechanism and a model for the electron and its neutrino'',} \textit{ Nucl.
  Phys. B} \textbf{ 90} (1975) 104,
\href{http://dx.doi.org/10.1016/0550-3213(75)90636-7}{\doi{10.1016/0550-3213(75)90636-7}}.

\bibitem{tHooft:1979rat}
\hrefCMSnoop {}{G.~'t~Hooft, ``Naturalness, chiral symmetry, and spontaneous
  chiral symmetry breaking'',} \textit{ NATO Sci. Ser. B} \textbf{ 59} (1980)
  135,
\href{http://dx.doi.org/10.1007/978-1-4684-7571-5_9}{\doi{10.1007/978-1-4684-7571-5_9}}.

\bibitem{Kaul:1981hi}
\hrefCMSnoop {}{R.~K. Kaul and P.~Majumdar, ``Cancellation of quadratically
  divergent mass corrections in globally supersymmetric spontaneously broken
  gauge theories'',} \textit{ Nucl. Phys. B} \textbf{ 199} (1982) 36,
\href{http://dx.doi.org/10.1016/0550-3213(82)90565-X}{\doi{10.1016/0550-3213(82)90565-X}}.

\bibitem{Nilles:1983ge}
\hrefCMSnoop {}{H.~P. Nilles, ``Supersymmetry, supergravity and particle
  physics'',} \textit{ Phys. Rept.} \textbf{ 110} (1984) 1,
\href{http://dx.doi.org/10.1016/0370-1573(84)90008-5}{\doi{10.1016/0370-1573(84)90008-5}}.

\bibitem{Martin:1997ns}
\hrefCMSnoop {}{S.~P. Martin, ``A supersymmetry primer'',} 1997.
\href{http://www.arXiv.org/abs/hep-ph/9709356}{\texttt{arXiv:hep-ph/9709356}}.

\bibitem{Farrar:1978xj}
\hrefCMSnoop {}{G.~R. Farrar and P.~Fayet, ``Phenomenology of the production,
  decay, and detection of new hadronic states associated with supersymmetry'',}
  \textit{ Phys. Lett. B} \textbf{ 76} (1978) 575,
\href{http://dx.doi.org/10.1016/0370-2693(78)90858-4}{\doi{10.1016/0370-2693(78)90858-4}}.

\bibitem{Witten:1981nf}
\hrefCMSnoop {}{E.~Witten, ``Dynamical breaking of supersymmetry'',} \textit{
  Nucl. Phys. B} \textbf{ 188} (1981) 513,
  \href{http://dx.doi.org/10.1016/0550-3213(81)90006-7}{\doi{10.1016/0550-3213(81)90006-7}}.

\bibitem{Dimopoulos:1981zb}
\hrefCMSnoop {}{S.~Dimopoulos and H.~Georgi, ``Softly broken supersymmetry and
  su(5)'',} \textit{ Nucl. Phys. B} \textbf{ 193} (1981) 150,
  \href{http://dx.doi.org/10.1016/0550-3213(81)90522-8}{\doi{10.1016/0550-3213(81)90522-8}}.

\bibitem{Sakai:1981gr}
\hrefCMSnoop {}{N.~Sakai, ``Naturalness in supersymmetric guts'',} \textit{ Z.
  Phys. C} \textbf{ 11} (1981) 153,
  \href{http://dx.doi.org/10.1007/BF01573998}{\doi{10.1007/BF01573998}}.

\bibitem{Hall:2011aa}
\hrefCMSnoop {}{L.~J. Hall, D.~Pinner, and J.~T. Ruderman, ``A natural {SUSY
  Higgs} near 126 {GeV}'',} \textit{ JHEP} \textbf{ 04} (2012) 131,
  \href{http://dx.doi.org/10.1007/JHEP04(2012)131}{\doi{10.1007/JHEP04(2012)131}},
  \href{http://www.arXiv.org/abs/1112.2703}{\texttt{arXiv:1112.2703}}.

\bibitem{Arbey:2011ab}
A.~Arbey\hrefCMSnoop {}{ {et~al.}, ``Implications of a 125 {GeV} {Higgs} for
  supersymmetric models'',} \textit{ Phys. Lett. B} \textbf{ 708} (2012) 162,
  \href{http://dx.doi.org/10.1016/j.physletb.2012.01.053}{\doi{10.1016/j.physletb.2012.01.053}},
  \href{http://www.arXiv.org/abs/1112.3028}{\texttt{arXiv:1112.3028}}.

\bibitem{Baer:1997yi}
H.~Baer\hrefCMSnoop {}{ {et~al.}, ``Collider phenomenology for supersymmetry
  with large tan$\beta$'',} \textit{ Phys. Rev. Lett.} \textbf{ 79} (1997) 986,
  \href{http://dx.doi.org/10.1103/PhysRevLett.79.986}{\doi{10.1103/PhysRevLett.79.986}},
  \href{http://www.arXiv.org/abs/hep-ph/9704457}{\texttt{arXiv:hep-ph/9704457}}.
[Erratum: \DOI{10.1103/PhysRevLett.80.642}].

\bibitem{Guchait:2002xh}
\hrefCMSnoop {}{M.~Guchait and D.~P. Roy, ``Using $\tau$ polarization as a
  distinctive {SUGRA} signature at {LHC}'',} \textit{ Phys. Lett. B} \textbf{
  541} (2002) 356,
  \href{http://dx.doi.org/10.1016/S0370-2693(02)02269-4}{\doi{10.1016/S0370-2693(02)02269-4}},
\href{http://www.arXiv.org/abs/hep-ph/0205015}{\texttt{arXiv:hep-ph/0205015}}.

\bibitem{Workman:2022ynf}
\hrefCMSnoop {}{{Particle Data Group}, ``Review of particle physics'',}
  \textit{ PTEP} \textbf{ 2022} (2022) 083C01,
  \href{http://dx.doi.org/10.1093/ptep/ptac097}{\doi{10.1093/ptep/ptac097}}.

\bibitem{Alwall:2008ag}
\hrefCMSnoop {}{J.~Alwall, P.~Schuster, and N.~Toro, ``Simplified models for a
  first characterization of new physics at the {LHC}'',} \textit{ Phys. Rev. D}
  \textbf{ 79} (2009) 075020,
  \href{http://dx.doi.org/10.1103/PhysRevD.79.075020}{\doi{10.1103/PhysRevD.79.075020}},
\href{http://www.arXiv.org/abs/0810.3921}{\texttt{arXiv:0810.3921}}.

\bibitem{Alves:2011wf}
\hrefCMSnoop {}{{LHC New Physics Working Group}, ``Simplified models for {LHC}
  new physics searches'',} \textit{ J. Phys. G} \textbf{ 39} (2012) 105005,
  \href{http://dx.doi.org/10.1088/0954-3899/39/10/105005}{\doi{10.1088/0954-3899/39/10/105005}},
\href{http://www.arXiv.org/abs/1105.2838}{\texttt{arXiv:1105.2838}}.

\bibitem{Sirunyan:2017xse}
\hrefCMSnoop {}{{CMS Collaboration}, ``Search for top squark pair production in
  pp collisions at $ \sqrt{s}=13 $ {TeV} using single lepton events'',}
  \textit{ JHEP} \textbf{ 10} (2017) 019,
  \href{http://dx.doi.org/10.1007/JHEP10(2017)019}{\doi{10.1007/JHEP10(2017)019}},
\href{http://www.arXiv.org/abs/1706.04402}{\texttt{arXiv:1706.04402}}.

\bibitem{Sirunyan:2017leh}
\hrefCMSnoop {}{{CMS Collaboration}, ``Search for top squarks and dark matter
  particles in opposite-charge dilepton final states at $\sqrt{s}=$ 13
  {TeV}'',} \textit{ Phys. Rev. D} \textbf{ 97} (2018) 032009,
  \href{http://dx.doi.org/10.1103/PhysRevD.97.032009}{\doi{10.1103/PhysRevD.97.032009}},
\href{http://www.arXiv.org/abs/1711.00752}{\texttt{arXiv:1711.00752}}.

\bibitem{Chatrchyan:2013xna}
\hrefCMSnoop {}{{CMS Collaboration}, ``Search for top-squark pair production in
  the single-lepton final state in pp collisions at $\sqrt{s}$ = 8 {TeV}'',}
  \textit{ Eur. Phys. J. C} \textbf{ 73} (2013) 2677,
  \href{http://dx.doi.org/10.1140/epjc/s10052-013-2677-2}{\doi{10.1140/epjc/s10052-013-2677-2}},
\href{http://www.arXiv.org/abs/1308.1586}{\texttt{arXiv:1308.1586}}.

\bibitem{Khachatryan:2016pup}
\hrefCMSnoop {}{{CMS Collaboration}, ``Search for direct pair production of
  scalar top quarks in the single- and dilepton channels in proton-proton
  collisions at $ \sqrt{s}=8 $ {TeV}'',} \textit{ JHEP} \textbf{ 07} (2016)
  027,
  \href{http://dx.doi.org/10.1007/JHEP07(2016)027}{\doi{10.1007/JHEP07(2016)027}},
  \href{http://www.arXiv.org/abs/1602.03169}{\texttt{arXiv:1602.03169}}.
[Erratum: \DOI{10.1007/JHEP09(2016)056}].

\bibitem{Khachatryan:2016pxa}
\hrefCMSnoop {}{{CMS Collaboration}, ``Search for top squark pair production in
  compressed-mass-spectrum scenarios in proton-proton collisions at $\sqrt{s}$
  = 8 {TeV} using the $\alpha_t$ variable'',} \textit{ Phys. Lett. B} \textbf{
  767} (2017) 403,
  \href{http://dx.doi.org/10.1016/j.physletb.2017.02.007}{\doi{10.1016/j.physletb.2017.02.007}},
\href{http://www.arXiv.org/abs/1605.08993}{\texttt{arXiv:1605.08993}}.

\bibitem{Sirunyan:2016jpr}
\hrefCMSnoop {}{{CMS Collaboration}, ``Searches for pair production of
  third-generation squarks in $\sqrt{s}=13$ {TeV} pp collisions'',} \textit{
  Eur. Phys. J. C} \textbf{ 77} (2017) 327,
  \href{http://dx.doi.org/10.1140/epjc/s10052-017-4853-2}{\doi{10.1140/epjc/s10052-017-4853-2}},
\href{http://www.arXiv.org/abs/1612.03877}{\texttt{arXiv:1612.03877}}.

\bibitem{Sirunyan:2017wif}
\hrefCMSnoop {}{{CMS Collaboration}, ``Search for direct production of
  supersymmetric partners of the top quark in the all-jets final state in
  proton-proton collisions at $ \sqrt{s}=13 $ {TeV}'',} \textit{ JHEP} \textbf{
  10} (2017) 005,
  \href{http://dx.doi.org/10.1007/JHEP10(2017)005}{\doi{10.1007/JHEP10(2017)005}},
\href{http://www.arXiv.org/abs/1707.03316}{\texttt{arXiv:1707.03316}}.

\bibitem{Sirunyan:2017pjw}
\hrefCMSnoop {}{{CMS Collaboration}, ``Search for supersymmetry in
  proton-proton collisions at 13 {TeV} using identified top quarks'',} \textit{
  Phys. Rev. D} \textbf{ 97} (2018) 012007,
  \href{http://dx.doi.org/10.1103/PhysRevD.97.012007}{\doi{10.1103/PhysRevD.97.012007}},
\href{http://www.arXiv.org/abs/1710.11188}{\texttt{arXiv:1710.11188}}.

\bibitem{Aaboud:2017nfd}
\hrefCMSnoop {}{{ATLAS Collaboration}, ``Search for direct top squark pair
  production in final states with two leptons in $\sqrt{s} = 13$ {TeV} pp
  collisions with the atlas detector'',} \textit{ Eur. Phys. J. C} \textbf{ 77}
  (2017) 898,
  \href{http://dx.doi.org/10.1140/epjc/s10052-017-5445-x}{\doi{10.1140/epjc/s10052-017-5445-x}},
\href{http://www.arXiv.org/abs/1708.03247}{\texttt{arXiv:1708.03247}}.

\bibitem{Aad:2015pfx}
\hrefCMSnoop {}{{ATLAS Collaboration}, ``{ATLAS} run 1 searches for direct pair
  production of third-generation squarks at the large hadron collider'',}
  \textit{ Eur. Phys. J. C} \textbf{ 75} (2015) 510,
  \href{http://dx.doi.org/10.1140/epjc/s10052-015-3726-9}{\doi{10.1140/epjc/s10052-015-3726-9}},
  \href{http://www.arXiv.org/abs/1506.08616}{\texttt{arXiv:1506.08616}}.
[Erratum: \DOI{10.1140/epjc/s10052-016-3935-x}].

\bibitem{Aad:2014kra}
\hrefCMSnoop {}{{ATLAS Collaboration}, ``Search for top squark pair production
  in final states with one isolated lepton, jets, and missing transverse
  momentum in $\sqrt s =$ 8 {TeV} pp collisions with the {ATLAS} detector'',}
  \textit{ JHEP} \textbf{ 11} (2014) 118,
  \href{http://dx.doi.org/10.1007/JHEP11(2014)118}{\doi{10.1007/JHEP11(2014)118}},
\href{http://www.arXiv.org/abs/1407.0583}{\texttt{arXiv:1407.0583}}.

\bibitem{Aad:2014qaa}
\hrefCMSnoop {}{{ATLAS Collaboration}, ``Search for direct top-squark pair
  production in final states with two leptons in pp collisions at $\sqrt{s} =$
  8 {TeV} with the {ATLAS} detector'',} \textit{ JHEP} \textbf{ 06} (2014) 124,
  \href{http://dx.doi.org/10.1007/JHEP06(2014)124}{\doi{10.1007/JHEP06(2014)124}},
\href{http://www.arXiv.org/abs/1403.4853}{\texttt{arXiv:1403.4853}}.

\bibitem{Aaboud:2016lwz}
\hrefCMSnoop {}{{ATLAS Collaboration}, ``Search for top squarks in final states
  with one isolated lepton, jets, and missing transverse momentum in
  $\sqrt{s}=13$ {TeV} pp collisions with the {ATLAS} detector'',} \textit{
  Phys. Rev. D} \textbf{ 94} (2016) 052009,
  \href{http://dx.doi.org/10.1103/PhysRevD.94.052009}{\doi{10.1103/PhysRevD.94.052009}},
\href{http://www.arXiv.org/abs/1606.03903}{\texttt{arXiv:1606.03903}}.

\bibitem{PhysRevD.98.032008}
\hrefCMSnoop {}{{ATLAS Collaboration}, ``Search for top squarks decaying to tau
  sleptons in pp collisions at $\sqrt{s}=13\text{ }\text{ }\text{Te}\mathrm{V}$
  with the {ATLAS} detector'',} \textit{ Phys. Rev. D} \textbf{ 98} (2018)
  032008,
  \href{http://dx.doi.org/10.1103/PhysRevD.98.032008}{\doi{10.1103/PhysRevD.98.032008}},
  \href{http://www.arXiv.org/abs/1803.10178}{\texttt{arXiv:1803.10178}}.

\bibitem{ATLAS:2021jyv}
\hrefCMSnoop {}{{ATLAS Collaboration}, ``Search for new phenomena in $pp$
  collisions in final states with tau leptons, b-jets, and missing transverse
  momentum with the {ATLAS} detector'',} \textit{ Phys. Rev. D} \textbf{ 104}
  (2021) 112005,
  \href{http://dx.doi.org/10.1103/PhysRevD.104.112005}{\doi{10.1103/PhysRevD.104.112005}},
  \href{http://www.arXiv.org/abs/2108.07665}{\texttt{arXiv:2108.07665}}.

\bibitem{CMS:2019lrh}
\hrefCMSnoop {}{{CMS Collaboration}, ``Search for top squark pair production in
  a final state with two tau leptons in proton-proton collisions at $ \sqrt{s}
  =$ 13 {TeV}'',} \textit{ JHEP} \textbf{ 02} (2020) 015,
  \href{http://dx.doi.org/10.1007/JHEP02(2020)015}{\doi{10.1007/JHEP02(2020)015}},
  \href{http://www.arXiv.org/abs/1910.12932}{\texttt{arXiv:1910.12932}}.

\bibitem{hepdata}
\hrefCMSnoop {}{}{{HEPD}ata record for this analysis}, 2023.
\newblock
  \href{http://dx.doi.org/10.17182/hepdata.138986}{\doi{10.17182/hepdata.138986}}.

\bibitem{Chatrchyan:2008zzk}
\hrefCMSnoop {}{{CMS Collaboration}, ``The {CMS} experiment at the {CERN}
  {LHC}'',} \textit{ JINST} \textbf{ 3} (2008) S08004,
  \href{http://dx.doi.org/10.1088/1748-0221/3/08/S08004}{\doi{10.1088/1748-0221/3/08/S08004}}.

\bibitem{Khachatryan:2016bia}
\hrefCMSnoop {}{{CMS Collaboration}, ``The {CMS} trigger system'',} \textit{
  JINST} \textbf{ 12} (2017) P01020,
  \href{http://dx.doi.org/10.1088/1748-0221/12/01/P01020}{\doi{10.1088/1748-0221/12/01/P01020}},
\href{http://www.arXiv.org/abs/1609.02366}{\texttt{arXiv:1609.02366}}.

\bibitem{CMS:2020cmk}
\hrefCMSnoop {}{{CMS Collaboration}, ``Performance of the {CMS} level-1 trigger
  in proton-proton collisions at $\sqrt{s} = 13$\,{TeV}'',} \textit{ JINST}
  \textbf{ 15} (2020) P10017,
  \href{http://dx.doi.org/10.1088/1748-0221/15/10/P10017}{\doi{10.1088/1748-0221/15/10/P10017}},
  \href{http://www.arXiv.org/abs/2006.10165}{\texttt{arXiv:2006.10165}}.

\bibitem{Oleari:2010nx}
\hrefCMSnoop {}{C.~Oleari, ``The {POWHEG-BOX}'',} \textit{ Nucl. Phys. B Proc.
  Suppl.} \textbf{ 205-206} (2010) 36,
  \href{http://dx.doi.org/10.1016/j.nuclphysbps.2010.08.016}{\doi{10.1016/j.nuclphysbps.2010.08.016}},
  \href{http://www.arXiv.org/abs/1007.3893}{\texttt{arXiv:1007.3893}}.

\bibitem{Nason:2004rx}
\hrefCMSnoop {}{P.~Nason, ``A new method for combining {NLO} {QCD} with shower
  {Monte Carlo} algorithms'',} \textit{ JHEP} \textbf{ 11} (2004) 040,
  \href{http://dx.doi.org/10.1088/1126-6708/2004/11/040}{\doi{10.1088/1126-6708/2004/11/040}},
\href{http://www.arXiv.org/abs/hep-ph/0409146}{\texttt{arXiv:hep-ph/0409146}}.

\bibitem{Frixione:2007vw}
\hrefCMSnoop {}{S.~Frixione, P.~Nason, and C.~Oleari, ``Matching {NLO} {QCD}
  computations with parton shower simulations: the {POWHEG} method'',} \textit{
  JHEP} \textbf{ 11} (2007) 070,
  \href{http://dx.doi.org/10.1088/1126-6708/2007/11/070}{\doi{10.1088/1126-6708/2007/11/070}},
\href{http://www.arXiv.org/abs/0709.2092}{\texttt{arXiv:0709.2092}}.

\bibitem{Alioli:2010xd}
\hrefCMSnoop {}{S.~Alioli, P.~Nason, C.~Oleari, and E.~Re, ``A general
  framework for implementing {NLO} calculations in shower {Monte Carlo}
  programs: the {POWHEG BOX}'',} \textit{ JHEP} \textbf{ 06} (2010) 043,
  \href{http://dx.doi.org/10.1007/JHEP06(2010)043}{\doi{10.1007/JHEP06(2010)043}},
\href{http://www.arXiv.org/abs/1002.2581}{\texttt{arXiv:1002.2581}}.

\bibitem{Frixione:2007nw}
\hrefCMSnoop {}{S.~Frixione, P.~Nason, and G.~Ridolfi, ``A positive-weight
  next-to-leading-order {Monte Carlo} for heavy flavour hadroproduction'',}
  \textit{ JHEP} \textbf{ 09} (2007) 126,
  \href{http://dx.doi.org/10.1088/1126-6708/2007/09/126}{\doi{10.1088/1126-6708/2007/09/126}},
\href{http://www.arXiv.org/abs/0707.3088}{\texttt{arXiv:0707.3088}}.

\bibitem{Alioli_2009}
\hrefCMSnoop {}{S.~Alioli, P.~Nason, C.~Oleari, and E.~Re, ``{NLO} single-top
  production matched with shower in {POWHEG}: $s$- and $t$-channel
  contributions'',} \textit{ JHEP} \textbf{ 09} (2009) 111,
  \href{http://dx.doi.org/10.1088/1126-6708/2009/09/111}{\doi{10.1088/1126-6708/2009/09/111}},
  \href{http://www.arXiv.org/abs/0907.4076}{\texttt{arXiv:0907.4076}}.
  [Erratum: JHEP 02, 011 (2010)].

\bibitem{Alwall:2014hca}
J.~Alwall\hrefCMSnoop {}{ {et~al.}, ``The automated computation of tree-level
  and next-to-leading order differential cross sections, and their matching to
  parton shower simulations'',} \textit{ JHEP} \textbf{ 07} (2014) 079,
  \href{http://dx.doi.org/10.1007/JHEP07(2014)079}{\doi{10.1007/JHEP07(2014)079}},
\href{http://www.arXiv.org/abs/1405.0301}{\texttt{arXiv:1405.0301}}.

\bibitem{Li:2012wna}
\hrefCMSnoop {}{Y.~Li and F.~Petriello, ``Combining {QCD} and electroweak
  corrections to dilepton production in {FEWZ}'',} \textit{ Phys. Rev. D}
  \textbf{ 86} (2012) 094034,
  \href{http://dx.doi.org/10.1103/PhysRevD.86.094034}{\doi{10.1103/PhysRevD.86.094034}},
  \href{http://www.arXiv.org/abs/1208.5967}{\texttt{arXiv:1208.5967}}.

\bibitem{Sjostrand:2014zea}
T.~Sj{\"o}strand\hrefCMSnoop {}{ {et~al.}, ``{An Introduction to PYTHIA
  8.2}'',} \textit{ Comput. Phys. Commun.} \textbf{ 191} (2015) 159,
  \href{http://dx.doi.org/10.1016/j.cpc.2015.01.024}{\doi{10.1016/j.cpc.2015.01.024}},
\href{http://www.arXiv.org/abs/1410.3012}{\texttt{arXiv:1410.3012}}.

\bibitem{CMS:2015wcf}
\hrefCMSnoop {}{{CMS Collaboration}, ``Event generator tunes obtained from
  underlying event and multiparton scattering measurements'',} \textit{ Eur.
  Phys. J. C} \textbf{ 76} (2016) 155,
  \href{http://dx.doi.org/10.1140/epjc/s10052-016-3988-x}{\doi{10.1140/epjc/s10052-016-3988-x}},
  \href{http://www.arXiv.org/abs/1512.00815}{\texttt{arXiv:1512.00815}}.

\bibitem{CMS-PAS-TOP-16-021}
\href {https://cds.cern.ch/record/2235192}{{CMS Collaboration},
  ``Investigations of the impact of the parton shower tuning in pythia8 in the
  modelling of $\mathrm{t\overline{t}}$ at $\sqrt{s}=8$ and 13 {TeV}'',} CMS
  Physics Analysis Summary CMS-PAS-TOP-16-021, 2016.

\bibitem{CMS:2019csb}
\hrefCMSnoop {}{{CMS Collaboration}, ``Extraction and validation of a new set
  of {CMS} pythia8 tunes from underlying-event measurements'',} \textit{ Eur.
  Phys. J. C} \textbf{ 80} (2020) 4,
  \href{http://dx.doi.org/10.1140/epjc/s10052-019-7499-4}{\doi{10.1140/epjc/s10052-019-7499-4}},
  \href{http://www.arXiv.org/abs/1903.12179}{\texttt{arXiv:1903.12179}}.

\bibitem{GEANT4:2002zbu}
\hrefCMSnoop {}{{GEANT4} Collaboration, ``{\GEANTfour} -- a simulation
  toolkit'',} \textit{ Nucl. Instrum. Meth. A} \textbf{ 506} (2003) 250,
  \href{http://dx.doi.org/10.1016/S0168-9002(03)01368-8}{\doi{10.1016/S0168-9002(03)01368-8}}.

\bibitem{bib-nlo-nll-01}
\hrefCMSnoop {}{W.~Beenakker, R.~H{\"o}pker, M.~Spira, and P.~M. Zerwas,
  ``Squark and gluino production at hadron colliders'',} \textit{ Nucl. Phys.
  B} \textbf{ 492} (1997) 51,
  \href{http://dx.doi.org/10.1016/S0550-3213(97)00084-9}{\doi{10.1016/S0550-3213(97)00084-9}},
\href{http://www.arXiv.org/abs/hep-ph/9610490}{\texttt{arXiv:hep-ph/9610490}}.

\bibitem{bib-nlo-nll-02}
\hrefCMSnoop {}{A.~Kulesza and L.~Motyka, ``Threshold resummation for
  squark-antisquark and gluino-pair production at the {LHC}'',} \textit{ Phys.
  Rev. Lett.} \textbf{ 102} (2009) 111802,
  \href{http://dx.doi.org/10.1103/PhysRevLett.102.111802}{\doi{10.1103/PhysRevLett.102.111802}},
\href{http://www.arXiv.org/abs/0807.2405}{\texttt{arXiv:0807.2405}}.

\bibitem{bib-nlo-nll-03}
\hrefCMSnoop {}{A.~Kulesza and L.~Motyka, ``Soft gluon resummation for the
  production of gluino-gluino and squark-antisquark pairs at the {LHC}'',}
  \textit{ Phys. Rev. D} \textbf{ 80} (2009) 095004,
  \href{http://dx.doi.org/10.1103/PhysRevD.80.095004}{\doi{10.1103/PhysRevD.80.095004}},
\href{http://www.arXiv.org/abs/0905.4749}{\texttt{arXiv:0905.4749}}.

\bibitem{bib-nlo-nll-04}
W.~Beenakker\hrefCMSnoop {}{ {et~al.}, ``Soft-gluon resummation for squark and
  gluino hadroproduction'',} \textit{ JHEP} \textbf{ 12} (2009) 041,
  \href{http://dx.doi.org/10.1088/1126-6708/2009/12/041}{\doi{10.1088/1126-6708/2009/12/041}},
\href{http://www.arXiv.org/abs/0909.4418}{\texttt{arXiv:0909.4418}}.

\bibitem{bib-nlo-nll-05}
W.~Beenakker\hrefCMSnoop {}{ {et~al.}, ``Squark and gluino hadroproduction'',}
  \textit{ Int. J. Mod. Phys. A} \textbf{ 26} (2011) 2637,
  \href{http://dx.doi.org/10.1142/S0217751X11053560}{\doi{10.1142/S0217751X11053560}},
\href{http://www.arXiv.org/abs/1105.1110}{\texttt{arXiv:1105.1110}}.

\bibitem{Giammanco:2014bza}
\hrefCMSnoop {}{A.~Giammanco, ``The fast simulation of the {CMS} experiment'',}
  \textit{ J. Phys. Conf. Ser.} \textbf{ 513} (2014) 022012,
\href{http://dx.doi.org/10.1088/1742-6596/513/2/022012}{\doi{10.1088/1742-6596/513/2/022012}}.

\bibitem{CMS-PRF-14-001}
\hrefCMSnoop {}{{CMS Collaboration}, ``Particle-flow reconstruction and global
  event description with the {CMS} detector'',} \textit{ JINST} \textbf{ 12}
  (2017) P10003,
  \href{http://dx.doi.org/10.1088/1748-0221/12/10/P10003}{\doi{10.1088/1748-0221/12/10/P10003}},
\href{http://www.arXiv.org/abs/1706.04965}{\texttt{arXiv:1706.04965}}.

\bibitem{CMS-TDR-15-02}
\href {http://cds.cern.ch/record/2020886}{{CMS Collaboration}, ``Technical
  proposal for the phase-ii upgrade of the compact muon solenoid'',} CMS
  Technical proposal CERN-LHCC-2015-010, CMS-TDR-15-02, 2015.

\bibitem{Cacciari:2008gp}
\hrefCMSnoop {}{M.~Cacciari, G.~P. Salam, and G.~Soyez, ``The anti-\kt jet
  clustering algorithm'',} \textit{ JHEP} \textbf{ 04} (2008) 063,
  \href{http://dx.doi.org/10.1088/1126-6708/2008/04/063}{\doi{10.1088/1126-6708/2008/04/063}},
  \href{http://www.arXiv.org/abs/0802.1189}{\texttt{arXiv:0802.1189}}.

\bibitem{Cacciari:2011ma}
\hrefCMSnoop {}{M.~Cacciari, G.~P. Salam, and G.~Soyez, ``{FastJet} user
  manual'',} \textit{ Eur. Phys. J. C} \textbf{ 72} (2012) 1896,
  \href{http://dx.doi.org/10.1140/epjc/s10052-012-1896-2}{\doi{10.1140/epjc/s10052-012-1896-2}},
\href{http://www.arXiv.org/abs/1111.6097}{\texttt{arXiv:1111.6097}}.

\bibitem{Khachatryan:2016kdb}
\hrefCMSnoop {}{{CMS Collaboration}, ``Jet energy scale and resolution in the
  {CMS} experiment in pp collisions at 8 {TeV}'',} \textit{ JINST} \textbf{ 12}
  (2017) P02014,
  \href{http://dx.doi.org/10.1088/1748-0221/12/02/P02014}{\doi{10.1088/1748-0221/12/02/P02014}},
\href{http://www.arXiv.org/abs/1607.03663}{\texttt{arXiv:1607.03663}}.

\bibitem{CMS-PAS-JME-16-003}
\href {https://cds.cern.ch/record/2256875}{{CMS Collaboration}, ``Jet
  algorithms performance in 13 {TeV} data'',} {CMS Physics Analysis Summary}
  CMS-PAS-JME-16-003, 2017.

\bibitem{CMS:2017wtu}
\hrefCMSnoop {}{{CMS Collaboration}, ``Identification of heavy-flavour jets
  with the {CMS} detector in pp collisions at 13 {TeV}'',} \textit{ JINST}
  \textbf{ 13} (2018) P05011,
  \href{http://dx.doi.org/10.1088/1748-0221/13/05/P05011}{\doi{10.1088/1748-0221/13/05/P05011}},
  \href{http://www.arXiv.org/abs/1712.07158}{\texttt{arXiv:1712.07158}}.

\bibitem{CMS-DP-2018-058}
\href {https://cds.cern.ch/record/2646773}{{CMS Collaboration}, ``{Performance
  of the DeepJet b tagging algorithm using 41.9/fb of data from proton-proton
  collisions at 13 TeV with Phase-I CMS detector}'',} CMS Detector Performance
  Note CMS-DP-2018-058, 2018.

\bibitem{Bols:2020bkb}
E.~Bols\hrefCMSnoop {}{ {et~al.}, ``{Jet flavour classification using
  DeepJet}'',} \textit{ JINST} \textbf{ 15} (2020) P12012,
  \href{http://dx.doi.org/10.1088/1748-0221/15/12/P12012}{\doi{10.1088/1748-0221/15/12/P12012}},
  \href{http://www.arXiv.org/abs/2008.10519}{\texttt{arXiv:2008.10519}}.

\bibitem{CMS:2020uim}
\hrefCMSnoop {}{{CMS Collaboration}, ``Electron and photon reconstruction and
  identification with the {CMS} experiment at the {CERN} {LHC}'',} \textit{
  JINST} \textbf{ 16} (2021) P05014,
  \href{http://dx.doi.org/10.1088/1748-0221/16/05/P05014}{\doi{10.1088/1748-0221/16/05/P05014}},
  \href{http://www.arXiv.org/abs/2012.06888}{\texttt{arXiv:2012.06888}}.

\bibitem{Khachatryan:2015hwa}
\hrefCMSnoop {}{{CMS Collaboration}, ``Performance of electron reconstruction
  and selection with the {CMS} detector in proton-proton collisions at
  $\sqrt{s} = 8$ {TeV}'',} \textit{ JINST} \textbf{ 10} (2015) P06005,
  \href{http://dx.doi.org/10.1088/1748-0221/10/06/P06005}{\doi{10.1088/1748-0221/10/06/P06005}},
\href{http://www.arXiv.org/abs/1502.02701}{\texttt{arXiv:1502.02701}}.

\bibitem{Sirunyan:2018_1804.04528}
\hrefCMSnoop {}{{CMS Collaboration}, ``Performance of the {CMS} muon detector
  and muon reconstruction with proton-proton collisions at $\sqrt{s} = 13$
  {TeV}'',} \textit{ JINST} \textbf{ 13} (2018) P06015,
  \href{http://dx.doi.org/10.1088/1748-0221/13/06/P06015}{\doi{10.1088/1748-0221/13/06/P06015}},
\href{http://www.arXiv.org/abs/1804.04528}{\texttt{arXiv:1804.04528}}.

\bibitem{Sirunyan:CMS-TAU-16-003}
\hrefCMSnoop {}{{CMS Collaboration}, ``Performance of reconstruction and
  identification of $\uptau$ leptons decaying to hadrons and v$_\uptau$ in pp
  collisions at $\sqrt{s}=13$ {TeV}'',} \textit{ JINST} \textbf{ 13} (2018)
  P10005,
  \href{http://dx.doi.org/10.1088/1748-0221/13/10/p10005}{\doi{10.1088/1748-0221/13/10/p10005}},
  \href{http://www.arXiv.org/abs/1809.02816}{\texttt{arXiv:1809.02816}}.

\bibitem{CMS:2022prd}
\hrefCMSnoop {}{{CMS Collaboration}, ``Identification of hadronic tau lepton
  decays using a deep neural network'',} \textit{ JINST} \textbf{ 17} (2022)
  P07023,
  \href{http://dx.doi.org/10.1088/1748-0221/17/07/P07023}{\doi{10.1088/1748-0221/17/07/P07023}},
  \href{http://www.arXiv.org/abs/2201.08458}{\texttt{arXiv:2201.08458}}.

\bibitem{Sirunyan:2019kia}
\hrefCMSnoop {}{{CMS Collaboration}, ``Performance of missing transverse
  momentum reconstruction in proton-proton collisions at $\sqrt{s} =$ 13 {TeV}
  using the {CMS} detector'',} \textit{ JINST} \textbf{ 14} (2019) P07004,
  \href{http://dx.doi.org/10.1088/1748-0221/14/07/P07004}{\doi{10.1088/1748-0221/14/07/P07004}},
\href{http://www.arXiv.org/abs/1903.06078}{\texttt{arXiv:1903.06078}}.

\bibitem{Lester:1999tx}
\hrefCMSnoop {}{C.~G. Lester and D.~J. Summers, ``Measuring masses of
  semiinvisibly decaying particles pair produced at hadron colliders'',}
  \textit{ Phys. Lett. B} \textbf{ 463} (1999) 99,
  \href{http://dx.doi.org/10.1016/S0370-2693(99)00945-4}{\doi{10.1016/S0370-2693(99)00945-4}},
  \href{http://www.arXiv.org/abs/hep-ph/9906349}{\texttt{arXiv:hep-ph/9906349}}.

\bibitem{Barr:2003rg}
\hrefCMSnoop {}{A.~Barr, C.~Lester, and P.~Stephens, ``{\mTii}: The truth
  behind the glamour'',} \textit{ J. Phys. G} \textbf{ 29} (2003) 2343,
  \href{http://dx.doi.org/10.1088/0954-3899/29/10/304}{\doi{10.1088/0954-3899/29/10/304}},
\href{http://www.arXiv.org/abs/hep-ph/0304226}{\texttt{arXiv:hep-ph/0304226}}.

\bibitem{Barr:2009wu}
\hrefCMSnoop {}{A.~J. Barr and C.~Gwenlan, ``The race for supersymmetry: Using
  {\mTii} for discovery'',} \textit{ Phys. Rev. D} \textbf{ 80} (2009) 074007,
  \href{http://dx.doi.org/10.1103/PhysRevD.80.074007}{\doi{10.1103/PhysRevD.80.074007}},
\href{http://www.arXiv.org/abs/0907.2713}{\texttt{arXiv:0907.2713}}.

\bibitem{CMS:2019eln}
\hrefCMSnoop {}{{CMS Collaboration}, ``Search for direct pair production of
  supersymmetric partners to the $\tau$ lepton in proton-proton collisions at
  $\sqrt{s}=$ 13 {TeV}'',} \textit{ Eur. Phys. J. C} \textbf{ 80} (2020) 189,
  \href{http://dx.doi.org/10.1140/epjc/s10052-020-7739-7}{\doi{10.1140/epjc/s10052-020-7739-7}},
  \href{http://www.arXiv.org/abs/1907.13179}{\texttt{arXiv:1907.13179}}.

\bibitem{Sirunyan:2645851}
\hrefCMSnoop {}{{CMS Collaboration}, ``Search for heavy neutrinos and
  third-generation leptoquarks in hadronic states of two $\tau$ leptons and two
  jets in proton-proton collisions at $\sqrt{s} =$ 13 {TeV}'',} \textit{ JHEP}
  \textbf{ 03} (2019) 170,
  \href{http://dx.doi.org/10.1007/JHEP03(2019)170}{\doi{10.1007/JHEP03(2019)170}},
\href{http://www.arXiv.org/abs/1811.00806}{\texttt{arXiv:1811.00806}}.

\bibitem{Sirunyan:2628769}
\hrefCMSnoop {}{{CMS Collaboration}, ``Search for supersymmetry in events with
  a $\tau$ lepton pair and missing transverse momentum in proton-proton
  collisions at $\sqrt{s} =$ 13 {TeV}'',} \textit{ JHEP} \textbf{ 11} (2018)
  151,
  \href{http://dx.doi.org/10.1007/JHEP11(2018)151}{\doi{10.1007/JHEP11(2018)151}},
\href{http://www.arXiv.org/abs/1807.02048}{\texttt{arXiv:1807.02048}}.

\bibitem{CMS-PAS-SUS-21-001}
\href {http://cds.cern.ch/record/2777046}{{CMS Collaboration}, ``Search for
  direct pair production of supersymmetric partners to the $\tau$ lepton in the
  all-hadronic final state at $\sqrt{s}=13~\mathrm{TeV}$'',} technical report,
  2021.

\bibitem{Sirunyan:2018nqx}
\hrefCMSnoop {}{{CMS Collaboration}, ``Measurement of the inelastic
  proton-proton cross section at $ \sqrt{s}=13 $ {TeV}'',} \textit{ JHEP}
  \textbf{ 07} (2018) 161,
  \href{http://dx.doi.org/10.1007/JHEP07(2018)161}{\doi{10.1007/JHEP07(2018)161}},
\href{http://www.arXiv.org/abs/1802.02613}{\texttt{arXiv:1802.02613}}.

\bibitem{Kalogeropoulos:2018cke}
\hrefCMSnoop {}{A.~Kalogeropoulos and J.~Alwall, ``The {SysCalc} code: A tool
  to derive theoretical systematic uncertainties'',} 2018.
  \href{http://www.arXiv.org/abs/1801.08401}{\texttt{arXiv:1801.08401}}.

\bibitem{CMS-LUM-17-003}
\hrefCMSnoop {}{{CMS Collaboration}, ``Precision luminosity measurement in
  proton-proton collisions at $\sqrt{s} =$ 13 {TeV} in 2015 and 2016 at
  {CMS}'',} \textit{ Eur. Phys. J. C} \textbf{ 81} (2021) 800,
  \href{http://dx.doi.org/10.1140/epjc/s10052-021-09538-2}{\doi{10.1140/epjc/s10052-021-09538-2}},
  \href{http://www.arXiv.org/abs/2104.01927}{\texttt{arXiv:2104.01927}}.

\bibitem{CMS-PAS-LUM-17-004}
\href {https://cds.cern.ch/record/2621960}{{CMS Collaboration}, ``{CMS}
  luminosity measurement for the 2017 data-taking period at $\sqrt{s} =
  13~\mathrm{TeV}$'',} CMS Physics Analysis Summary CMS-PAS-LUM-17-004, 2018.

\bibitem{CMS-PAS-LUM-18-002}
\href {http://cds.cern.ch/record/2676164}{{CMS Collaboration}, ``{CMS}
  luminosity measurement for the 2018 data-taking period at $\sqrt{s} =
  13~\mathrm{TeV}$'',} CMS Physics Analysis Summary CMS-PAS-LUM-18-002, 2019.

\bibitem{Sirunyan:2018owv}
\hrefCMSnoop {}{{CMS Collaboration}, ``Measurement of the differential
  {Drell-Yan} cross section in proton-proton collisions at $ \sqrt{\mathrm{s}}
  $ = 13 {TeV}'',} \textit{ JHEP} \textbf{ 12} (2019) 059,
  \href{http://dx.doi.org/10.1007/JHEP12(2019)059}{\doi{10.1007/JHEP12(2019)059}},
  \href{http://www.arXiv.org/abs/1812.10529}{\texttt{arXiv:1812.10529}}.

\bibitem{Sirunyan:2018ucr}
\hrefCMSnoop {}{{CMS Collaboration}, ``Measurements of $\mathrm{t\overline{t}}$
  differential cross sections in proton-proton collisions at $\sqrt{s}=$ 13
  {TeV} using events containing two leptons'',} \textit{ JHEP} \textbf{ 02}
  (2019) 149,
  \href{http://dx.doi.org/10.1007/JHEP02(2019)149}{\doi{10.1007/JHEP02(2019)149}},
\href{http://www.arXiv.org/abs/1811.06625}{\texttt{arXiv:1811.06625}}.

\bibitem{Aaboud:2017qkn}
\hrefCMSnoop {}{{ATLAS Collaboration}, ``Measurement of the {$\mathrm{W^+W^-}$}
  production cross section in pp collisions at a centre-of-mass energy of
  $\sqrt{s}$ = 13 {TeV} with the {ATLAS} experiment'',} \textit{ Phys. Lett. B}
  \textbf{ 773} (2017) 354,
  \href{http://dx.doi.org/10.1016/j.physletb.2017.08.047}{\doi{10.1016/j.physletb.2017.08.047}},
\href{http://www.arXiv.org/abs/1702.04519}{\texttt{arXiv:1702.04519}}.

\bibitem{CMS:2019too}
\hrefCMSnoop {}{{CMS Collaboration}, ``Measurement of top quark pair production
  in association with a {Z} boson in proton-proton collisions at $\sqrt{s}=$ 13
  {TeV}'',} \textit{ JHEP} \textbf{ 03} (2020) 056,
  \href{http://dx.doi.org/10.1007/JHEP03(2020)056}{\doi{10.1007/JHEP03(2020)056}},
  \href{http://www.arXiv.org/abs/1907.11270}{\texttt{arXiv:1907.11270}}.

\bibitem{Sirunyan:2019bez}
\hrefCMSnoop {}{{CMS Collaboration}, ``Measurements of the pp $\to$ {WZ}
  inclusive and differential production cross section and constraints on
  charged anomalous triple gauge couplings at $\sqrt{s} =$ 13 {TeV}'',}
  \textit{ JHEP} \textbf{ 04} (2019) 122,
  \href{http://dx.doi.org/10.1007/JHEP04(2019)122}{\doi{10.1007/JHEP04(2019)122}},
\href{http://www.arXiv.org/abs/1901.03428}{\texttt{arXiv:1901.03428}}.

\bibitem{Sirunyan:2017wgx}
\hrefCMSnoop {}{{CMS Collaboration}, ``Measurement of the differential cross
  sections for the associated production of a $w$ boson and jets in
  proton-proton collisions at $\sqrt{s}=13$ {TeV}'',} \textit{ Phys. Rev. D}
  \textbf{ 96} (2017) 072005,
  \href{http://dx.doi.org/10.1103/PhysRevD.96.072005}{\doi{10.1103/PhysRevD.96.072005}},
\href{http://www.arXiv.org/abs/1707.05979}{\texttt{arXiv:1707.05979}}.

\bibitem{CMS-NOTE-2011-005}
\href {https://cds.cern.ch/record/1379837}{{ATLAS and CMS Collaborations, and
  the LHC Higgs Combination Group}, ``Procedure for the {LHC} {Higgs} boson
  search combination in summer 2011'',} Technical Report CMS-NOTE-2011-005,
  ATL-PHYS-PUB-2011-11, 2011.

\bibitem{JUNK1999435}
\hrefCMSnoop {}{T.~Junk, ``Confidence level computation for combining searches
  with small statistics'',} \textit{ Nucl. Instrum. Meth. A} \textbf{ 434}
  (1999) 435,
  \href{http://dx.doi.org/10.1016/S0168-9002(99)00498-2}{\doi{10.1016/S0168-9002(99)00498-2}},
  \href{http://www.arXiv.org/abs/hep-ex/9902006}{\texttt{arXiv:hep-ex/9902006}}.

\bibitem{Read_2002}
\hrefCMSnoop {}{A.~L. Read, ``Presentation of search results: the {\CLs}
  technique'',} \textit{ J. Phys. G} \textbf{ 28} (2002) 2693,
  \href{http://dx.doi.org/10.1088/0954-3899/28/10/313}{\doi{10.1088/0954-3899/28/10/313}}.

\bibitem{Cowan2011}
\hrefCMSnoop {}{G.~Cowan, K.~Cranmer, E.~Gross, and O.~Vitells, ``Asymptotic
  formulae for likelihood-based tests of new physics'',} \textit{ Eur. Phys. J.
  C} \textbf{ 71} (2011) 1554,
  \href{http://dx.doi.org/10.1140/epjc/s10052-011-1554-0}{\doi{10.1140/epjc/s10052-011-1554-0}},
  \href{http://www.arXiv.org/abs/1007.1727}{\texttt{arXiv:1007.1727}}.
[Erratum: \DOI{10.1140/epjc/s10052-013-2501-z}].

\end{thebibliography}\endgroup

\cleardoublepage \appendix\section{The CMS Collaboration \label{app:collab}}\begin{sloppypar}\hyphenpenalty=5000\widowpenalty=500\clubpenalty=5000
\cmsinstitute{Yerevan Physics Institute, Yerevan, Armenia}
{\tolerance=6000
A.~Tumasyan\cmsAuthorMark{1}\cmsorcid{0009-0000-0684-6742}
\par}
\cmsinstitute{Institut f\"{u}r Hochenergiephysik, Vienna, Austria}
{\tolerance=6000
W.~Adam\cmsorcid{0000-0001-9099-4341}, J.W.~Andrejkovic, T.~Bergauer\cmsorcid{0000-0002-5786-0293}, S.~Chatterjee\cmsorcid{0000-0003-2660-0349}, K.~Damanakis\cmsorcid{0000-0001-5389-2872}, M.~Dragicevic\cmsorcid{0000-0003-1967-6783}, A.~Escalante~Del~Valle\cmsorcid{0000-0002-9702-6359}, P.S.~Hussain\cmsorcid{0000-0002-4825-5278}, M.~Jeitler\cmsAuthorMark{2}\cmsorcid{0000-0002-5141-9560}, N.~Krammer\cmsorcid{0000-0002-0548-0985}, L.~Lechner\cmsorcid{0000-0002-3065-1141}, D.~Liko\cmsorcid{0000-0002-3380-473X}, I.~Mikulec\cmsorcid{0000-0003-0385-2746}, P.~Paulitsch, J.~Schieck\cmsAuthorMark{2}\cmsorcid{0000-0002-1058-8093}, R.~Sch\"{o}fbeck\cmsorcid{0000-0002-2332-8784}, D.~Schwarz\cmsorcid{0000-0002-3821-7331}, M.~Sonawane\cmsorcid{0000-0003-0510-7010}, S.~Templ\cmsorcid{0000-0003-3137-5692}, W.~Waltenberger\cmsorcid{0000-0002-6215-7228}, C.-E.~Wulz\cmsAuthorMark{2}\cmsorcid{0000-0001-9226-5812}
% ...
\par}
\vskip\cmsinstskip
\dag:~Deceased\\
$^{1}$Also at Yerevan State University, Yerevan, Armenia\\
$^{2}$Also at TU Wien, Vienna, Austria\\
% ...
\end{sloppypar}
\end{document}